\documentclass{statsoc}
\usepackage{mathrsfs}
\usepackage{graphicx,psfrag,epsf}
\usepackage{enumerate}
\usepackage{natbib}
\usepackage{comment}
\usepackage{booktabs}
\usepackage{multirow}
\usepackage{xcolor}
\usepackage{calligra}
\usepackage{tikz}
\usepackage{adjustbox}
\usepackage{amsbsy}
\newcommand{\blind}{0}
\usepackage{anyfontsize}

\newcommand{\E}{\mathbb{E}}
\newcommand{\Var}{\mathbb{V}\textrm{ar}}

\newcommand{\mb}[1]{\boldsymbol{#1}}

\newtheorem{remark}{Remark}

\usepackage[normalem]{ulem}

\DeclareFontFamily{U}{BOONDOX-calo}{\skewchar\font=45 }
\DeclareFontShape{U}{BOONDOX-calo}{m}{n}{
  <-> s*[1.05] BOONDOX-r-calo}{}
\DeclareFontShape{U}{BOONDOX-calo}{b}{n}{
  <-> s*[1.05] BOONDOX-b-calo}{}
\DeclareMathAlphabet{\mathcalboondox}{U}{BOONDOX-calo}{m}{n}
\SetMathAlphabet{\mathcalboondox}{bold}{U}{BOONDOX-calo}{b}{n}
\DeclareMathAlphabet{\mathbcalboondox}{U}{BOONDOX-calo}{b}{n}

\usepackage[a4paper]{geometry}
\usepackage{graphicx}
\usepackage[textwidth=8em,textsize=small]{todonotes}
\usepackage{amsmath}
\usepackage{natbib}
\usepackage{url}

\title[A functional regression model for heterogeneous BioGeoChemical Argo data]{A functional regression model for heterogeneous BioGeoChemical Argo data in the Southern Ocean}

\author[Moritz Korte-Stapff {\it et al.}]{Moritz Korte-Stapff, Stilian Stoev, Tailen Hsing}
\address{University of Michigan,
Ann Arbor, MI
United States America.}
\email{kortest@umich.edu, sstoev@umich.edu, thsing@umich.edu}
\author[Moritz Korte-Stapff {\it et al.}]{Drew Yarger}
\address{Purdue University,
150 N. University Street,
West Lafayette, IN, United States of America, 47907}
\email{anyarger@purdue.edu}

\DeclareMathOperator*{\argmax}{arg\,max}

\usepackage{amssymb}
\renewcommand{\thetable}{\arabic{table}}
\begin{document}

\begin{abstract}

Leveraging available measurements of our environment can help us understand complex processes. 
One example is Argo Biogeochemical data, which aims to collect measurements of oxygen, nitrate, pH, and other variables at varying depths in the ocean.
We focus on the oxygen data in the Southern Ocean, which has implications for ocean biology and the Earth's carbon cycle. 
Systematic monitoring of such data has only recently begun to be established, and the data is sparse.
In contrast, Argo measurements of temperature and salinity are much more abundant. 
In this work, we introduce and estimate a functional regression model describing dependence in oxygen, temperature, and salinity data at all depths covered by the Argo data simultaneously.
Our model elucidates important aspects of the joint distribution of temperature, salinity, and oxygen.
Due to fronts that establish distinct spatial zones in the Southern Ocean, we augment this functional regression model with a mixture component. 
By modelling spatial dependence in the mixture component and in the data itself, we provide predictions onto a grid and improve location estimates of fronts.
Our approach is scalable to the size of the Argo data, and we demonstrate its success in cross-validation and a comprehensive interpretation of the model.

\end{abstract}

\noindent%
{\it Keywords:}  %3 to 6 keywords, that do not appear in the title
clustering, functional regression, penalized splines, physical and biogeochemical oceanography, spatially-dependent data
\vfill

\newpage

\section{Introduction}

% \begin{table}
%     \centering
%     \caption{Caption1}
%     \begin{tabular}{c|c}
%          &  \\
%          & 
%     \end{tabular}
%         \label{tab:my_label}
% \end{table}

% \begin{table}
%     \centering
%     \caption{Caption2}
%     \begin{tabular}{c|c}
%          &  \\
%          & 
%     \end{tabular}
%     \label{tab:my_label2}
% \end{table}

% \ref{tab:my_label2}

\begin{figure}[ht]
    \centering
        \includegraphics[width = .705\textwidth]{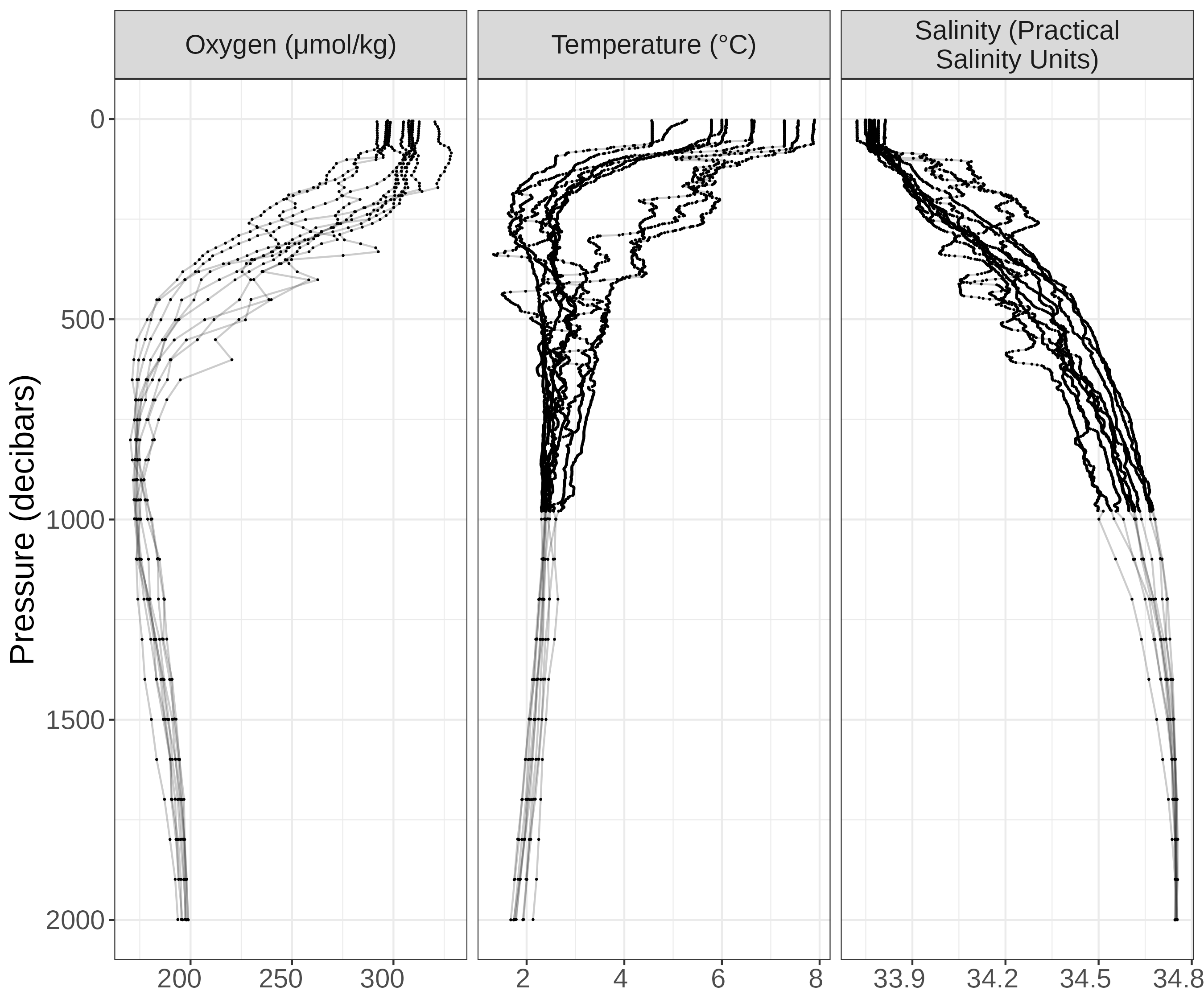}
    \includegraphics[width = .275\textwidth]{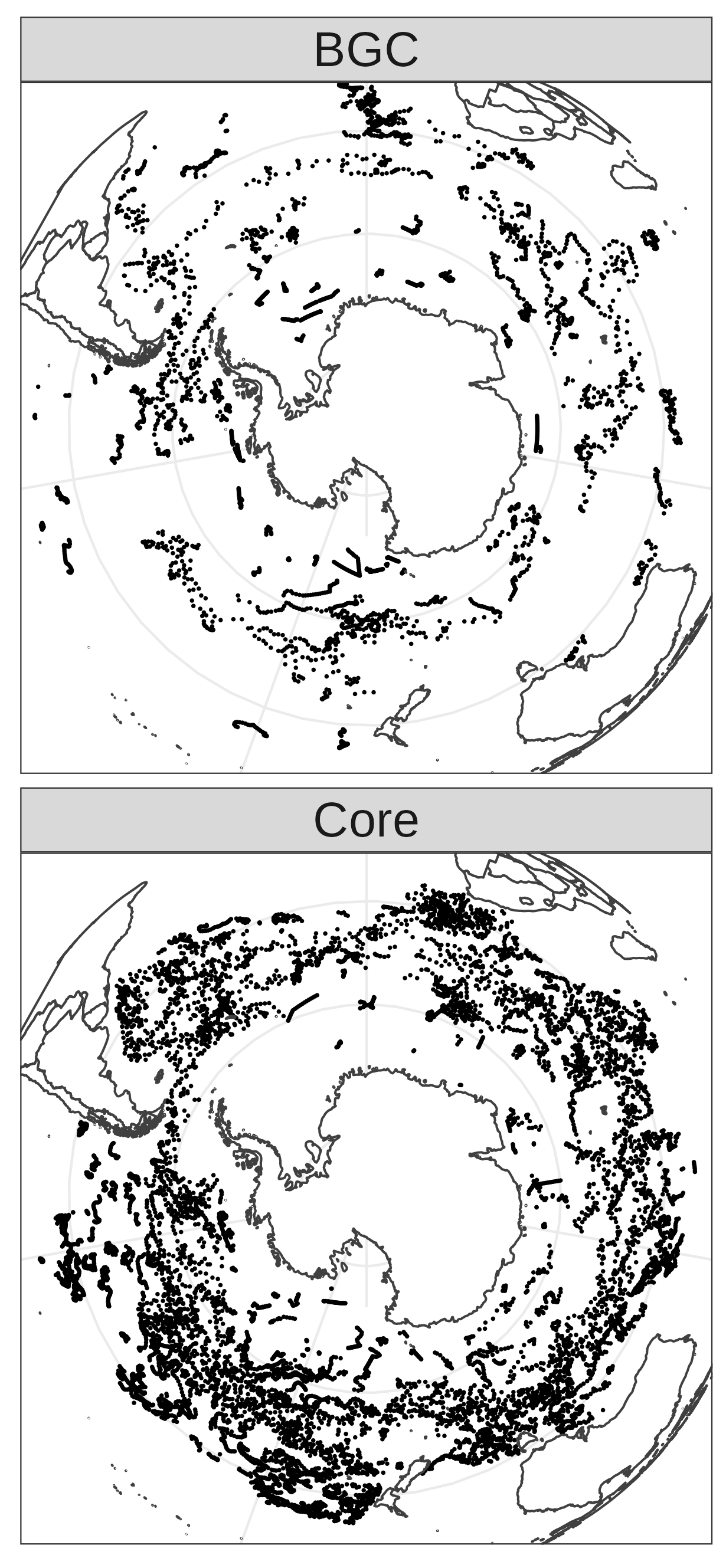}
    \caption{(Left) Oxygen, temperature, and salinity data of ten profiles collected from a BGC Argo float (WMO \#5904679) as a function of pressure (1 decibar of pressure is approximately 1 meter of depth). (Top Right) Locations of BGC profiles collected in the Southern Ocean in 2020 by the Southern Ocean Carbon and Climate Observations and Modeling (SOCCOM) Project and (Bottom Right) locations of nearby Core Argo profiles collected in 2020. }
    \label{fig:argo_data}
\end{figure}

The rapid growth of high-resolution data motivates continued advances in flexible and efficient statistical methodology. 
One prime example is the Argo data, which is collected by autonomous devices drifting through the world oceans called \emph{floats} that measure temperature and salinity.
Argo floats collect their data in so-called \emph{profiles}, collections of measurements at various depths for fixed points in time and space \citep[see Figure~\ref{fig:argo_data} and][]{argo2020}.
The Argo data has greatly enhanced our understanding of the physical nature of the upper 2000 meters of the oceans with near-uniform spatial sampling of these variables \citep{johnson2022argo}.
Argo floats determine their own depth through pressure measurements as an increase in 1 decibar in pressure occurs for each additional meter of depth (approximately). 
From here on, we will thus often use pressure rather than depth to refer to a float's vertical position in the ocean.

An important expansion of the Argo Project's goals is Biogeochemical (BGC) Argo, which has added a limited number of floats that also measure oxygen and nitrate, among other variables. 
We focus on Argo data in the Southern Ocean because the majority of the BGC Argo data has been collected under the Southern Ocean Carbon and Climate Observations and Modeling (SOCCOM) project \citep[][\url{soccom.princeton.edu}]{riser_soccom_2023}. 
The increased availability of biogeochemical data is important to improve our understanding of vital biogeochemical processes such as the biological carbon pump and air-sea CO$_2$ exchanges, monitor changes such as ocean deoxygenation and acidification, and improve estimates of the carbon budget \citep{bittig_bgc-argo_2019, claustre_observing_2020, gray_autonomous_2018}.
However, BGC floats use additional sensors, leading to higher fixed and maintenance costs than Core Argo floats (that is, those that measure only temperature and salinity). 
Thus, it is not practical to introduce the same coverage as achieved by Core Argo floats. Indeed, in the Southern Ocean during 2020, Core Argo floats collected approximately 12000 profiles while BGC Argo floats collected just 3600 over the same time period (see Figure~\ref{fig:argo_data}). 
When modelling biogeochemical data, it is thus desirable to utilize the more prevalently available Core Argo data \emph{in addition to} the available BGC data to improve prediction and estimates and reduce uncertainty.
%information plays a vital role in the understanding of the world's carbon cycle as well as ocean ecosystems \citep{claustre_observing_2020}. 
This paper is aimed towards this goal, and our main contributions are as follows.
%\moritz{Why are these letters?}
%\drew{automatically. It says in the style guide that that's the order. }
\begin{enumerate}
    \item We develop a descriptive statistical model on the joint distribution of oxygen, temperature, and salinity in the Southern Ocean. The model fully takes into account the functional nature and space-time information of the Argo data.
     
    \item Based on this model, and leveraging the more readily available temperature and salinity data, we address methodological and computational issues related to the functional prediction and interpolation of oxygen levels at arbitrary locations.
\end{enumerate}

\noindent While both (a) and (b) are of significant scientific as well as statistical interests, we are unaware of any existing work that considers them comprehensively. Below, we explain our approaches in more detail.

The basic building block for our model is a spatio-temporal, function-on-function regression model that encodes the functional relationship between the temperature, salinity and oxygen profiles across all pressure levels ranging from $0$ to $2000$ decibars. This enables us to predict oxygen profiles using temperature and salinity profiles. It also facilitates the estimation and prediction of other quantities of scientific interests such as derivatives and integrals of the profiles \citep{kuusela_2023}.

One major challenge of modelling Argo data in the Southern Ocean is that the transition from warmer subtropical waters in the north to colder Antarctic waters does not occur smoothly. 
Instead, ocean properties in that region are characterized by a series of sudden changes, occurring at so-called fronts --- sharp transition zones generally aligned east–west \citep{chapman_defining_2020, orsi_meridional_1995}.
In particular, the contrasting behavior of oxygen in various regimes of the Southern Ocean has been well established in \cite{bushinsky_oxygen_2017}. 
Consequently, any stationarity assumption on the data across different fronts would be violated. 
However, it is established that ocean properties in any region between fronts tend to be homogeneous \citep{chapman_defining_2020}.
To address this challenge, we augment our functional regression model with a latent mixture component aimed at capturing regional information.
Our model thus describes aspects of the joint distribution of oxygen, temperature, and salinity modulated by different regimes in the Southern Ocean.
In our analyses, we demonstrate that the dependence structure in pressure of these three variables varies critically across the regimes estimated through the mixture model.
As a byproduct we obtain estimates for the locations of ocean fronts and how they vary with time.
These are of considerable scientific interest in their own right and have been the subject of research in the oceanographic literature \citep{chapman_defining_2020}. 
Consequently, they constitute an additional key contribution of this work.

Identifying ocean fronts through the use of clustering and mixture models in oceanography has recently gained significant traction. 
For example, \cite{jones2023unsupervised, jones_unsupervised_2019} focus on the Southern Ocean region; \cite{rosso_water_2020}, \cite{boehme_classifying_2021}, and \cite{fonvieille2023swimming} consider subregions of the Southern Ocean, while \cite{maze_coherent_2017}, \cite{poropat2023unsupervised}, and \cite{houghton_nino_2020} analyze other oceanographic areas.
%In short, clustering has successfully determined or described ocean properties in a variety of tasks and data situations. 
Most of these works use standard Gaussian mixture models.
The primary difference between this literature and our work here is that we account for the spatio-temporal dependence for the cluster memberships and the data within each cluster while these critical aspects are ignored in the above cited works. 
We accomplish this by encoding the spatio-temporal dependence of cluster memberships using a Markov random field model \citep{liang_modeling_2020, jiang_clustering_2012, besag_pseudo_75}. In particular, this enables predictions at any location whether temperature and salinity data are available or not. An important application of this is {\it mapping}, or the creation of gridded products, where data sampled at irregular locations in space and time are used to provide estimates at a standard grid \citep[for example,][]{roemmich_20042008_2009, gray_global_2014}.

When temperature and salinity measurements are available, predictions of oxygen may be straightforwardly made using off-the-shelf methodology. 
A well-known example in that regard is the random forest approach of \cite{giglio2018}, which produces highly accurate predictions of oxygen at fixed pressure levels using temperature and salinity data.
For this scenario, we provide comparisons of our mode-based approach with the random forest approach of \cite{giglio2018}.
However, when estimating the oxygen level at a location or a pressure level where temperature and/or salinity information is not available, the random forest approach requires an additional step of interpolating predicted oxygen data onto this location. Our model-based approach, on the other hand, readily provides direct predictions as well as estimated uncertainty of oxygen levels at arbitrary locations and pressure levels using nearby temperature and salinity data without interpolations.

% In this two-step procedure, it is not clear how uncertainty in oxygen predictions may be propagated. 
% Our model instead provides direct predictions and estimated uncertainty of oxygen at arbitrary locations and pressures. 

Our approach also involves advancements in both statistical methodology and implementation.  
There is a rich literature on function-on-function regression, including the seminal paper \cite{yao2005} and \cite{zhou_pca}, while modelling spatio-temporal dependence in univariate functional data has been studied extensively by, e.g., \cite{zhou_reduced_2010, jiang_clustering_2012, liang_modeling_2020}.
However, we are unaware of any development of function-on-function regression models that accounts for spatio-temporal correlation in the functional response and predictor variables.
Furthermore, to the best of our knowledge, functional kriging and cokriging problems
%(in the Argo data application, the spatial prediction/interpolation of oxygen from temperature and salinity) 
in the context of functional mixture models have not been explored at all. The SOCCOM Argo data considered here alone consists of more than fifteen thousand locations and millions of observations. 
Consequently, reducing the runtime and memory usage of the model estimation and prediction procedure is essential. 
This work is accompanied by an R package implementing our methodology where crucial parts of the estimation and prediction procedure are written in C++.
% Our software seamlessly integrates a Vecchia approximation of the likelihood when modeling spatial dependence \citep{guinness_gaussian_2021}.

% In this paper, our principal contribution is to provide a comprehensive functional-data mixture model that enables functional predictions of biogeochemical data by modelling their correlation across pressure with the more pervasively available temperature and salinity data as well their spatio-temporal dependence structure.
% The mixture component accounts for tries to the lack of stationarity but is of scientific interest in its own right.
% A second contribution of our work is thus the identification of ocean fronts.
%Furthermore, we provide a tractable AIC criterion in the presence of latent variables for model selection.

We finally briefly describe the organization of the rest of the paper. 
In Section~\ref{sec:model_framework}, we propose a novel mixture model for multivariate, functional data with spatial dependence. 
In Section~\ref{sec:model_estimation}, we discuss parameter estimation, prediction, and model selection. 
In Section \ref{sec:data_analysis}, we describe our analysis on the Argo data, including the description of the estimated model, evaluation of oceanographic front estimates, and cross-validation experiments. 
Finally, we discuss limitations and potential improvements to our work in Section~\ref{sec:conclusion}.

\section{The Data and the Model}\label{sec:model_framework} 
%\drew{Should probably have a plot in this Section? Maybe something demonstrating nonstationarity?}
%\moritz{Maybe profiles from different locations, i.e. 10 from cluster 1 and 10 from cluster 5 and next to it a map with locations by color?}

%The Argo data consists of so-called profiles, a collection of measurements taken at different depths in the ocean at a fixed points in space and time, see Figure \ref{fig:argo_data}.
Argo data is publicly available from the Argo Global Data Assembly \citep[GDAC,][]{argo2020} after it undergoes various data control measures \citep{maurer_delayed-mode_2021}.
%Data quality control for oxygen is described in \cite{maurer_delayed-mode_2021}.
%Here, we work with a 2021 high-resolution MLR snapshot of the SOCCOM data after preprocessing in order to guarantee sufficient quality of the data.
A precise description of the data under consideration is given in the supplement Section S1.

Notationally, our data consists of profiles at $n$ locations $\{x_i\}_{i = 1}^n$ in the Southern Ocean and corresponding points in time $\{t_i\}_{i = 1}^n$.
We use $\mathsf{D}$ to denote the spatial domain and use the positive real axis $\mathbb{R}_+$ as time domain.
To simplify notation, we will denote a space-time coordinate $(x_i, t_i) \in \mathsf{D} \times \mathbb{R}_+$ by $s_i$.
At each point $s_i$, we observe profiles of a biogeochemical quantity $Y$ (oxygen) as well as temperature $T$ and salinity $S$ each at pressure levels $\{p_{ij}\}_{j = 1}^n$.
Consequently, oxygen data for one profile for $Y$ is then $\{Y_{s_i}(p_{ij})\}_{j = 1}^{n_i}$ and for temperature and salinity $\{T_{s_i}(p_{ij})\}_{j = 1}^{n_i}$ and $\{S_{s_i}(p_{ij})\}_{j = 1}^{n_i}$.
Even though the number of observations as well as the pressure indexes may vary between temperature, salinity and oxygen, we suppress this in our notation for clarity of presentation.
Motivated by this structure in the data, we model the observations as point evaluations from unknown random functions of pressure, observed with additive measurement errors:
\begin{align*}
    Y_{s_i}(p_{ij}) = \mathcal{Y}_{s_i}(p_{ij}) + \epsilon_{ij}^Y, \quad T_{s_i}(p_{ij}) = \mathcal{T}_{s_i}(p_{ij}) + \epsilon_{ij}^T, \quad S_{s_i}(p_{ij}) = \mathcal{S}_{s_i}(p_{ij}) + \epsilon_{ij}^S.
\end{align*}
We use calligraphic letters to denote the underlying functions and regular letters for the observations.
For each index $s$, we model the functions $\mathcal{Y}_s, \mathcal{T}_{s}$ and $\mathcal{S}_s$ as elements of a Hilbert space $\mathbb{H}$ which we choose to be $\mathbb{H} = L_2([0,2000])$, the Hilbert space of square integrable functions on the closed interval $[0,2000]$ because most Argo profiles consist of measurements between 0 and 2000 decibars.
In addition, we assume sufficient regularity of the functions to ensure pointwise evaluations, and we later impose a smoothness penalty on the 2nd derivative.
The measurement errors $\{\epsilon_{ij}^Y\}, \{\epsilon_{ij}^T\}$, and $\{\epsilon_{ij}^S\}$ are assumed to be i.i.d.\ Gaussian random variables with zero mean and variances $\sigma_{Y}^2, \sigma_{T}^2$ and $\sigma_S^2$ to be estimated from the data.
The main focus of our model is the biogeochemical quantity $Y$, so we collect temperature and salinity in a covariate vector of functions $\mathcal{X}_s(\cdot, \cdot) = (\mathcal{T}_s(\cdot), \mathcal{S}_s(\cdot)) \in \mathbb{H} \times \mathbb{H}$.
Combined, this yields two space-time, function-valued random fields $\{\mathcal{Y}_s : s \in \mathsf{D} \times \mathbb{R}_+\}$ and $\{\mathcal{X}_s : s \in \mathsf{D} \times \mathbb{R}_+\}$. 

\textbf{Modelling Ocean Fronts via a Latent Mixture Component} As stated in the introduction, one main challenge with the Argo data is its nonstationary.
To account for this, much of the previous work on the Argo data uses local approaches for example via sliding windows; see for example, \cite{kuusela_locally_2018, argofda_to_appear, kuusela_2023}.
A locally-stationary model is estimated for each window, and since parameters are estimated anew for each window, the model adapts to varying covariance structures in different locations. %general flexbility is given to 
However, in the Southern Ocean, this approach may not be suitable due to the sharp and sudden changes in water properties at the ocean fronts. 
Moreover, due to the circumpolar nature of the ocean fronts, water properties may stay homogeneous across long distances which any local approach will fail to fully utilize. 
%\drew{could/should the above start to the paragraph go to the intro?}
%\moritz{Commented out a sentence including references for ocean fronts here because we have mentioned this many times by now}
%The existence of these characteristics in the Southern Ocean are widely accepted in the oceanography community \citep{jones_unsupervised_2019, rosso_water_2020, chapman_defining_2020} and has been used in a previous analysis of oxygen in the Southern Ocean \citep{bushinsky_oxygen_2017}. 
In order to explicitly model the ocean fronts, we partition the data into $G$ clusters, with ocean fronts representing the boundaries between clusters.
We use a latent field $\{Z(s)\}$ where each $Z(s)$ is a random variable taking value in $\{1, \dots, G\}$ for some $G$ to be specified.
We then model 
\begin{align*}
    \mathcal{Y}_s(p) = \sum_{g = 1}^G \mathbb{I}(Z(s) = g)\mathcal{Y}_s^g(p), \quad \mathcal{X}_{s}(p_1, p_2) = \sum_{g = 1}^G \mathbb{I}(Z(s) = g) \mathcal{X}_{s}^g(p_1, p_2)
\end{align*}
where the random fields $\{\mathcal{Y}_s^g\}$ and $\{\mathcal{X}_s^g\}$ govern the spatio-temporal covariance structure in cluster $g$.
Here $\mathbb{I}$ is the usual indicator function.

For $\{Z(s)\}$, we employ a commonly used Markov random field model \citep[cf.][]{jiang_clustering_2012, liang_modeling_2020}, where the marginal distribution $Z(s_i)$ given appropriately defined neighbors $\partial s_i$ is given by
\begin{align}
    \mathbb{P}\left(Z(s_i) = g \vert \partial s_i \right) \propto \exp \bigg( \xi \sum_{s_j \in \partial s_i}  \frac{\mathbb{I}(Z(s_j) = g)}{\mathrm{dist}(s_i, s_j)}\bigg).\label{eq:Potts}
\end{align}
We use $k$-nearest neighbor graphs to define the neighborhood structure based on the distance $\mathrm{dist}(s_i, s_j) = \mathrm{dist}_{\mathrm{long}} + 3\mathrm{dist}_{\mathrm{lat}} + 12\mathrm{dist}_{\mathrm{day of year}}$ to account for the anisotropy in the Argo data. 
The additional weighting by the inverse distance follows the intuition that profiles that are close should provide better cluster information than profiles that are further away.
This also reduces the importance of the choice of the number of nearest neighbors to construct the underlying graph which we chose to be 15.
Consequently, we found that reasonably varying this number has negigible impact.
The Markov random field model allows cluster memberships to be interpolated between the irregularly-sampled profile locations.
Moreover the Markov structure yields computational advantages by using a pseudo-likelihood approximation as in \cite{besag_pseudo_75}. 
%However, it has the disadvantage that it requires redrawing the graph to incorporate new locations for prediction.

We found that incorporating a temporal component in the distance is important. Indeed, we implemented a Markov random field that used only space (without a seasonal component) and found that in many locations this gave conflicting information on cluster membership from winter versus summer profiles.

\textbf{Spatio-Temporal Function on Function Regression} Within each cluster, we use the following spatial regression model to specify the joint distribution of $\mathcal{X}^g$ and $\mathcal{Y}^g$:
\begin{align}
    \mathcal{Y}_s^g(\cdot) = \mu^g_{\mathcal{Y},t}(\cdot) + \mathscr{F}^g (\mathcal{X}_s^g - \mu^g_{\mathcal{X}, t})(\cdot) + \mathcal{E}_s^g(\cdot),\label{eq:abstract_model}
\end{align}
where the $\mathscr{F}^g : \mathbb{H} \times \mathbb{H} \to \mathbb{H}$ are linear operators and the $\mathcal{E}^g_s$ are functional, spatio-temporally correlated residuals that are independent of $\{\mathcal{X}^g_s\}$.
Here, $\mu^g_{\mathcal{Y},t}(\cdot)$ and $\mu^g_{\mathcal{X},t}(\cdot)$ are mean functions of pressure for the response and predictors, respectively, in the $g$-th group and depend on time $t$.
%Similar spatial regression models are standard in finite dimensions but we are unaware of any usage for similar spatio-temporal function on function regression models.
In the mean function $\mu^g_{ \mathcal{Y}, t}(\cdot)$, we include seasonality using a Fourier basis as in \cite{roemmich_20042008_2009}: 
%similar to \cite{roemmich_20042008_2009}: 
\begin{align*}
    \mu^g_{\mathcal{Y},t}(\cdot) &= \gamma^g_{\mathcal{Y}, 0}(\cdot) + \sum_{l=1}^R \sin\left(2 \pi l \frac{t}{365.25}\right) \gamma^g_{\mathcal{Y},l}(\cdot) +\sum_{l=1}^R\cos\left(2 \pi l \frac{t}{365.25}\right)\gamma^g_{\mathcal{Y},R + l}(\cdot).
\end{align*}
The evaluation $\gamma^g_{\mathcal{Y},0}(p)$ then represents the average (across the year) mean of oxygen for group $g$ at pressure $p$, while the $\gamma^g_{\mathcal{Y}, l }(p)$ for $l  = 1, \dots, 2R$ describe seasonal variations in that mean based on sines and cosines. 
We take $R=3$ in this case to capture the main seasonality component, which is half of the number basis functions used in \cite{roemmich_20042008_2009} because in Antarctica there are only two seasons \citep{seaons_antarctica}.
The mean functions of the predictors have identical form. 
Next, we employ a Karhunen-Lo\`eve type expansion along the pressure dimension, %that is
\begin{align}
    \mathcal{X}_s^g (\cdot, \cdot) =  \mu^g_{\mathcal{X},t}(\cdot, \cdot) + \sum_{l  = 1}^\infty \alpha^g_{l }(s) \psi^g_{l }(\cdot, \cdot), \quad \mathcal{E}^g_s(\cdot) = \sum_{l  = 1}^\infty \eta^g_{l }(s) \phi^g_{l }(\cdot) \label{eq:karhunen_loeve}
\end{align}
in order to model the spatio-temporal correlation structure through the principal component random fields $\{\alpha^g_{l }(s) : s \in \mathsf{D} \times \mathbb{R}_+ \}$ and $\{\eta^g_{l }(s) : s \in \mathsf{D} \times \mathbb{R}_+ \}$ \citep{zhou_reduced_2010}.
The functions $\{\psi^g_{l }\}_{l  = 1}^\infty$ and $\{\phi^g_{l }\}_{l  = 1}^\infty$ form an orthonormal basis for $\mathbb{H} \times \mathbb{H}$ and $\mathbb{H}$ respectively for each $g$ and are called the principal component functions (PCFs).
In practice, the infinite sums have to be truncated appropriately.
Through the joint principal component decomposition of $\mathcal{X}_s^g(\cdot, \cdot)$, dependence between predictors (temperature and salinity) is modeled. 
Quantifying this dependence in the model is critical, since these variables are known to be dependent.
For example, temperature, salinity, and pressure jointly determine potential 
density which is known to be generally monotone as a function of pressure \cite[Section 3.5 of][]{talley2011descriptive}.
The coefficient random fields $\{\alpha^g_{l }(s) : s \in \mathsf{D} \times \mathbb{R}_+ \}$ and $\{\eta^g_{l }(s) : s \in \mathsf{D} \times \mathbb{R}_+ \}$ are assumed to be independent across cluster $g$ and component $l $. 
Using \eqref{eq:karhunen_loeve} in \eqref{eq:abstract_model}, and afterwards truncating the infinite sums after $Q_1$ and $Q_2$ terms, we obtain 
\begin{align*}
    \mathcal{Y}^g_s(\cdot) = \mu^g_{\mathcal{Y}, t}(\cdot) + \sum_{l  = 1}^{Q_1}\alpha^g_{l  }(s) (\mathscr{F}^g\psi^g_{l })(\cdot) + \sum_{l  = 1}^{Q_2}\eta^g_{l }(s) \phi^g_{l }(\cdot).
\end{align*}
%\drew{We should decide if we want $\ell$ or $l$ and make it consistent throughout the document.}

We model the random fields constituting the principal component scores as Gaussian random fields as is common in spatial statistics.
Because they have zero mean, it remains to specify their covariance structure. 
A popular choice is the Mat\'ern covariance function as advocated for by \cite{stein_interpolation_2013}.
However, for the Argo data an isotropic covariance function such as the standard Mat\'ern covariance is likely not appropriate.
For example, \cite{kuusela_locally_2018} found that, typically, the correlation range scales are different for longitude and latitude.
We follow \cite{guinness_gaussian_2021} who, when modelling Argo data, used spatial deformation of the sphere  via gradients of the spherical harmonics to model anisotropies.
Concretely, we deform the sphere by mapping a point $x \mapsto x + \Phi_{g, \alpha, l }(x)$ where $\Phi_{g, \alpha, l }(x)$ is the weighted sum of gradients of the spherical harmonics of degree 2 at $x$. 
The weights will be learned from the data. 
We then use the following covariance
\begin{align}
    \mathbb{C}\textrm{ov}\{\alpha_{g,l }(s), \alpha_{g,l }(s^\prime)\} = \sigma^2_{g,\alpha, l } \left\{d_{g, \alpha, l }(s, s^\prime)\right\}^{\nu^{g, \alpha, l }} K_{\nu^{g,\alpha, l }} \{d_{g, \alpha, l }(s, s^\prime)\}, \label{eq:matern}
\end{align}
where $\nu \geq 0$, $K_{\nu}$ is the modified Bessel function of the second type and for $s = (x,t)$,
\begin{align*}
    d_{g, \alpha, l }(s, s^\prime) = \frac{\lVert x + \Phi_{g, \alpha, l }(x) - \{x^\prime + \Phi_{g, \alpha, l }(x^\prime) \}\rVert}{\kappa_{g, \alpha, l , x}} + \frac{\lvert t - t^\prime \rvert}{\kappa_{g, \alpha, l , t}},
\end{align*}
where $\| \cdot \|$ is the standard Euclidean norm (chordal distance).
An \texttt{R} implementation of this covariance function is readily available in the \texttt{GpGp} package \citep{GpGp}.
%The space-time covariance structure of each latent principal component fields is thus described by comparatively many parameters.
Generally, we found that allowing more flexibility in the covariance function improved prediction performance. 
The smoothness parameter $\nu^{g, \alpha, l }$, variance parameter $\sigma_{g, \alpha, l }^2$, and range parameter $\kappa_{g, \alpha, l , x}$ and $\kappa_{g, \alpha, l , t}$ govern properties of the Mat\'ern model for the $l $-th predictor principal component of the $g$-th group. 
%{\color{red}When we tried simpler models such as fixing $\nu^{g, \alpha, l }= 1/2$ in \eqref{eq:matern} corresponding to an exponential covariance function or a standard Mat\'ern model with varying range scale parameters for longitude, latitude and time, we always noticed a uniform worsening of the prediction performance although the changes were not dramatic.
%Because a primary focus is prediction, we used the covariance function that gave the best predictions.
%However, simpler models might provide more stable estimates of the spatio-temporal covariance structure. }
We use the same covariance form for $\{\eta_{g, l }(s)\}$.
 %to establish the same flexibility in predictions that the Markov random field provides. 

\section{Model Estimation and Prediction}\label{sec:model_estimation}

The mean functions, the PCFs, and the functions $\{\mathscr{F}^g\psi^g_{l}\}$ are unknown and have to be estimated from the data.
To do so, we use a cubic B-spline basis, a common choice in functional data analysis with favourable computational and provable optimality properties \citep{wahba_splines,hsing:eubank:2015}. 
Specifically, if $\{b_i\}_{i = 1}^P$ are the elements of the cubic B-spline basis on [0, 2000] write $B(p) = (b_1(p), \dots, b_P(p))$.
Then we use 
\begin{align*}
    \gamma_{\mathcal{Y}, l}^g(p) = B(p)\Upsilon_{\mathcal{Y},l}^g, \quad \phi^g_{l}(p) = B(p) \Theta^g_{\mathcal{E}, l}, \quad (\mathscr{F}^g\psi^g_{l})(p) = B(p) \Lambda^g_l,\\
    \gamma_{\mathcal{X},l}^g(p_1, p_2) = (B(p_1), B(p_2))\Upsilon_{\mathcal{X}, l}^g, \quad \psi^g_l(p_1, p_2) = (B(p_1), B(p_2)) \Theta^g_{\mathcal{X}, l},
\end{align*}
where $\Upsilon_{\mathcal{Y}, l}^g$ is the $l$-th column of the $P\times 7$ matrix of spline coefficients for the seasonal means of $\mathcal{Y}_g$ and $\Upsilon_{\mathcal{X}, l}^g$ is the corresponding matrix for the covariates, $\Theta^g_{\mathcal{E},l}$ is the $l$-th column of the $P \times Q_2$ matrix containing the coefficients for the PCFs for $\mathcal{E}$, and the $P \times Q_1$ matrix $\Lambda^g$ contains the coefficients for the $\{\mathscr{F}^g\psi^g_{l}\}$.
Similarly, $\Theta^g_{\mathcal{X}}$ is a $2P \times Q_1$ matrix.
For the PCFs, we impose the following constraints to enforce orthonormality:
\begin{align}
    \left(\Theta_{\mathcal{X}}^g\right)^\top  \left( I_2\otimes \int B(t)^\top B(t)\, \mathrm{d} t \,\right)  \Theta_{\mathcal{X}}^g  = I_{Q_1}, ~~~~  \left(\Theta_{\mathcal{E}}^g\right)^\top  \int B(t)^\top B(t)\, \mathrm{d} t \,\Theta_{\mathcal{E}}^g = I_{Q_2} \label{eq:ortho}.
\end{align}
Here $\otimes$ is the standard Kronecker product and $I_q$ is the $q\times q$ identity matrix.
For identifiability reasons, we require the first entries in each column of $\Theta_{\mathcal{X}}^g$ and $\Theta^g_{\mathcal{E}}$ to be positive and order the PCFs by the marginal variances of the random fields constituting the principal component scores.
\begin{remark}\label{remark:bsplines}
A common approach to enforce the orthogonality constraints in \eqref{eq:ortho} is to use an orthogonalized spline basis $\Tilde{B}(t)$ and then require $\smash{(\Theta_{\mathcal{X}}^g)^\top \Theta_{\mathcal{X}}^g = I_{Q_1}}$ and $\smash{(\Theta_{\mathcal{E}}^g)^\top \Theta_{\mathcal{E}}^g = I_{Q_2}}$; see \cite{zhou_reduced_2010} and \cite{liang_modeling_2020}, among others.
However, for datasets as large as the Argo data, this becomes impractical because the evaluation matrices $\Tilde{B}(t)$ (at observed data points) for orthogonalized splines are less sparse compared to regular B-splines.
We outline in the supplement Section S9 that regular B-splines can be used to obtain orthonormal principal components with a proper orthonormalization step.
\end{remark}

\subsection{Monte Carlo EM Algorithm}\label{sec:mcem_algo}
Notationally, we collect all observations $\{\{Y_{s_i}(p_{ij})\}_{j = 1}^{n_i}\}_{i = 1}^n$ in the vector $\mb{Y}$ and all temperature and salinity observations in the vector $\mb{X}$.
Data from a single profile will be denoted by $\mb{Y}_i$ and $\mb{X}_i$ respectively.
The corresponding latent principal component scores and the cluster memberships are collected in the vectors $\mb{\alpha}, \mb{\eta}$, and $\mb{Z}$ respectively.
We collect the spline coefficient matrices, the covariance parameters, and the MRF parameter in the set $\Omega$.
To fit the model, we aim to maximize the penalized log-likelihood of the observed data
 \begin{align}
    \ell(\Omega, \mb{X}, \mb{Y}) - \mathscr{P}(\Omega) =\log \int L(\Omega, \mb{X}, \mb{Y}, \mb{\alpha}, \mb{\eta}, \mb{Z}) \, \mathrm{d}\mb{\alpha} \mathrm{d}\mb{\eta} \mathrm{d}\mb{Z} - \mathscr{P}(\Omega) \label{eq:win},
\end{align}
where $\mathscr{P}(\Omega)$ denotes a roughness penalty and $L$ denotes the likelihood function. 
Corresponding to our choice of cubic B-splines, we use integrated squared second derivatives of the respective functions as penalty \citep[cf. Theorem 6.6.9;][]{hsing:eubank:2015}. 

Because the random vectors $\mb{\alpha}, \mb{\eta}$ and $\mb{Z}$ are unobserved, direct maximization of \eqref{eq:win} is not feasible.
A useful tool for such scenarios is the EM algorithm \citep{dempster_EM_1977}.
It consists of initializing the parameters to $\Omega^{(1)}$ and alternating between the following two steps for $m=1, 2, \dots$ until convergence:
\begin{enumerate}
    \item \textbf{E-step}: Compute 
    %\begin{align}
        $\mathcal{Q}\left(\Omega \vert \Omega^{(m)}\right) = \E\left\{\ell\left(\Omega, \mb{X}, \mb{Y}, \mb{\alpha}, \mb{\eta}, \mb{Z}\right) \middle \vert \mb{X}, \mb{Y}, \Omega^{(m)}\right\}$.
    %\end{align}
    \item \textbf{M-step}: Update the parameters via $\Omega^{(m+1)} = \argmax_{\Omega}\mathcal{Q}\left(\Omega \vert \Omega^{(m)}\right)- \mathscr{P}(\Omega)$.
\end{enumerate}
Direct application in of the EM algorithm in our case is not possible because $\mathcal{Q}(\Omega \vert \Omega^{(m)})$ is not tractable.
Instead, we use a Monte Carlo EM algorithm \citep{wei_tanner_mcem_1990}.

%To tackle this we use a Monte Carlo EM algorithm \citep{wei_tanner_mcem_1990}.

{\bf Monte Carlo E-Step.} We use importance sampling to estimate $\mathcal{Q}\left(\Omega \vert \Omega^{(m)}\right)$.
That is, we fix $T \in\mathbb{N}$ and create samples $\smash{\mb{\alpha}^{(\tau)}, \mb{\eta}^{(\tau)}, \mb{z}^{(\tau)}}$ for $\tau = 1, \dots, T$ from a proposal distribution $h$ and approximate $\mathcal{Q}\left(\Omega \vert \Omega^{(m)}\right)$ via
\begin{equation}
\begin{aligned}
    \hat{\mathcal{Q}}\left(\Omega \middle\vert \Omega^{(m)}\right) &= \left(\sum_{\tau = 1}^T w^{(\tau)}\right)^{-1}\sum_{\tau = 1}^T w^{(\tau)}\ell\big(\Omega, \mb{X}, \mb{Y}, \mb{\alpha}^{(\tau)}, \mb{\eta}^{(\tau)}, \mb{z}^{(\tau)}\big) \\
    &= \sum_{\tau = 1}^T \bar{w}^{(\tau)}\ell\big(\Omega, \mb{X}, \mb{Y}, \mb{\alpha}^{(\tau)}, \mb{\eta}^{(\tau)}, \mb{z}^{(\tau)}\big)
\end{aligned}
\end{equation}
where the quantities $\smash{w^{(\tau)} = f(\mb{\alpha}^{(\tau)}, \mb{\eta}^{(\tau)}, \mb{z}^{(\tau)} \vert \mb{X}, \mb{Y}, \Omega^{(m)}) / h(\mb{\alpha}^{(\tau)}, \mb{\eta}^{(\tau)}, \mb{z}^{(\tau)} \vert\Omega^{(m)})}$ are called the importance sampling weights and $\bar{w}^{(\tau)}$ are the self-normalized weights.
For details on importance sampling, see \cite{robert_monte_2004}. 
In a similar situation, \cite{liang_modeling_2020} use a Markov Chain Monte Carlo EM algorithm by employing a Gibbs sampler to generate samples of the latent random variables.
When we applied their sampler to the Argo data, we found that the resulting Markov Chain mixes slowly and thus obtained suboptimal results.
Both Monte Carlo approaches are motivated by the fact that, due to the dependence in the data and the discrete nature of the cluster memberships, the conditional distribution $\mb{Z} \vert \mb{X}, \mb{Y}$ is not tractable in closed form.
%We believe this is due tare motio the comparatively large amount of observations per profile. \moritz{Imprecise, we are speculating so maybe not even say anything here?}
A brief discussion and comparison as well as details on our proposal $h$ can be found in the supplement Section S4.

{\bf M-step.} To update the parameters set $\Omega^{(m+1)} = \argmax_{\Omega} \hat{\mathcal{Q}}\left(\Omega \middle\vert \Omega^{(m)}\right)- \mathscr{P}(\Omega)$. 
The log-likelihood of the complete data factors into separate terms
\begin{align*}
    \hat{\mathcal{Q}}\big(\Omega \big\vert \Omega^{(m)}\big) = \sum_{\tau = 1}^T \bar{w}^{(\tau)}\Big\{&\ell(\mb{Y} \vert \mb{z}^{(\tau)}, \mb{\alpha}^{(\tau)}, \mb{\eta}^{(\tau)}) + \ell(\mb{X} \vert \mb{z}^{(\tau)}, \mb{\alpha}) \\
    & + \ell(\mb{\alpha}^{(\tau)} \vert \mb{z}^{(\tau)}) + \ell(\mb{\eta} ^{(\tau)}\vert \mb{z}^{(\tau)}) + \ell(\mb{z}^{(\tau)}) \Big\}. 
\end{align*}
allowing for most parameters to be updated separately.
A closed form solution exists for most of the parameters, except for the spatial parameters and the MRF parameter for which numerical maximization is necessary. 
The updating formulas are given in the supplement Section S8.
For EM algorithms, initialization can play an important role. 
We provide a detailed initialization procedure that we found to work well in the supplement Section S10.

\subsection{Model Selection}\label{sec:model_selection}
We choose the number of clusters $G = 5$ by domain knowledge 
\citep[see e.g.][]{chapman_defining_2020} as indicated in the introduction 
and have already specified our MRF structure.
It remains to choose the number of principal components and the smoothing penalties.
The number of knots for the spline basis plays a small role in model fit as long as there are sufficiently many to approximate relevant functions because we impose a smoothing penalty \citep[cf.][]{zhou_pca}. 
For the remaining tuning parameters, we use an AIC criterion where we use again importance sampling to approximate the intractable observed data likelihood.
Let $\mathscr{D}(\Omega)$ denote the model degrees of freedom, $\mb{z}^{(1)}, \dots, \mb{z}^{(T)}$ samples of the latent field from a proposal $\tilde{h}$ and $\tilde{w}^{(\tau)}$ the corresponding weights. 
Then our AIC criterion is 
\begin{align}
    %\hat{\mathrm{AIC}} =
    -2\log \big( \hat{f}(\mb{X}, \mb{Y})\big) + 2\mathscr{D}(\Omega) = -2\log \left( \sum_{\tau = 1}^T \tilde{w}^{(\tau)} f(\mb{X}, \mb{Y} \vert \mb{Z} = \mb{z}^{(\tau)})\right) + 2\mathscr{D}(\Omega) .\label{eq:model_selection}
\end{align}
The proposal $\tilde{h}$ is slightly different from the one used during the EM-algorithm; details are found in the supplement Section S4.
To determine $\mathscr{D}(\Omega)$, we take into account the smoothing penalty imposed on functions as in \cite{WEI20101918}.
We choose different smoothing parameters for $\{\mu^g_{\mathcal{Y}, t}\}$, $\{\mu^g_{\mathcal{X},t}\}$, $\{\phi^g_{l}\}$, $\{\psi^g_l\}$, and $\{\mathscr{F}^g\psi^g_l\}$; for simplicity, we assume that these parameters do not vary over the cluster or principal component index. 

A frequently used alternative choice is to use the output of the Monte-Carlo E-step $\hat{\mathcal{Q}}$ from the last EM-iteration as an approximation to the true likelihood \citep{li_cocaine_joint_2014, liang_modeling_2020}.
This did not work well for the Argo data as anticipated by \cite{Ibrahim_2008}, and we found that the cost of one additional E-step of \eqref{eq:model_selection} was offset by a more reliable model selection.

\subsection{Prediction}\label{sec:pred_sec}
To provide functional predictions at new locations $\check{\mb{s}}= \{\check{s}_i\}$ using the observed data $\mb{Y}$ and $\mb{X}$ we use the conditional expectation $\hat{\mathcal{Y}}_{\check{\mb{s}}}(\cdot) = \E \left\{\mathcal{Y}_{\check{\mb{s}}}(\cdot) \middle \vert \mb{X}, \mb{Y}\right\}$. 
In this setting, we may include additional data (for example, temperature and salinity data at or nearby $\check{\mb{s}}$) in $\mb{X}$ not included in model training. 
Furthermore, we stress that one does not need temperature and salinity at $\check{\mb{s}}$ to construct predictions at these locations. 
Similar to the E-step, closed-form expressions of the conditional expectation $\E \left\{\mathcal{Y}_{\check{\mb{s}}}(\cdot) \middle \vert \mb{X}, \mb{Y}\right\}$ are not available due to the latent field $\mb{Z}$.
Therefore, we again use importance sampling.
Concretely, we generate samples of the latent field (which now includes the new locations $\check{\mb{s}}$) $\mb{z}^{(1)}, \dots, \mb{z}^{(T)}$ from the Potts model \eqref{eq:Potts} and then set
\begin{align}
    \hat{\mathcal{Y}}_{\check{s}}(\cdot) = \sum_{i = 1}^T \bar{w}^{(\tau)} \E \left\{ \mathcal{Y}_{\check{s}} (\cdot) \middle \vert \mb{X}, \mb{Y}, \mb{z}^{(\tau)}\right\},\label{eq:pred}
\end{align}where $\bar{w}^{(\tau)}$ are the importance weights. 
The conditional (co)variances for oxygen at varying pressures or locations are formed using the importance samples and the law of total (co)variance.
For example, for one location $\check{s}_1$ and pressure $p$, we compute:
\begin{align}\begin{split}
    \Var \left\{\hat{\mathcal{Y}}_{\check{s}_1}(p) \middle\vert  \mb{X}, \mb{Y} \right\} &= \sum_{\tau = 1}^T \bar{w}^{(\tau)} \bigg[\Var \left\{\mathcal{Y}_{\check{s}_1}(p) \middle \vert \mb{X}, \mb{Y}, \mb{z}^{(\tau)} \right\} \\
    &~~~~~~~~~~~~~~~~~~~~~~+ \left\{ \E \left( \mathcal{Y}_{\check{s}_1}(p) \middle \vert \mb{X}, \mb{Y}, \mb{z}^{(\tau)}\right) - \hat{\mathcal{Y}}_{\check{s}_1}(p)\right\}^2\bigg]. \label{eq:pred_var_formula_2}
\end{split}\end{align}
Due to the Gaussianity of the latent fields $\{\alpha_{g,l}(s)\}$ and $\{\eta_{g,l}(s)\}$, closed from expressions for the quantities appearing in \eqref{eq:pred} and \eqref{eq:pred_var_formula_2} are available, see the supplement Section S5.
Inference for the latent field $\mb{Z}(\check{\mb{s}})$ at new locations is based on the samples $\mb{z}^{(1)}, \dots, \mb{z}^{(T)}$. 

Modeling functional dependence enables us to construct confidence bands for $\mathcal{Y}_{\check{s}_i}(\cdot)$ for a location $\check{s}_i$ such that we expect $\mathcal{Y}_{\check{s}_i}(p)$ to lie entirely (for all $p$) within the band. 
We construct bands based on the estimates of $\textrm{Var}(\alpha_{g, l }(s)\vert \mb{X}, \mb{Y})$ and $\textrm{Var}(\eta_{g, l }(s) \vert \mb{X}, \mb{Y})$ using the approach of \cite{choi2018geometric}. 
When we predict that $Z(\check{s}_i)$ belongs to a certain cluster with probability very close to 1, these bands may be directly applied; otherwise, bands from different groups should be considered individually. 
Our functional approach also yields principled predictions for linear functionals such integrals and derivatives of $\mathcal{Y}_s$ that are frequently of interest for Argo data.

\subsection{Computational Details}
Due to the size of the Argo data considered here -- alone the SOCCOM data consists of more than fifteen thousand locations and millions of observations -- naive implementation of the above EM-algorithm is computationally expensive.
The most computationally demanding steps are the covariance parameter estimation of the latent random fields $\{\alpha_g(s) \}$ and $\{\eta_g(s)\}$ during the M-step as well as computing their conditional distribution for the E-step and for predictions.
The main tool we employ for these steps is Vecchia's approximation \cite{katzfuss_vecchia_2020} but additional steps are necessary.
We outline them in the supplement Section S7.

\section{Data Analysis Results}\label{sec:data_analysis}

Here, we present results from our analysis of the Argo data, beginning with the estimated model in Section \ref{sec:clustering_product}.
We then provide a cross-validation study to assess prediction performance in Section \ref{sec:preds} and present our data product of comprehensive functional predictions in Section \ref{sec:pred_product}.

\subsection{Description of estimated model}\label{sec:clustering_product}
Our final model uses 18 principal components for the response and 18 for the predictors even though our AIC criterion (provided in the supplement Section S2) indicates that even more principal components could be used.
However, we found that additional principal components did not improve prediction performance and increased computation time.

\begin{figure}[t]
    \centering
    \includegraphics[width = .98\textwidth]{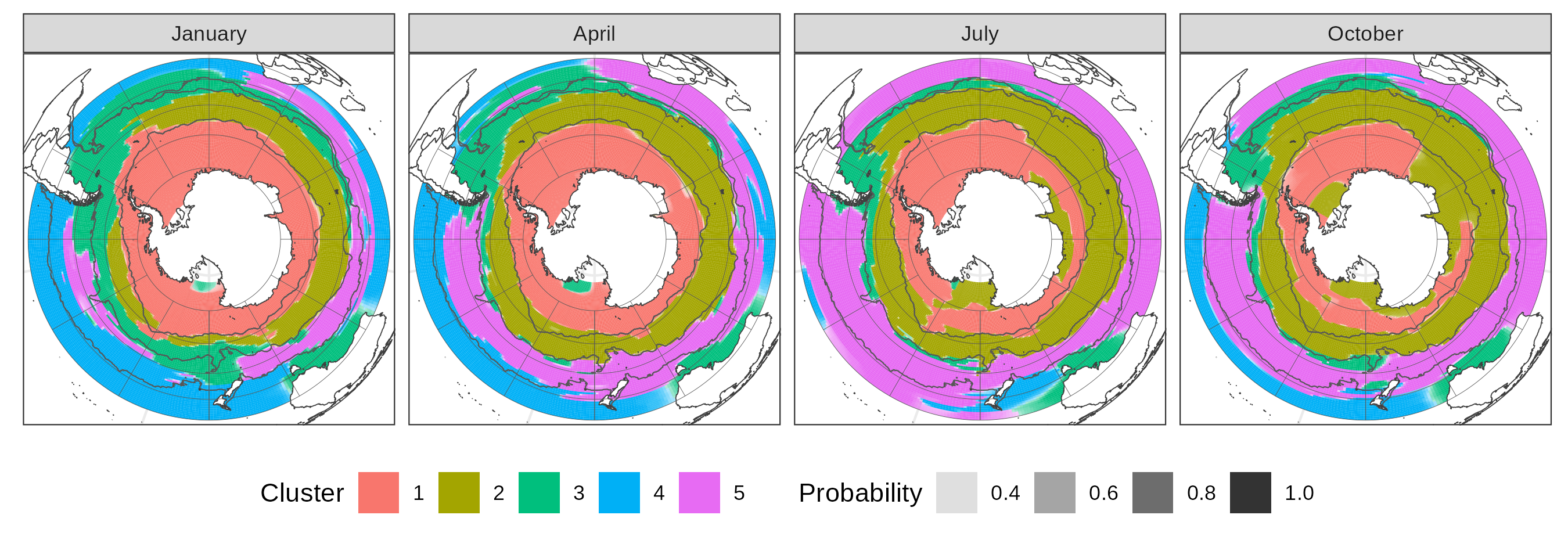}

    \includegraphics[width = .98\textwidth]{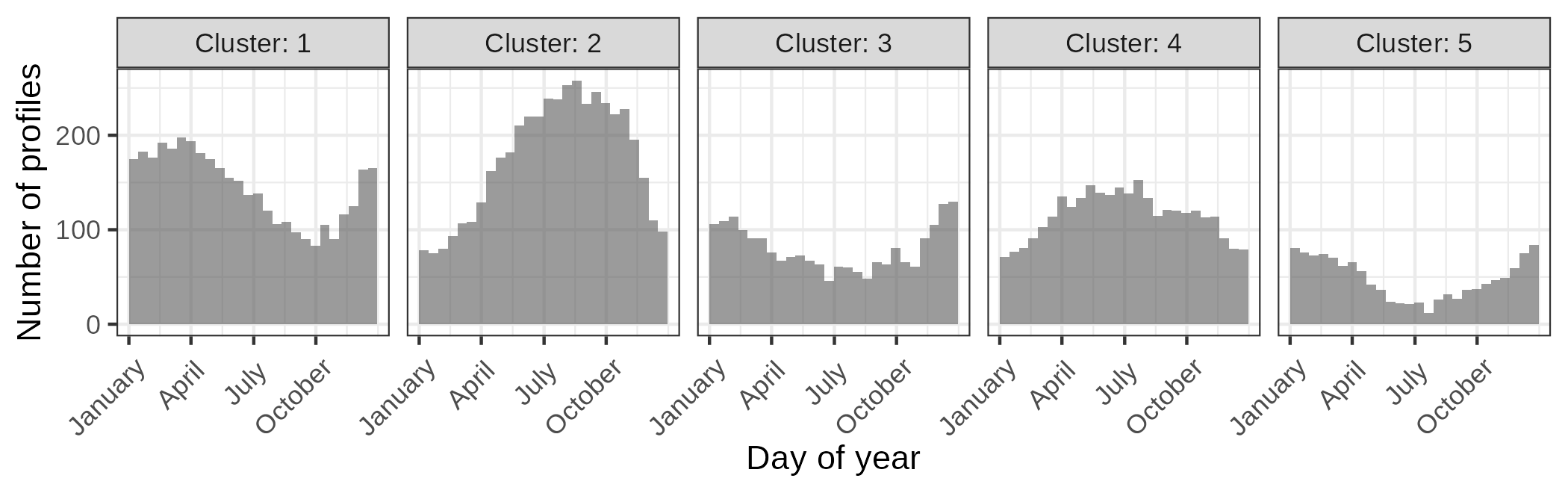}

    \caption{(Top) Cluster predictions for the 15th day of four months. We plot in black lines traditional estimates of fronts using the same criteria for fronts as \cite{bushinsky_oxygen_2017} based on the 2004-2018 \cite{roemmich_20042008_2009} estimate. (Bottom) Cluster membership of BGC Argo profiles, in a histogram summarized by the day of the year.}
    \label{fig:mrf_seasonality}
\end{figure}

In Figure \ref{fig:mrf_seasonality}, we plot the predicted clusters for four months of the year.
Our clusters are ordered by their mean temperature function's value at 1000 decibars. 
We also plot in black lines previous estimates of fronts using the same criteria as \cite{bushinsky_oxygen_2017} based on the 2004-2018 \cite{roemmich_20042008_2009} estimate, see also \cite{orsi_meridional_1995}. 
These can be described as follows.
South of the southern-most line is the Seasonal Ice Zone, whose northern boundary is defined by the maximum extent of sea ice. 
The next zone is defined by the maximum extent of sea ice in the south and the Subantarctic front in the north, which contains the Antarctic Circumpolar Current (ACC), a strong ocean current that encircles Antarctica \citep{bushinsky_accurate_2016}. 
The final plotted front is the Subtropical front. 
These traditional estimates of fronts are generally represented well by our clustering plotted in Figure \ref{fig:mrf_seasonality}.
For example, our first cluster matches the Seasonal Ice Zone for most months. 
Clusters 2 and 3 make up the main component of the ACC, with their northern boundary generally aligning with the Subantarctic front. 
Clusters 4 and 5 define waters above the Subantarctic front, with cluster 4 more concentrated in the Pacific and Atlantic. 
There are also substantial seasonal variations, underscoring the importance of modeling the contrast between winter and summer in the Southern Ocean.
For example, in the austral winter (July) clusters 2 and 4 are disproportionately represented while cluster 5 is nearly absent. 
Between cluster boundaries, the predicted cluster has relatively high probability, while there is some uncertainty on their boundaries.
In the supplement Section S3, we also compare with \cite{jones_unsupervised_2019}'s analysis of 6 clusters from temperature data.

In general, the clusters are influenced by a variety of processes occurring in the Southern Ocean. 
Near the ocean surface, interactions with the atmosphere result in higher levels of oxygen dissolved near oxygen's saturation point in water.
In this level, the primary determinant of dissolved oxygen is the water's capacity to hold oxygen, which increases as temperature or salinity decreases \citep[see Section 4.5 of][]{talley2011descriptive}. 
Oxygen is generally decreasing for a water mass as a function of ``age,'' that is, the time since it has interacted with the atmosphere \citep[Section 3.6 of][]{talley2011descriptive}, due to the utilization of oxygen by ocean life.
In the Southern Ocean, upwelling (water at lower depths rising to the surface) of older, low-oxygen water occurs near around the Southern  ACC front (approximately, our cluster 2), after which it splits both poleward and toward the equator at the surface \citep{gray_autonomous_2018}. 
During the winter, ice cover blocks the potential uptake of oxygen to the upper layer of the ocean in the Seasonal Ice Zone; freezing water also ejects its salinity as it freezes, increasing the salinity of the surrounding water \citep{chamberlain2018observing, talley2011descriptive}.  %resulting in lower oxygen levels seen in the right panel of Figure \ref{fig:cluster_means}. 
Finally, photosynthesis from plants increases oxygen, while remineralization of organic matter decreases the oxygen concentration \citep{stramma_spatial_2021}.

\begin{figure}[ht]
    \centering

    \begin{minipage}{ .47\textwidth}
        \includegraphics[width = .95\textwidth]{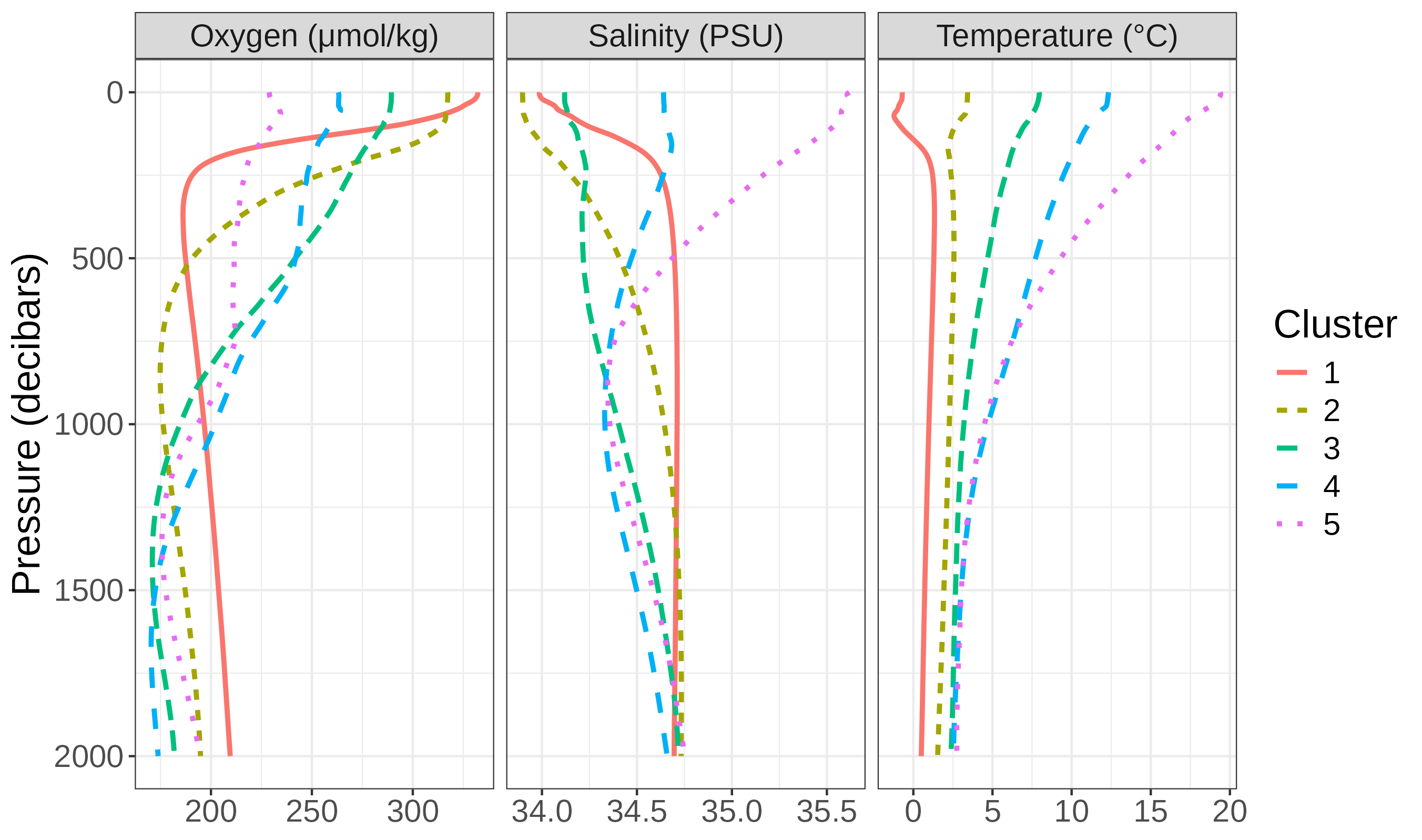}

                \includegraphics[width = .95\textwidth]{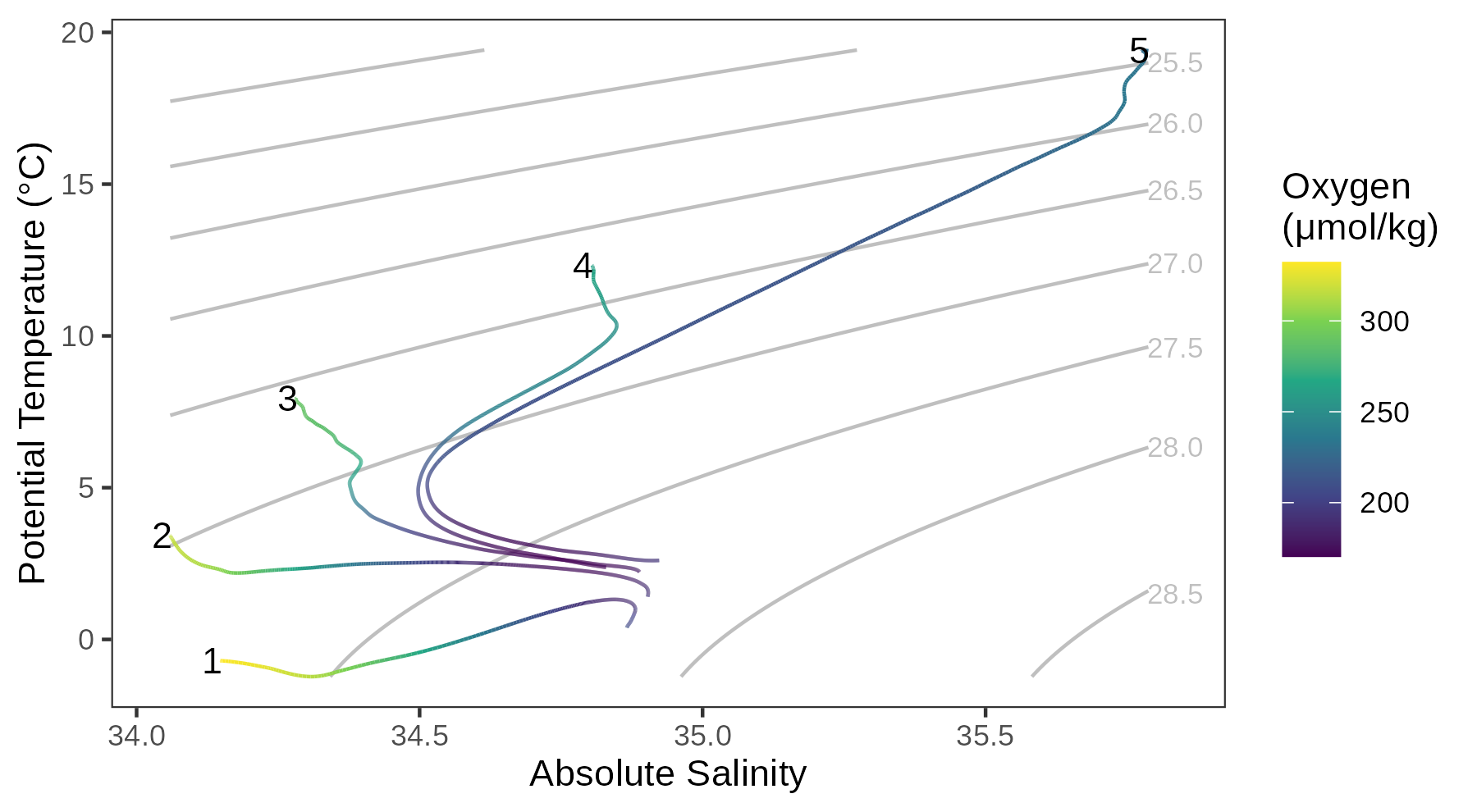}
    \end{minipage}\begin{minipage}{ .53\textwidth}
        \includegraphics[width = .95\textwidth]{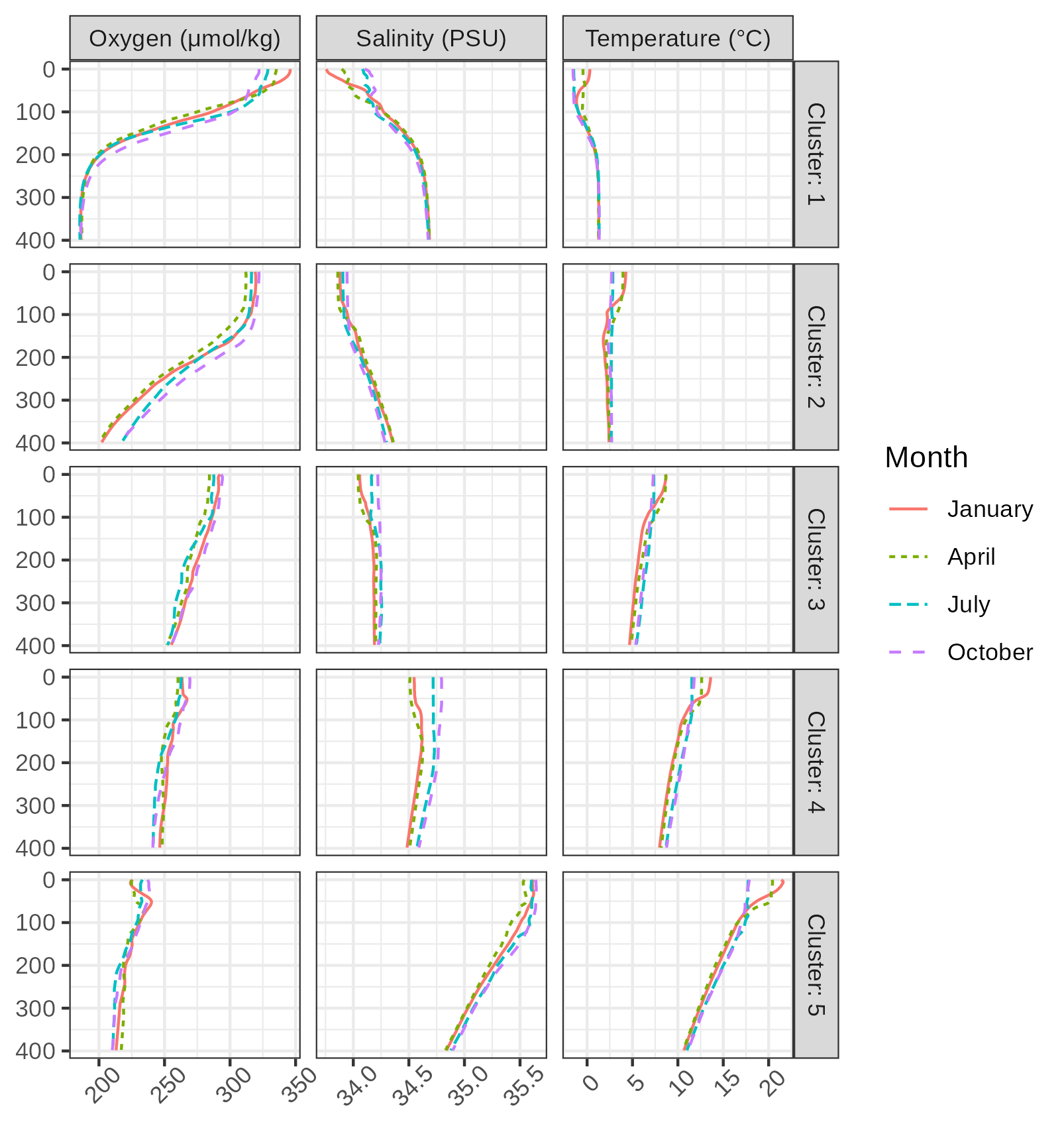}

    \end{minipage}
    \caption{(Top Left) Functional time-averaged mean functions $\gamma_{\mathcal{Y}, 0}$ (for oxygen) and $\gamma_{\mathcal{X}, 0}$ (for temperature and salinity); (Bottom Left) these mean functions are summarized in a temperature-salinity diagram. 
    Contours of potential density are plotted in the background. (Right) Seasonal variation in mean functions for four months of the year and $0$ to $400$ decibars.}
    \label{fig:cluster_means}
\end{figure}

We find evidence of such processes in the mean functions for each of the clusters in Figure \ref{fig:cluster_means}.
% We find evidence for these processes within our clusters. 
For example, since temperature and salinity is low for cluster 1, the water's capacity to hold oxygen is very high.
Furthermore, there is no ice blocking the flux of oxygen into the ocean during the summer.
As a result, dissolved oxygen for cluster 1 at the surface has the highest oxygen levels across all pressures and clusters in Figure \ref{fig:cluster_means}. 
The sharp gradient in cluster 1's mean between low and high oxygen represents the contrast and mixing of the old oxygen-sparse water with water that is highly-saturated in oxygen at the surface. 
In general, we associate cluster 2 with upwelling water. 
During the winter, the ice cover blocks the upwelling water flowing in the Seasonal Ice Zone from interacting with the atmosphere, and thus its oxygen levels are lower compared to the summer.
We attribute the extent of cluster 2 in Figure \ref{fig:mrf_seasonality} during the winter in the Seasonal Ice Zone to this phenomenon. 
%The increase in oxygen in each group at further depths are believed to occur from multiple separate processes \citep[Section 4.5 of][]{talley2011descriptive}. 
Clusters 4 and 5 generally have similar mean functions for depths of more than 800 meters, representing shared deep bottom water across the Southern Ocean. 
In addition, the seasonal variations in the right panel of Figure \ref{fig:cluster_means} follow the expected changes: a decrease in oxygen and increase in salinity for winter months in cluster 1 at the surface, and increases in temperature during summer months at the surface for all clusters. 
We also provide a temperature and salinity plot, a visualization tool commonly used in oceanography, of the mean functions in Figure \ref{fig:cluster_means}. 
Comparing this to Figure 13.7 of \cite{talley2011descriptive}, clusters 4 and 5 correspond to the Subantarctic zone, cluster 3 corresponds to the Polar Frontal Zone, clusters 2 and 1 correspond to the Antarctic Zone north and south of the Southern ACC front, respectively.

\begin{figure}[t]
    \centering
    \includegraphics[width = .233\textwidth]{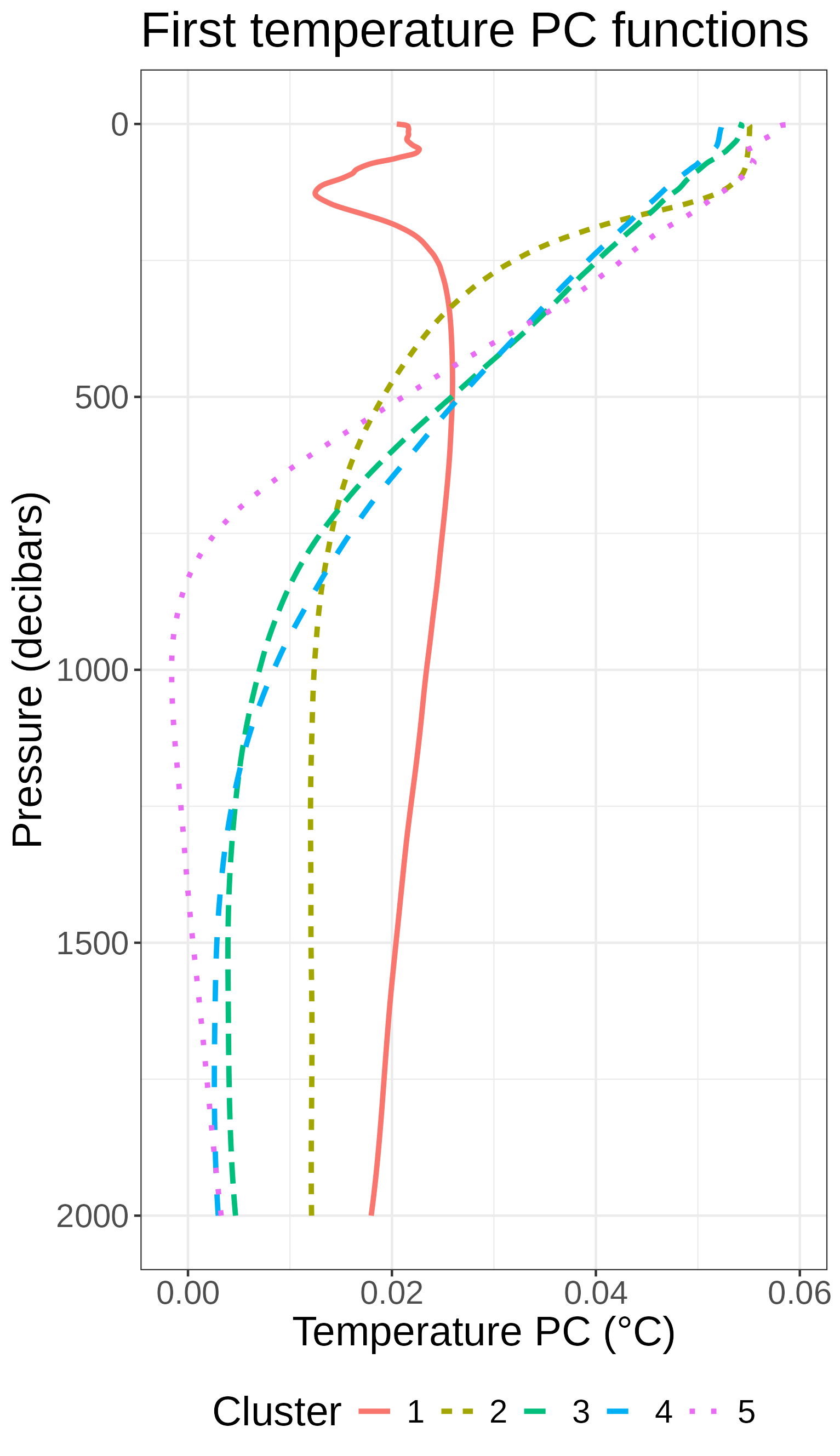}
    \includegraphics[width = .233\textwidth]{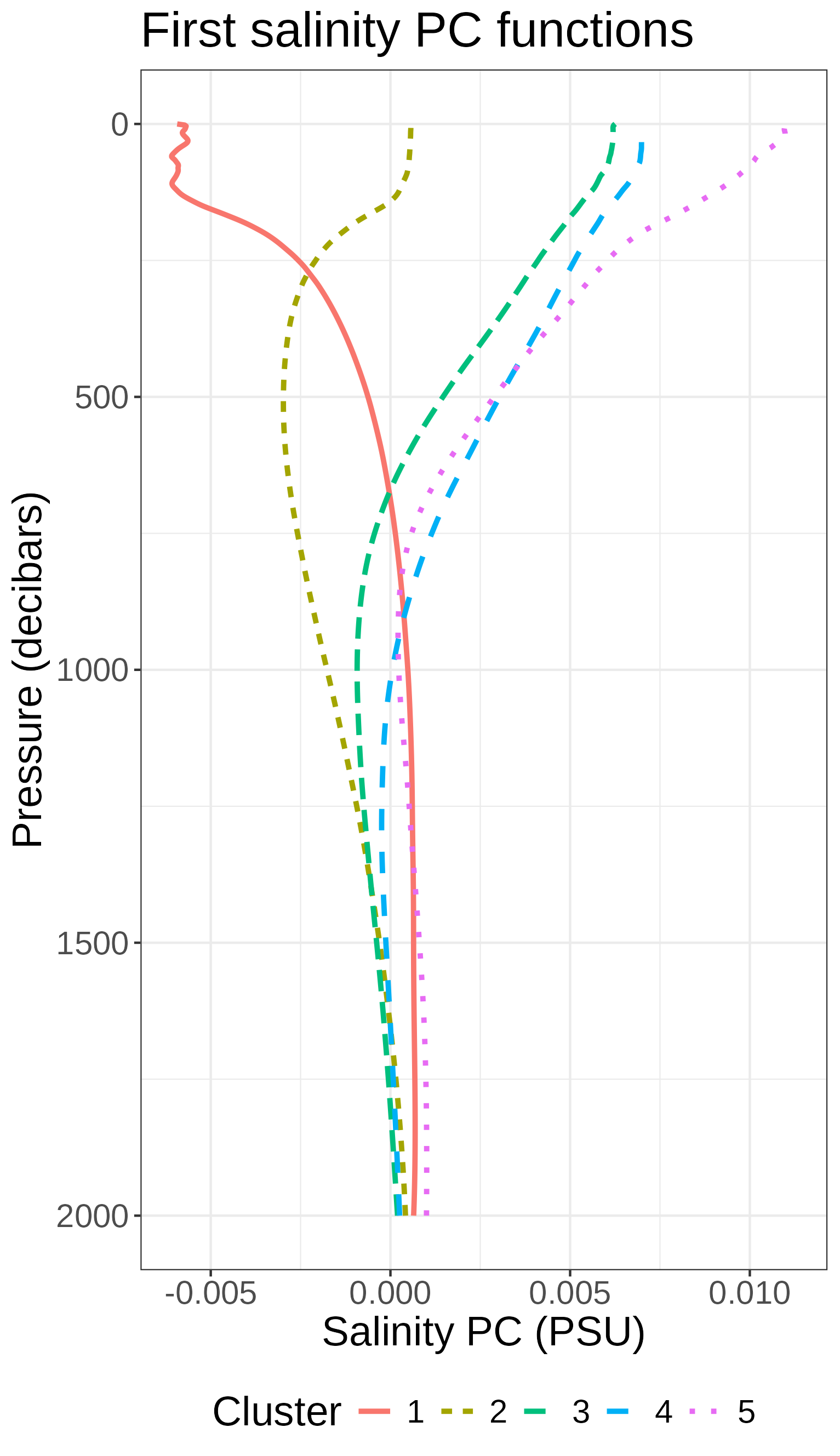}
    \includegraphics[width = .233\textwidth]{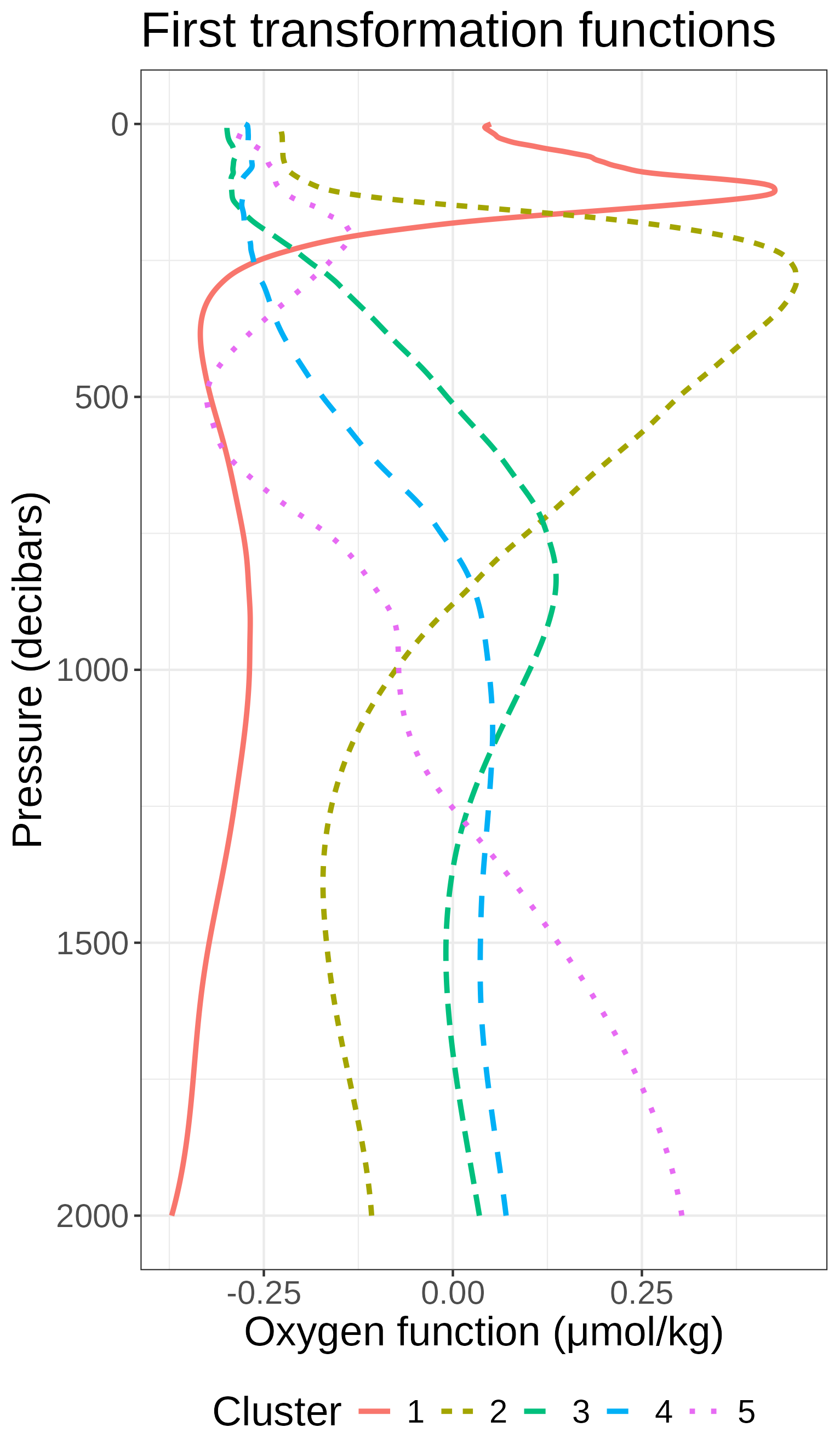}
    \includegraphics[width = .233\textwidth]{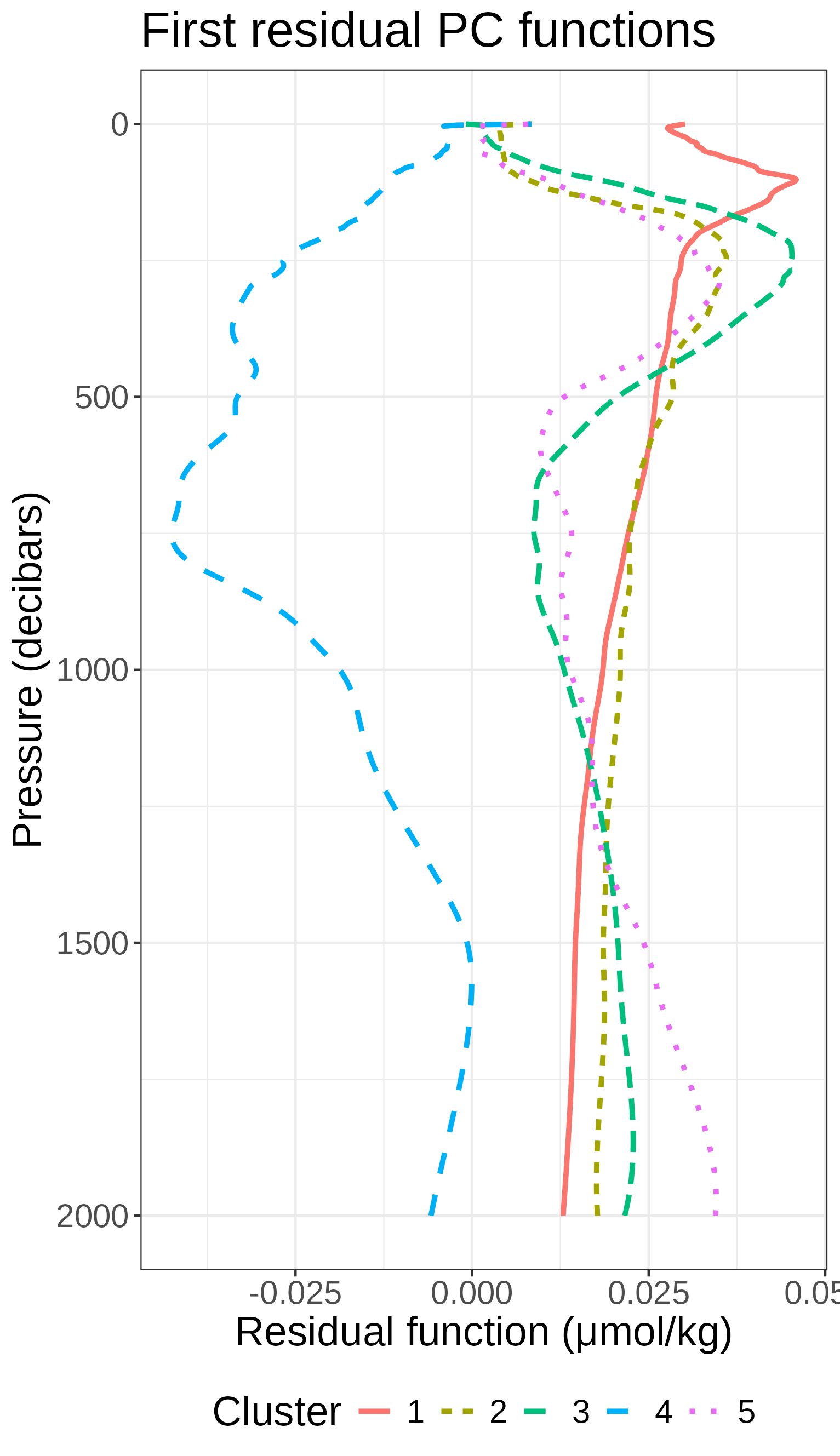}
    \caption{The first estimated functions for each group, for a) temperature PCFs, b) salinity PCFs, c) linear transformation functions, d) principal components of the residual $\mathcal{E}$. }
    \label{fig:pc_functions}
\end{figure}  

We plot the first functional principal components that describe the variance of the data in Figure \ref{fig:pc_functions}. 
We found that the leading principal components are relatively smooth compared to higher-order principal components.
In general, the principal components' higher values (in absolute terms) near the surface indicate more variability there. 
The first principal component of the predictors measures a general increase or decrease in temperature across nearly all pressures. 
Based on the first transformation functions, a higher value for the first predictors PCFs leads to different responses in oxygen by group and pressure. 
At the surface, clusters 2 through 5 have decreasing oxygen for higher temperature, the expected direction based on a decrease in oxygen solubility.
For cluster 1, this relationship is likely confounded by higher temperatures that are associated with less ice cover, which would increase oxygen levels. 
For greater depths, oxygen's response to changes in the first predictor component is varied across groups, emphasizing the importance of accounting for heterogeneity in the Southern Ocean.

\begin{figure}[t]
    \centering
    \includegraphics[width = .98\textwidth]{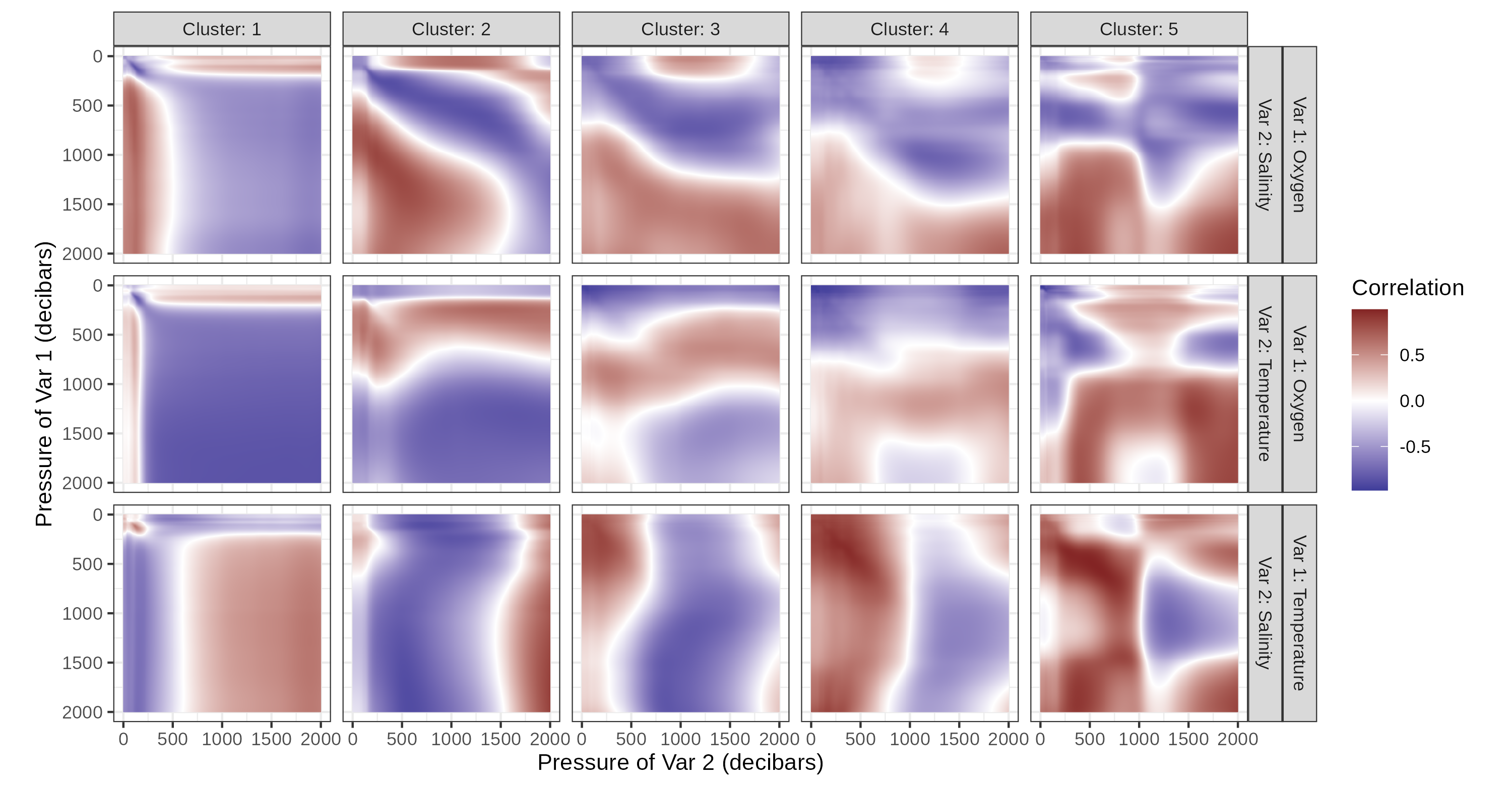}
    \caption{Estimated correlation between functional processes at different pressures by cluster. The diagonal of each entry represents the correlation at the same pressure. 
    }
    \label{fig:correlation}
\end{figure}

In Figure \ref{fig:correlation}, we plot the functional cross-correlation estimated between each of the processes, again demonstrating cluster heterogeneity. 
At the surface, the dependence between oxygen and temperature as well as oxygen and salinity again follow the expected (negative) direction in terms of solubility \citep{bushinsky_oxygen_2017}.
The thickness of this layer in the oxygen and temperature correlations matches the general increase in the depth of the mixed layer, the area at the ocean surface with relatively constant ocean properties, as one moves north in the Southern Ocean \citep{holte2017argo}. 
This also generally matches well with water mass analysis of the Southern Ocean \citep[for example, Figure 13.4 of][]{talley2011descriptive}. 
Finally, between temperature and salinity, the overall signs of the relationship match with exploratory analysis in \cite{salvana20223d}, where there is a band of positive correlations for measurements at the same pressure. 

Our model thus comprehensively estimates the complex nonstationarities and dependence between temperature, salinity, and oxygen at different pressures through the clusters and estimated model properties for each cluster. 
Our results are in accordance with known processes and characteristics of these variables in the Southern Ocean. 
We further elucidate seasonal variations in the clusters and the variables across the entire pressure dimension. 
These results motivate more detailed analysis of the seasonal variations in clusters which is beyond the scope of the current paper.
\subsection{Prediction Performance}\label{sec:preds}

We now turn to the prediction of oxygen based on temperature and salinity.
We assess the prediction performance of our model in two separate cross-validation scenarios:  leave-profile-out (LPO) and leave-float-out (LFO). 
For LPO, we hold out oxygen data from 1/5 of BGC \emph{profiles} using a random partition, train the model on the remaining data, predict the left-out oxygen data, and repeat this process for of the five sets in the partition. 
For LFO, we follow a similar process, but we instead partition {\it floats} into 10 sets, so that we hold out all oxygen data from all profiles from selected floats at a time.
In the LPO setting, nearby oxygen data from the same float is available while in the LFO setting only data from other floats can be used.
We find that as a consequence, prediction errors are generally smaller in the LPO setting.
Since Argo floats either have or do not have BGC sensors, LFO is closer to the practical prediction setting. 

First, we consider predictions of oxygen at locations where temperature and salinity data is available mimicking the scenario in which one is interested in reconstructing the Biogeochemical data at Core Argo floats.
In this setting, we can compare to the random forest approach of \cite{giglio2018}.
Afterwards, we consider the more challenging scenario of predicting oxygen at locations where no data is available and only nearby data can be used.

\subsubsection{Oxygen Reconstruction with Temperature and Salinity}\label{sec:comparison}
\renewcommand{\thetable}{\arabic{table}}

\begin{figure}[ht]
    \centering
    \includegraphics[width = .24\textwidth]{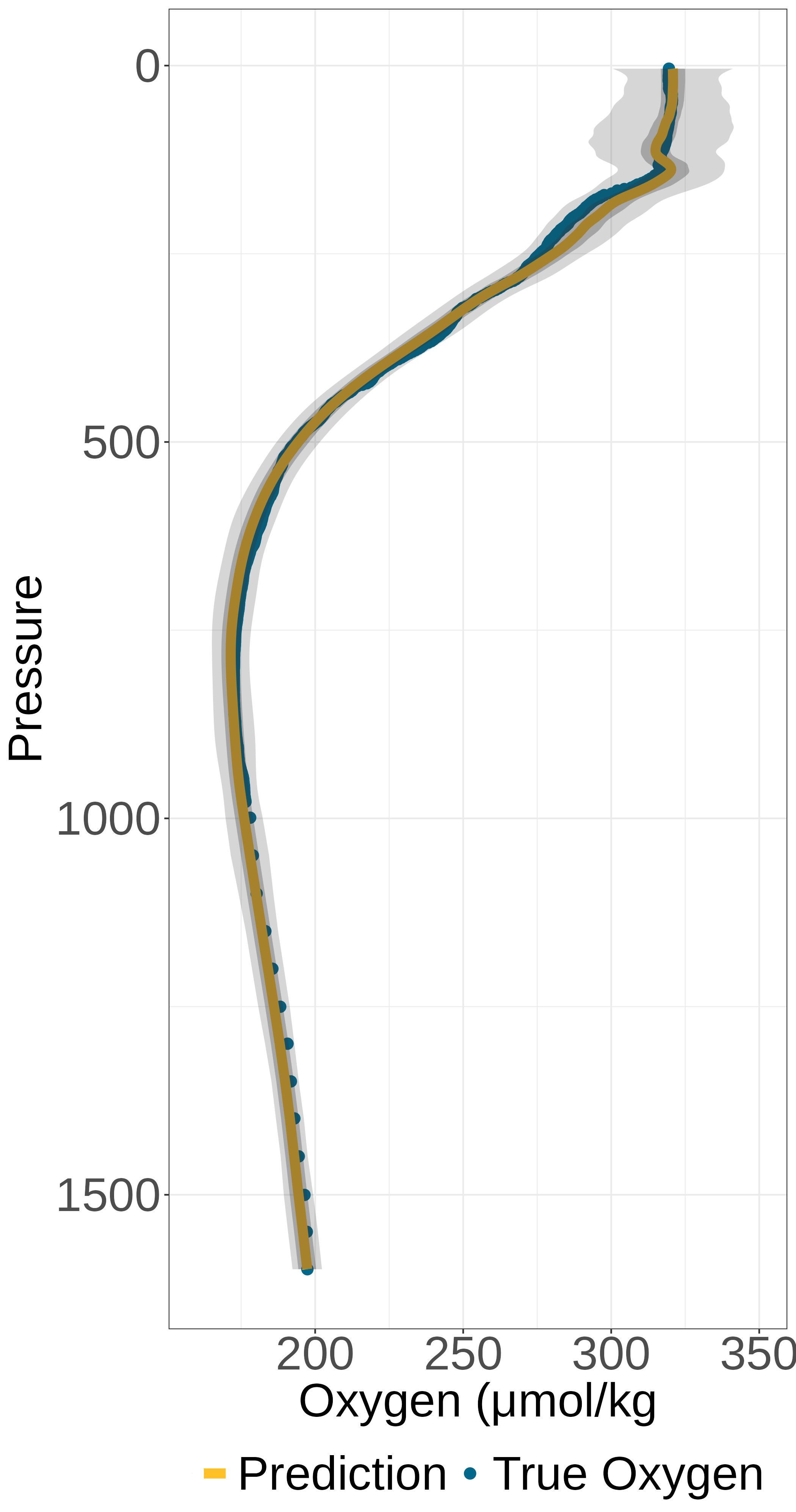}
    \includegraphics[width = .24\textwidth]{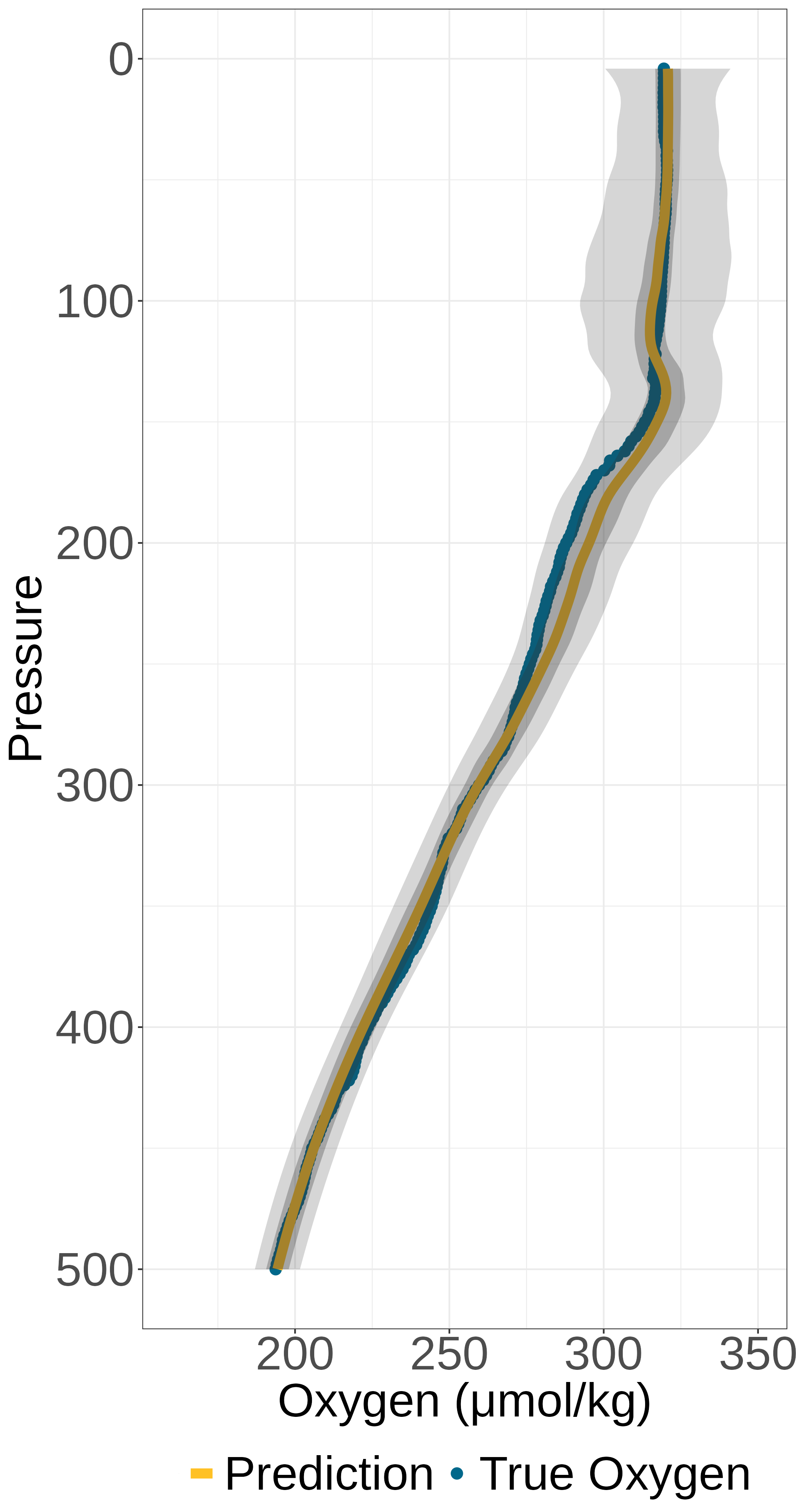}
    \includegraphics[width = .24\textwidth]{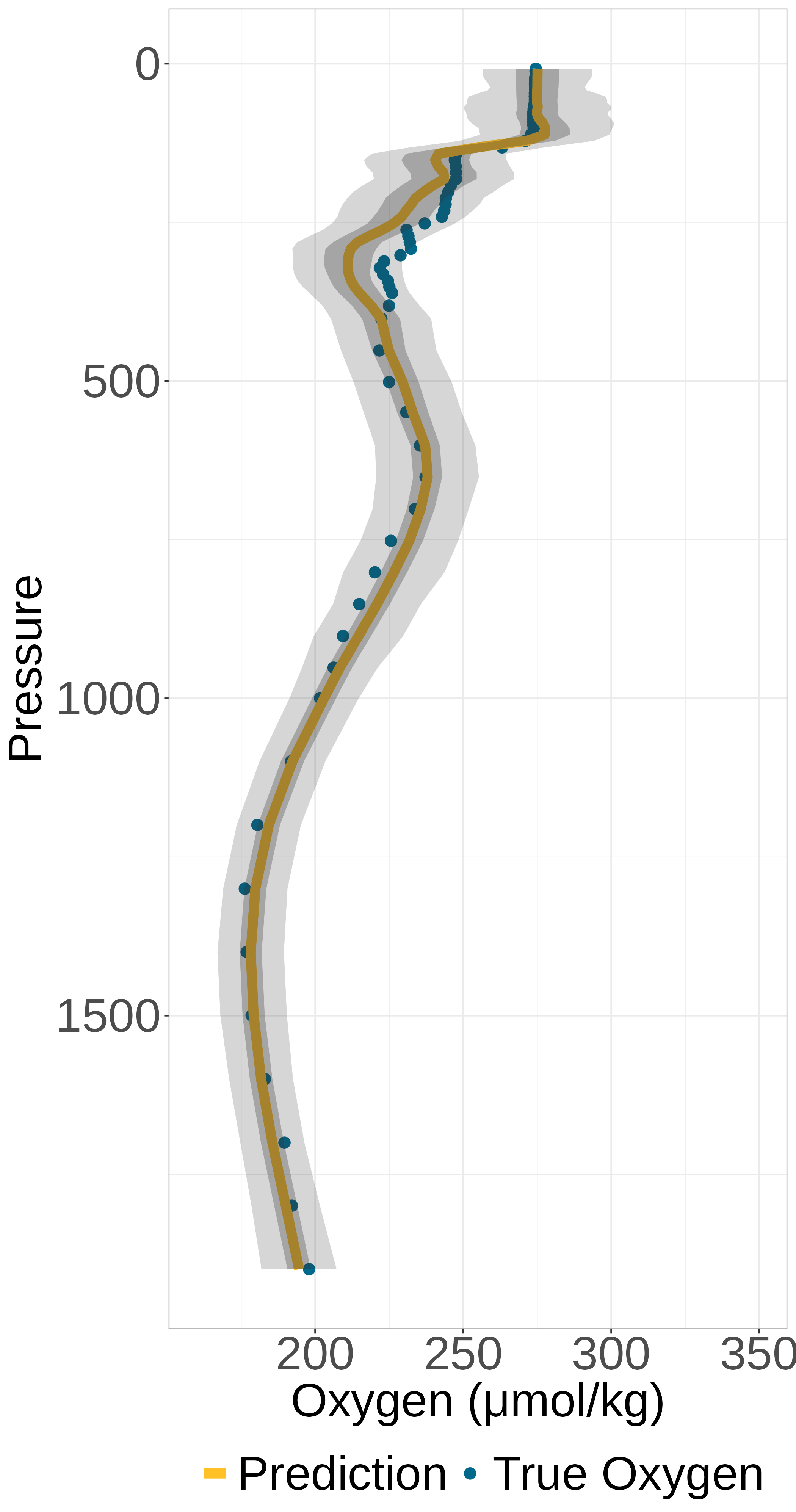}
    \includegraphics[width = .24\textwidth]{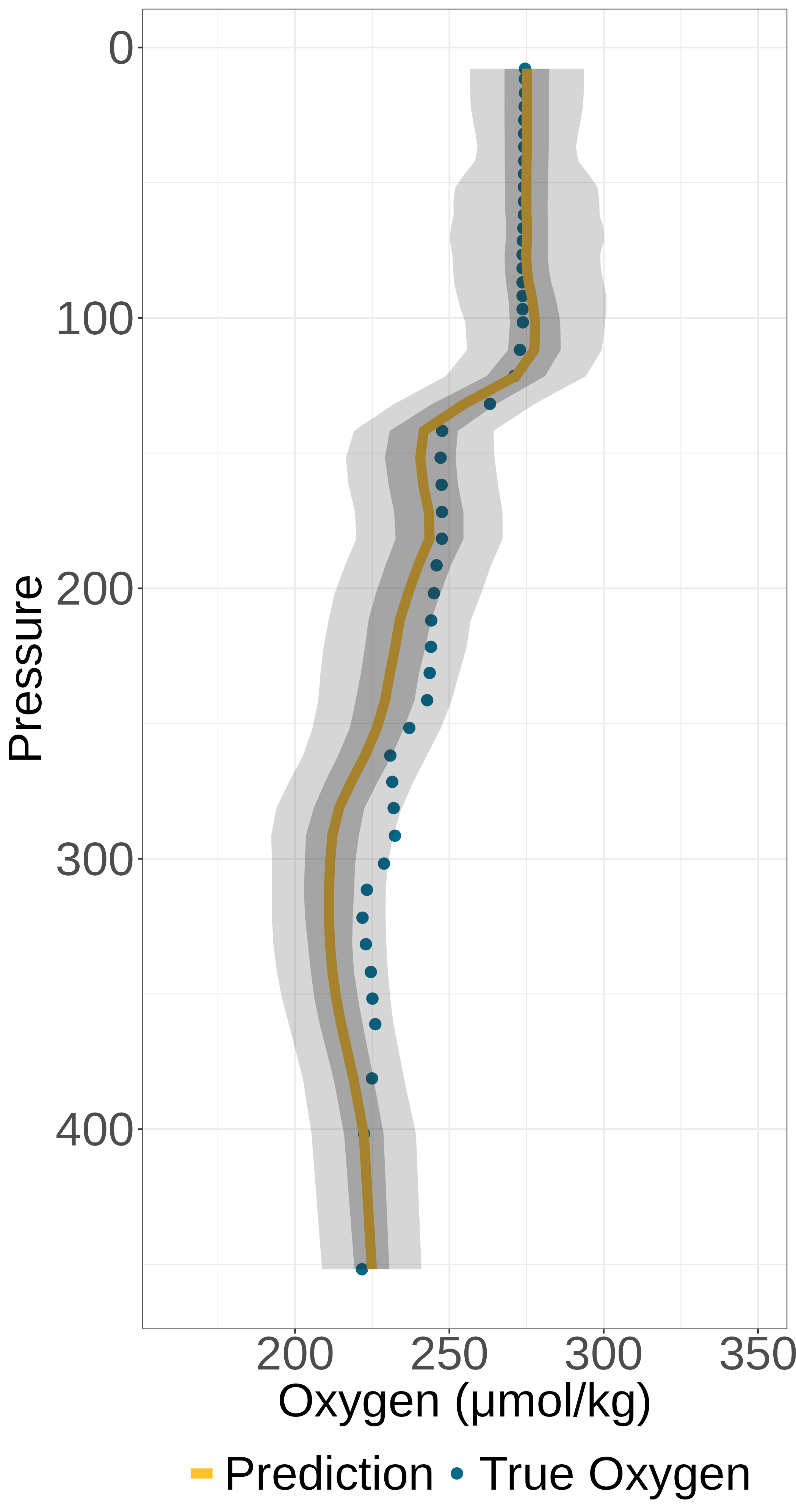}
    \caption{Example profile prediction in the leave-profile-out (Left two panels, float $5905072$ cycle $1$, looking at different pressure ranges) and leave-float-out (Right two panels, float $5906249$ cycle $13$) settings. The dark gray shaded area is based on 95\% pointwise intervals, while the light gray shared area is a 95\% confidence band. MAD for measurements between 145 and 155 dbars are 3.99 and 6.6 for these profiles, respectively.}
    \label{fig:cv_example}
\end{figure}

The results of the cross-validation exercise can be found in Tables \ref{tab:rf_comparison} and \ref{tab:no_TS} and plots of example predictions can be found in Figure \ref{fig:cv_example}.
The shaded areas in Figure \ref{fig:cv_example} correspond to pointwise prediction intervals (dark grey) and confidence bands for the entire function (light grey).
We found that for held-out locations with temperature and salinity data available, there is virtually no variation in cluster assignment across Monte Carlo samples used for prediction.
Consequently, each profile is assigned entirely into a particular cluster, and it is straightforward to construct confidence intervals and bands using the resulting Gaussian distribution using solely that cluster for that profile. 
A description on the validity of our uncertainty quantification can be found in Table~\ref{tab:leave_out_coverages}.

\begin{figure}[t]
    \centering
    \includegraphics[width = .85\textwidth]{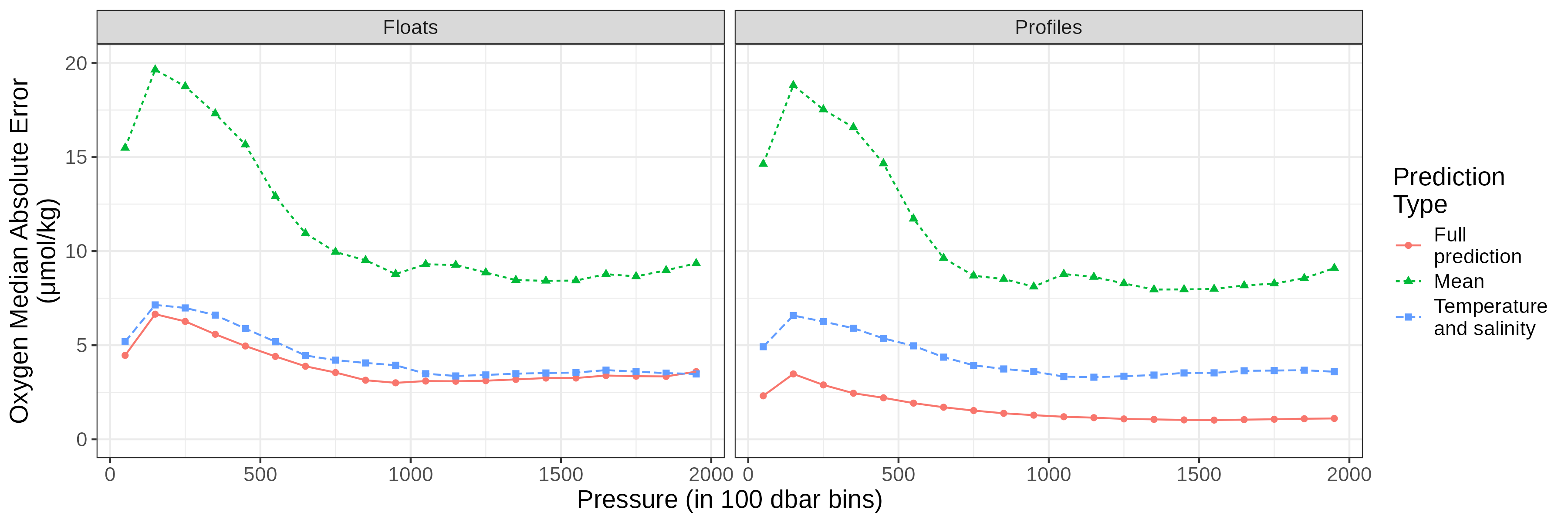}
    \caption{Median absolute error of oxygen (in $\mu$mol/kg) for leave-floats-out (Left) and leave-profiles-out (Right), summarized in 100 decibar bins. We compare predictions based on the cluster means only (``Mean''), the cluster means combined with temperature and salinity data (``Temperature and salinity''), and the full prediction (``Full prediction'').}
    \label{fig:leave_out}
\end{figure}

\begin{table}
    %\centering
    \caption{Comparison with random forest (RF) and  our functional cokriging (FC) approach at 145-155 decibars. The median absolute deviation (MAD) and root-mean-squared-error (RMSE) are the metrics used for each of the methods and leave-out settings.     \label{tab:rf_comparison}}

    %\centering

 %\begin{tabular}{c  c  c  c  c}
  \begin{tabular}{ccccc}
     Approach & \begin{tabular}{c}LPO MAD\\ ($\mu$mol/kg)\end{tabular} & \begin{tabular}{c}LPO RMSE\\ ($\mu$mol/kg)\end{tabular}  & \begin{tabular}{c}LFO MAD\\ ($\mu$mol/kg)\end{tabular}  &\begin{tabular}{c}LFO RMSE\\ ($\mu$mol/kg)\end{tabular}  \\ 
    RF & 3.01 & 6.98 & 6.29 & 13.22 \\
    FC & 3.73 & 8.68 & 7.02 & 13.99 \\ 
    \end{tabular}

\end{table}

To compare prediction performance with the random forest employed in \cite{giglio2018}, we replicate their setup and consider Argo measurements between 145 and 155 decibars and compare the point predictions for each held-out data experiment in Table~\ref{tab:rf_comparison}. 
The hyperparameters of the random forest are as in \cite{giglio2018}.  
In the LPO setting, the resulting median prediction error is close to the instrument measurement error around 3 $\mu$mol/kg \citep{maurer_delayed-mode_2021,mignot_quantifying_2019} for our functional prediction error as well as the random forest model.
The random forest obtains slightly better pointwise prediction results, though we believe they are qualitatively similar: relative to the range over which oxygen ranges for pressures between 145 and 155 dbar, the median errors are 1.2\% (random forest) vs. 1.5\% based our functional model and both these errors are very near the instrument error.
More importantly, our model produces predictions across the \emph{entire} pressure dimension which the random forest \emph{cannot} do.
It needs all covariates (including temperature and salinity) available in order to produce predictions.

\begin{table}
\caption{Prediction performance when no temperature or salinity data is available at the location of prediction.\label{tab:no_TS}}

%%% THE TEMPLATE SAYS %% Caption MUST come immediately after \begin{table}
\centering
\begin{tabular}{c  c  c  c  c}
%\toprule
&\begin{tabular}{c}LPO MAD\\ ($\mu$mol/kg)\end{tabular} & \begin{tabular}{c}LPO RMSE\\ ($\mu$mol/kg)\end{tabular}  & \begin{tabular}{c}LFO MAD\\ ($\mu$mol/kg)\end{tabular}  &\begin{tabular}{c}LFO RMSE\\ ($\mu$mol/kg)\end{tabular}  \\ %\midrule
\underline{\textbf{Pressure in}  $[145,155]$} \\
Overall & 6.83  & 16.96 & 12.06 & 27.32 \\
Core Data Nearby & 6.80 & 16.70 & 10.55  & 23.27\\ 
No Core Data Nearby & 6.88 & 17.40 & 15.77  & 33.04\\
%\midrule
\underline{\textbf{All Pressure Levels}} \\
Overall & 3.10  & 10.07 & 6.80 & 16.38 \\
Core Data Nearby & 3.23 & 9.83 & 6.23  & 14.13\\ 
No Core Data Nearby & 2.88 & 10.48 & 8.02  & 19.75\\
%\bottomrule
\end{tabular}

\end{table}

\subsubsection{Prediction at Locations without Temperature and Salinity}
We next assess the prediction performance of our model when no temperature and salinity data is available at the location of prediction.
We use the same set-up and data splits as in the previous cross-validation experiment, however, this time we predict without the available temperature and salinity data at the prediction location.
On the other hand, we use approximately 15000 additional Core profiles to illustrate how our model can utilize additional Core data to improve prediction performance.
The result are displayed in Table \ref{tab:no_TS}.
Table \ref{tab:no_TS} also provides a comparison of prediction errors when a Core Argo profile is nearby vs.~when no Core Argo data is nearby.
Here, nearby means that there is at least one Core Argo profile collected within 30 days and no further than 2 degrees in longitude or latitude away.
This distance yields $\approx 65$\% of the observations having nearby Core data. 
Unsurprisingly, the difference is particularly pronounced in the leave-float-out setting where there may be few BGC profiles nearby.
However, even in the leave-profile-out setting there is marginal improvement.

%\begin{comment}

\begin{table}
%\centering
% \centering
\caption{Leave-out uncertainty quantification performance. The values represent average coverage of 95\% confidence intervals. \label{tab:leave_out_coverages}}
\begin{tabular}{ccccccc}
Leave-out type &\multicolumn{5}{c}{Pointwise prediction intervals, by pressure} & Functional Band \\ 
 & All &  0--500 &  500--1000 & 1000--1500 &  1500--2000 &\\ 
LFO &0.91 &   0.91 &0.91 & 0.88 & 0.88 & 0.91\\
LPO  &0.95 & 0.94  &0.96 & 0.97 & 0.97 & 0.92 \\
\end{tabular}
\end{table}

%\end{comment}
% You must place all tables at the end of the main text. 
%Numbers in tables should be right justified or with decimal points lined up where appropriate.

\subsection{Data Product: Gridded predictions}\label{sec:pred_product}

\begin{figure}[t]
    \centering
    \includegraphics[width = .48\textwidth]{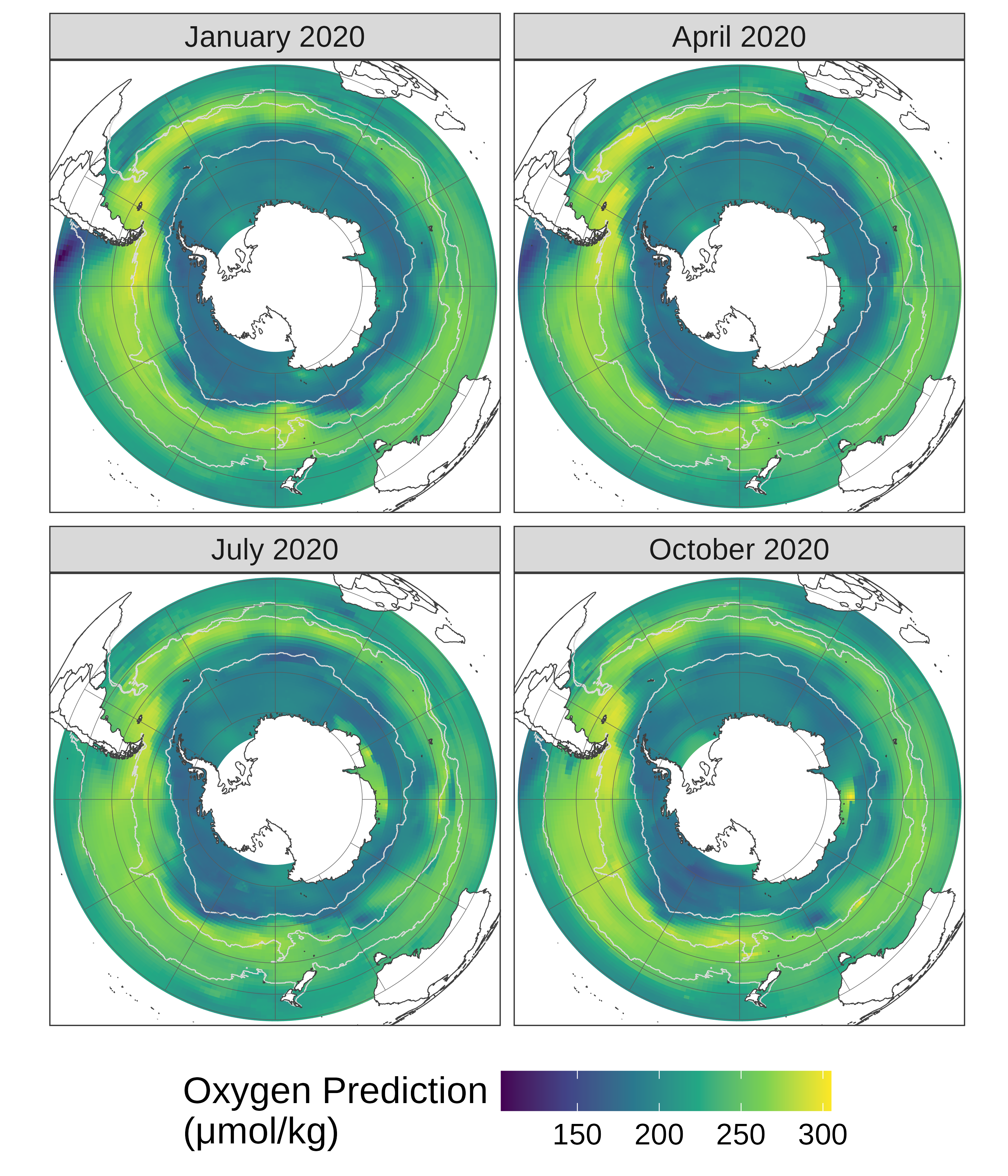}
    \includegraphics[width = .48\textwidth]{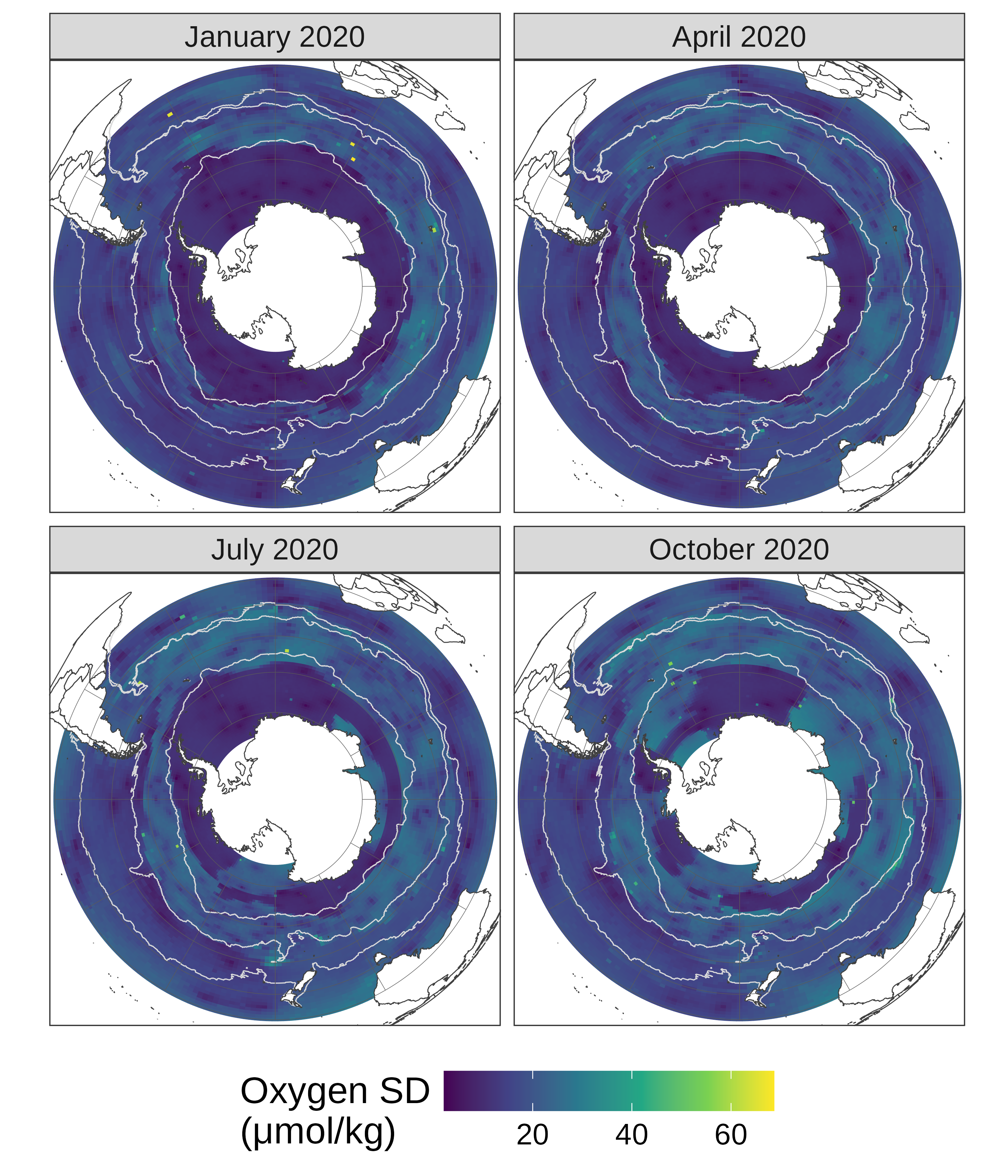}
    \caption{(Left) Oxygen prediction $\hat{\mathcal{Y}}_{\check{\mb{s}}}(p = 300 \textrm{ decibars})$ and various $t$. (Right) Their standard deviations. We also plot in white lines traditional estimates of fronts using the same criteria as \cite{bushinsky_oxygen_2017} based on the 2004-2018 \cite{roemmich_20042008_2009} estimate. }
    \label{fig:oxy_predictions}
\end{figure}

We now present overall prediction results using the entire BGC data onto a grid in the Southern Ocean. 
As an example in Figure \ref{fig:oxy_predictions}, we plot oxygen predictions at 300 decibars for four different months encompassing the seasons. 
The clustering approach results in relatively hard boundaries in the oxygen predictions.
By plotting four months of the year, we demonstrate that the Markov random field and predicted functions model some seasonality in the data, though this is less pronounced at 300 decibars compared to the surface compared to the surface. 
%In the ACC current, where oxygen is highest for each of the seasons, we see that oxygen is at its lowest level in April compared to the other months
We additionally plot the oxygen standard deviations for the same level and months.
%Uncertainty in the cluster assignment at each gridpoint filters down to the predictions of oxygen; we can see boundaries between clusters where the uncertainty in oxygen is higher. 
At 300 decibars, the prediction standard devations are generally higher at and North of the ACC. 
There are also smaller areas where the prediction standard deviation is lower, representing areas where a float collected data nearby in space and time.

\begin{figure}[t]
    \centering
    \centering
    \includegraphics[width = .32\textwidth]{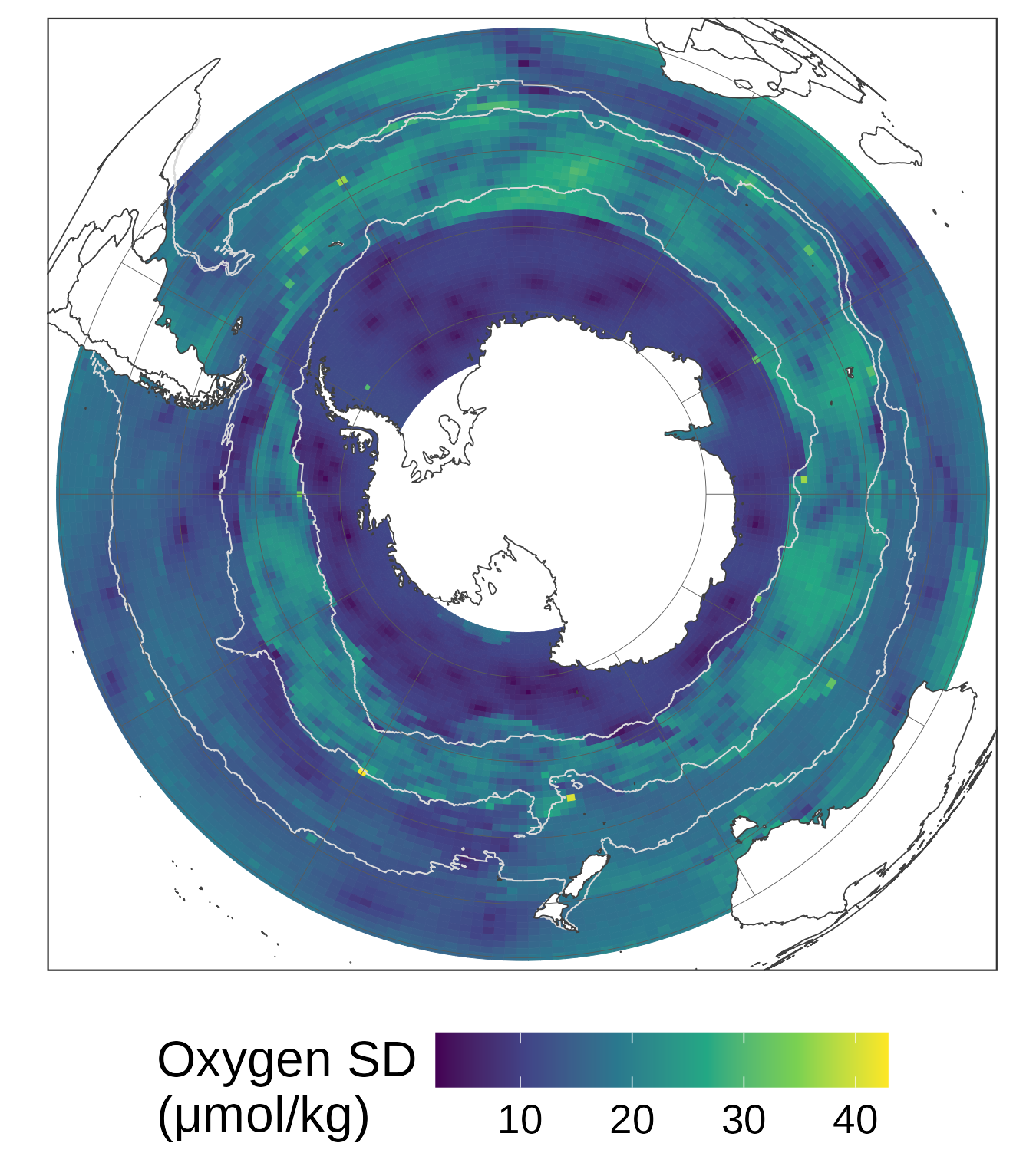}
    \includegraphics[width = .32\textwidth]{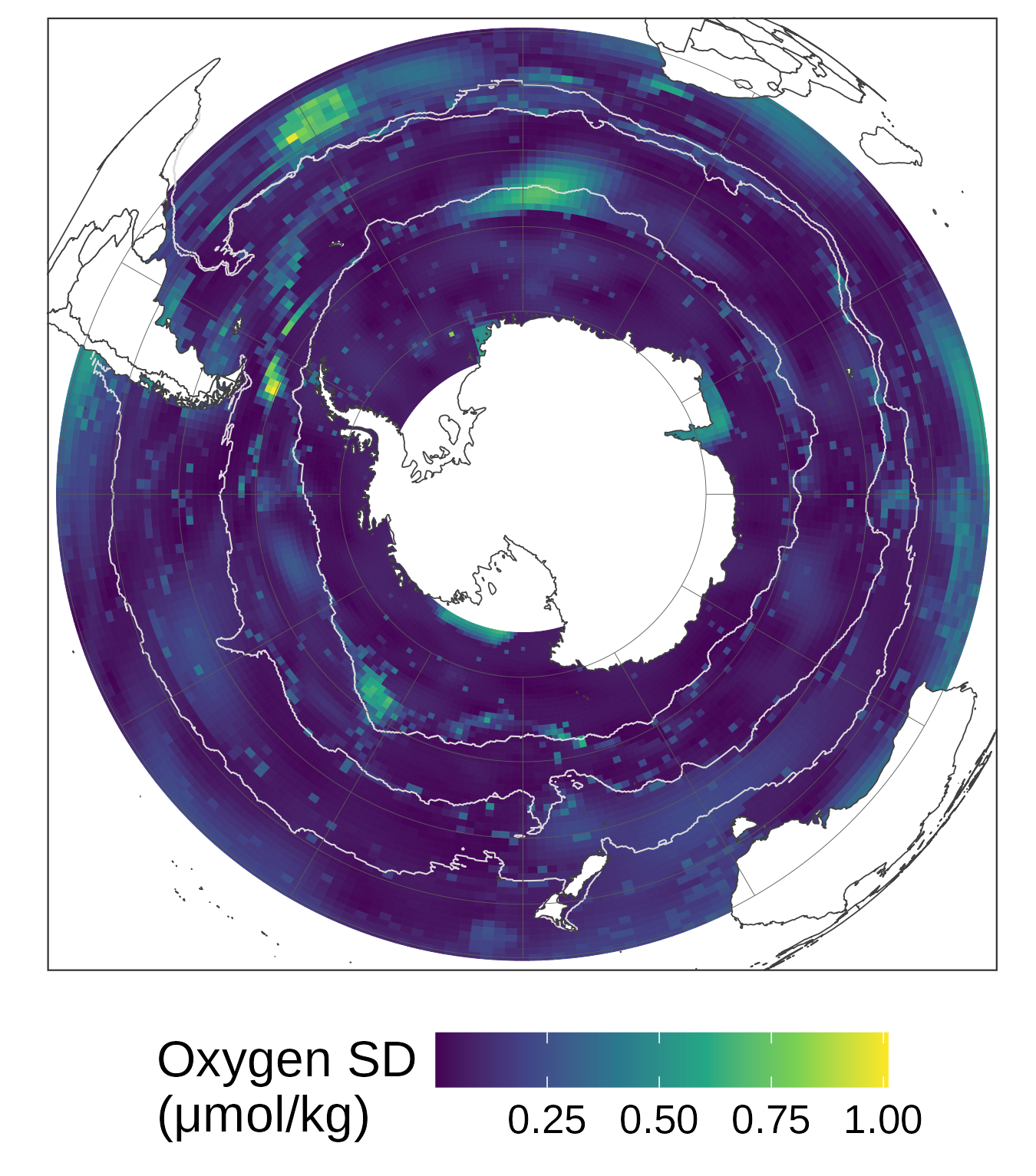}
    \includegraphics[width = .32\textwidth]{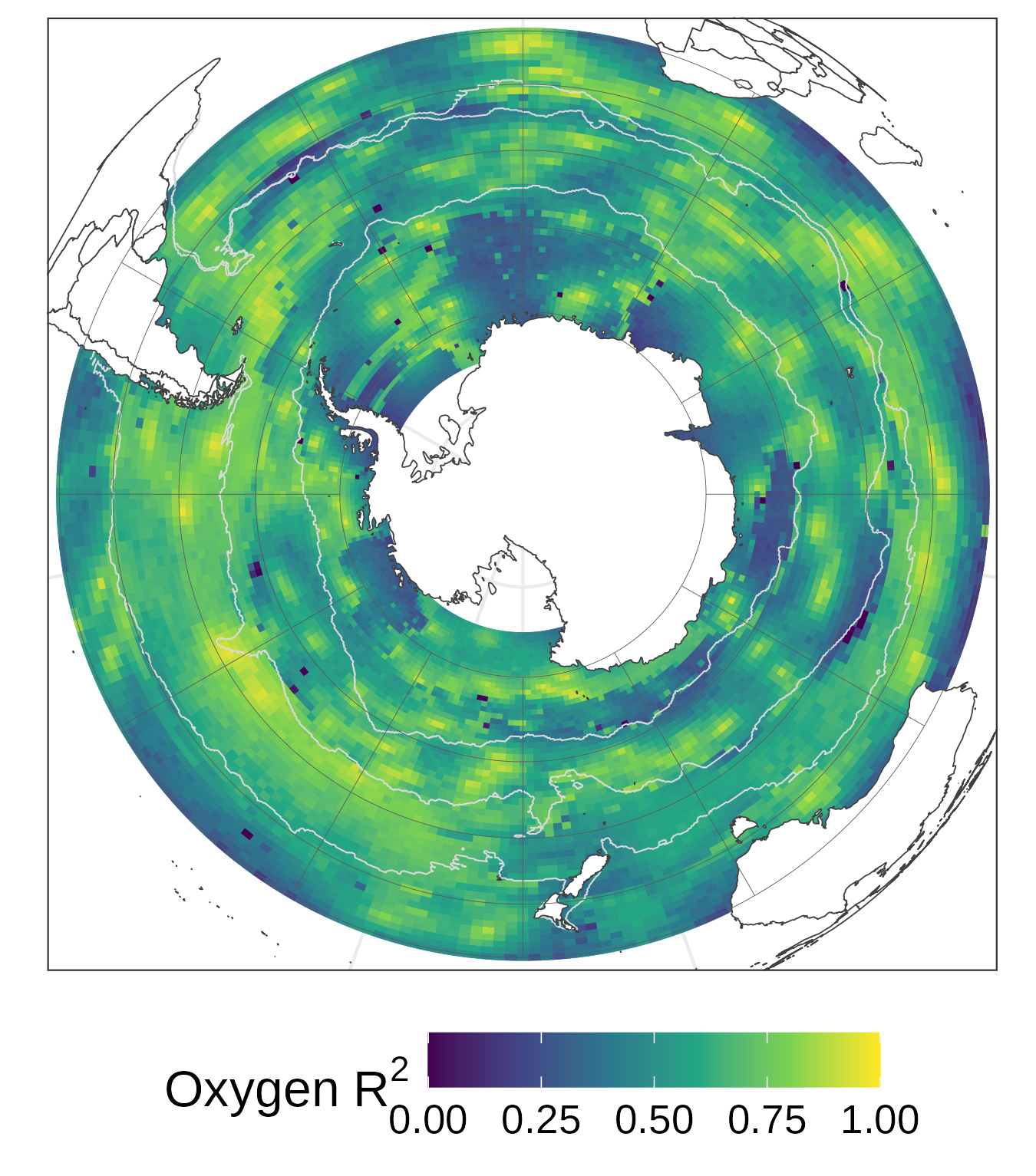}
    \caption{(Left) Model-based uncertainty of oxygen based on the scores in term \eqref{eq:pred_var_formula_2} for April 15th, 2020 at 300 decibars. (Middle) Model-uncertainty based on the clustering in term \eqref{eq:pred_var_formula_2}. The combination of this uncertainty represents the overall uncertainty in the oxygen estimates (that is, the right panel of Figure \ref{fig:oxy_predictions}. (Right) Proportion of oxygen variance explained by the covariates and spatial correlation in prediction, April 15th, 2020. %\drew{I will remake plots with less text in them. } 
    }
    \label{fig:uncertainty_decomp}
\end{figure}

In Figure \ref{fig:uncertainty_decomp}, we decompose the variance into the two terms of \eqref{eq:pred_var_formula_2} for one month. 
In general, the second term is smaller, and captures variability due to uncertainty in the cluster membership with elevated levels around fronts, while the first term captures variability to the uncertainty in the principal component scores that is decreased when floats are nearby. 
By comparing the total variance before and after prediction on the right panel of Figure \ref{fig:uncertainty_decomp}, we provide a estimate of an R-squared statistic per location, which is high in most areas. 

\begin{figure}
    \centering
    \includegraphics[width = .48\textwidth]{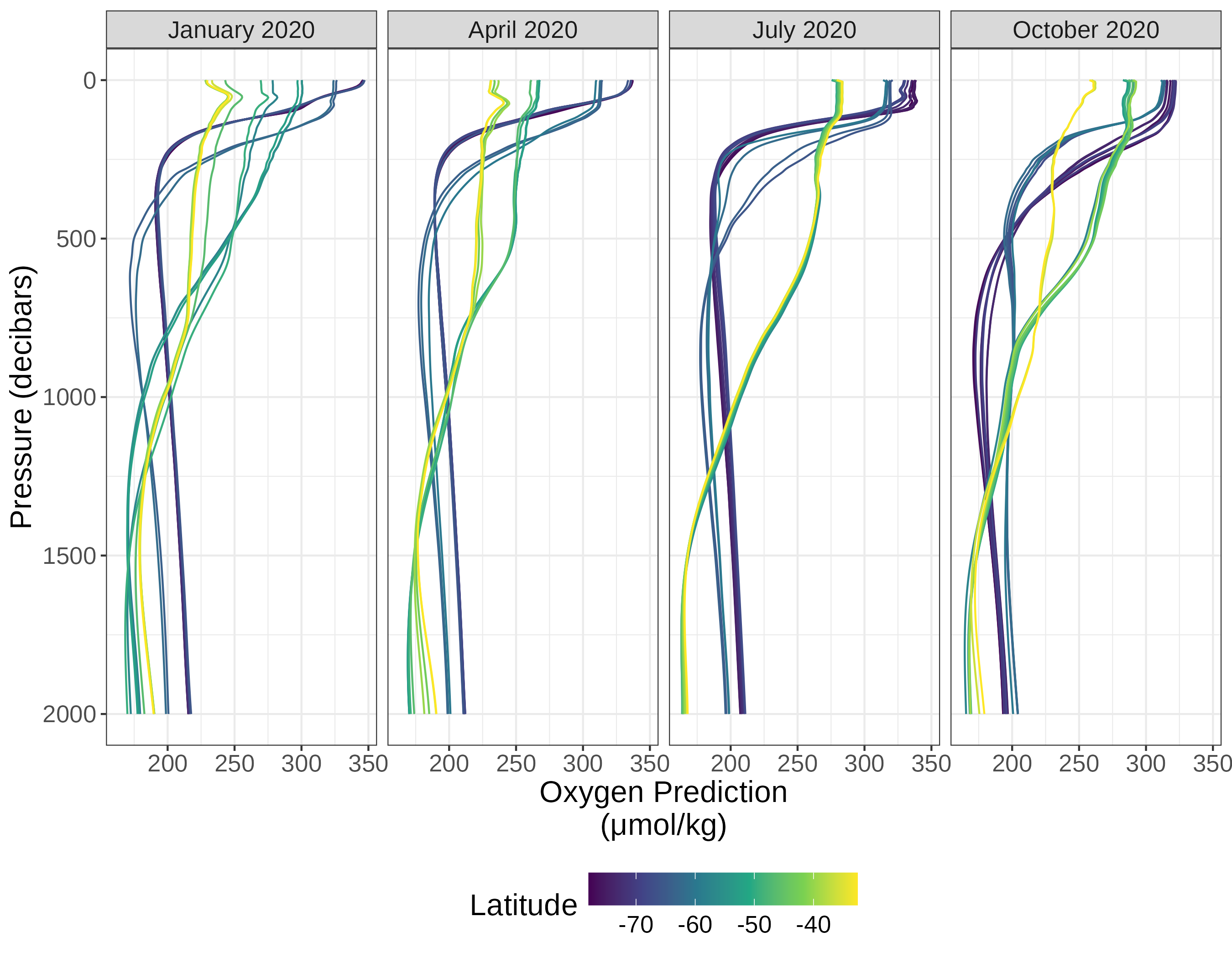}
    \includegraphics[width = .48\textwidth]{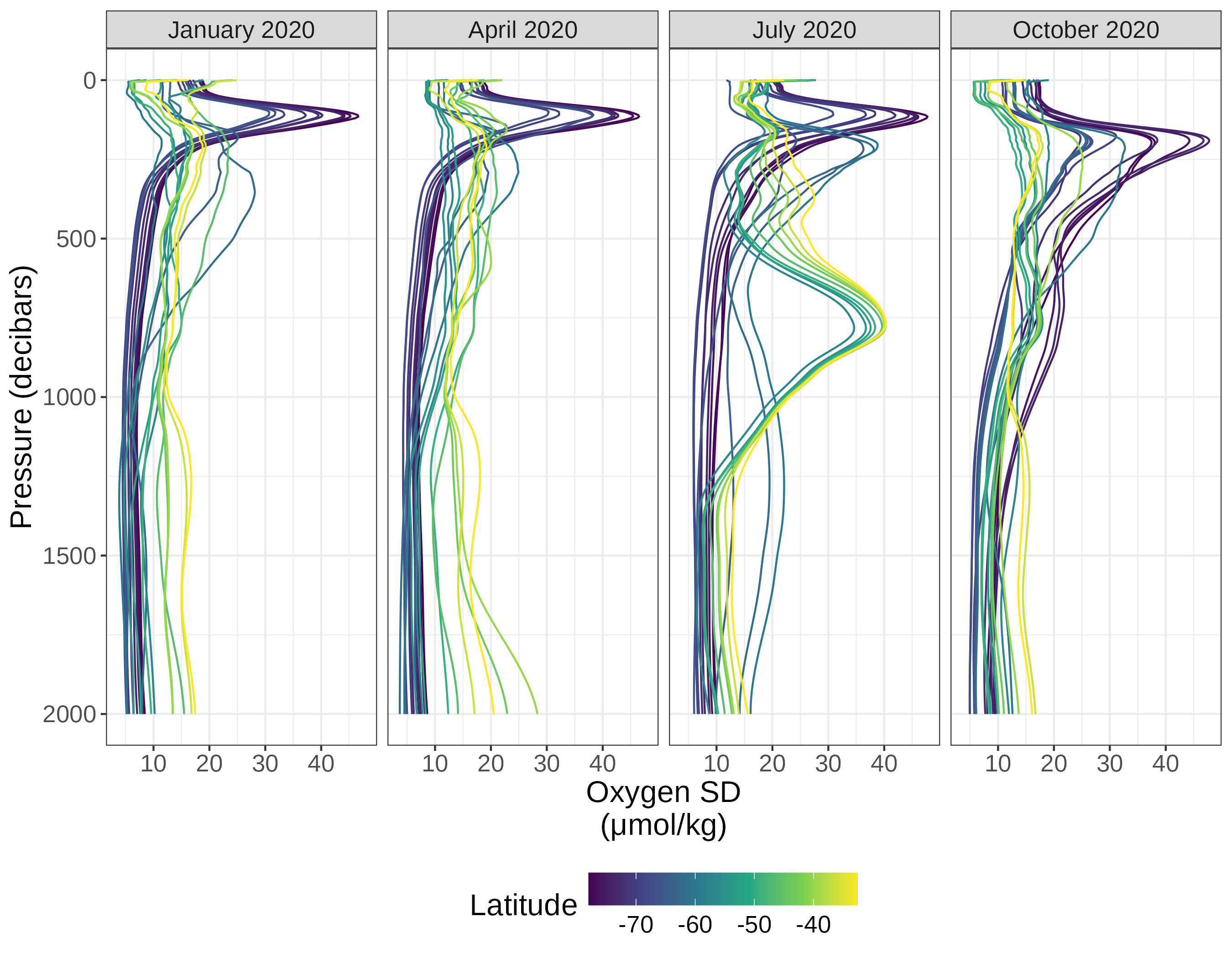}
    \caption{Estimated oxygen (left) and oxygen standard deviation (right) for longitude -90$^\circ$ for varying latitude and pressure. As one moves from south (dark) to north (light), the predicted oxygen and their uncertainties take different shapes. }
    \label{fig:section_lines_plot}
\end{figure}

%This predictions or standard deviations?)
A principal advantage of our approach is the ability to provide {\it functional} predictions. 
We plot an example in Figure \ref{fig:section_lines_plot} for locations with the same longitude and varying latitude.
As suggested by the previous plots, we can identify which cluster contributes the most to each of the locations for the most part, with few functions lying in between the clustered groups.
We also provide functional standard deviation estimates, which have different shape depending on latitude. 
Oxygen standard deviation is highest in the Southern-most sections of the Southern Ocean around 125 decibars. 

\section{Discussion}\label{sec:conclusion}

We have developed and implemented a comprehensive space-time functional regression mixture model for temperature, salinity, and oxygen.
The model yields a functional cokriging framework, where the oxygen concentration as a function of depth and the season can be estimated over the entire Southern Ocean along with (functional) uncertainties. The model has 
been validated against the most accurate existing non-functional approaches, and the mixture component of the model identifies contrasting distributions of the measured variables.

While our analysis introduces a comprehensive study of the dependence structure and prediction of oxygen in the Southern Ocean, there are a number of areas where the richness and complexity of the Argo data calls for improvements. 
As the principal components are entirely based on the pressure domain, predictions may be better if the components considered spatial information as well \citep[cf.,][]{kuenzer_principal_2020}. 
A nonfunctional prediction approach, as done in \cite{salvana20223d}, might also fare well, but requires more computing than readily available at the scale of the entire Southern Ocean. 
Our model also does not take into account the monotonicity of density or the soft constraint on the saturation of oxygen near the ocean surface, which could improve predictions; however, applying standard functional data analysis approaches were not successful due to the limited domain in the density dimension for some profiles. 
Addressing this problem with the recent research of (for example) \cite{lin2022mean} may be possible.
The current model is also developed with independence assumed between different clusters; incorporating dependence between clusters may better represent the available data. 
This, as well as evaluating whether independence should be assumed for different principal components  the principal component scores may be assumed independent \citep[as in ][]{liang2023test} would more comprehensively justify the data analysis.
Including scalar covariates like satellite data in the model is straightforward and may be of practical interest. 
Finally, using more sophisticated model selection for the Markov random field might improve the cluster predictions. 

Statistical approaches like the one developed here will play an important role for the analysis of data from the planned global deployment of BGC Argo floats (see NSF Award 1946578).

%\newpage

\section*{Data Availability Statement}

All data used in this work is publicly available. 
The Argo data is available at \cite{argo2020}, and we use the SOCCOM-specific data files from \cite{riser_soccom_2023}. 
For plotting of traditional estimates of fronts, we use the sea ice data \cite{fetterer_f_knowles_k_meier_w_savoie_m_windnagel_a_sea_2017} and the Roemmich-Gilson mean \cite{roemmich_20042008_2009}. 
Finally, we use the grid for prediction from the Southern Ocean State Estimate \citep{verdy_data_2017}.
Code to produce results presented in the paper are available at \url{https://github.com/dyarger/argo_fstmr_oxygen/}.

\if\blind0{
\section*{Acknowledgements}

Data were collected and made freely available by the Southern Ocean Carbon and Climate Observations and Modeling (SOCCOM) Project funded by the National Science Foundation, Division of Polar Programs (NSF PLR-1425989 and OPP-1936222), supplemented by NASA, and by the International Argo Program and the NOAA programs that contribute to it. (\url{https://argo.ucsd.edu},  \url{https://www.ocean-ops.org}). The Argo Program is part of the Global Ocean Observing System. 
Computational resources for the SOSE \citep{verdy_data_2017} were provided by NSF XSEDE resource grant OCE130007 and SOCCOM NSF award PLR-1425989. This research was supported in part through computational resources and services provided by Advanced Research Computing at the University of Michigan, Ann Arbor. 
Authors Korte-Stapff, Stoev, and Hsing are supported for this work by NSF Grant DMS-1916226. Author Yarger is supported for this work by NSF Grant DGE-1841052.} \fi

\bibliographystyle{rss}
%\bibliography{example}
\bibliography{cokriging.bib}

\end{document}

\maketitle

% \begin{center}
% {\large\bf SUPPLEMENTARY MATERIAL}
% \end{center}

\tableofcontents

\section{Data Description}\label{sec:data_description}

We describe the data used in the analysis here. We use:

\begin{itemize}
    \item the August 28, 2023 high-resolution MLR snapshot of the SOCCOM data:\\ \href{http://doi.org/10.6075/J0T43SZG}{https://doi.org/10.6075/J0542NS9} \citep{riser_soccom_2023},
    \item the June 6, 2021 snapshot of the Argo data for nearby core data:\\ \href{http://doi.org/10.17882/42182\#85023}{http://doi.org/10.17882/42182\#85023} \citep{argo2020}.
\end{itemize}
We remove observations with pressure less than $0$ decibar or greater than $2{,}000$ decibars.
For the SOCCOM data, we remove measurements with: a) latitude greater than $-25$ degrees, b) oxygen quality control denotes ``bad'' (8) c) temperature, salinity, or pressure quality control is not good (0) d) missing pressure or location data (this keeps linearly interpolated locations while under ice part of the data).

Following these reductions, we remove profiles with: \begin{itemize}
    \item any measurements of salinity less than $33$ PSU or greater than $37.2$ PSU
    \item a pressure gap of $200$ decibars or more between consecutive measurements
    \item a minimum pressure greater than $100$ decibars or a maximum pressure less than $1{,}450$ decibars
    \item fewer than 15 measurements
    \item had a potential density estimated derivative less than $-0.005$ kg/L/decibar for either of its core or BGC measurements
    \item there are fewer than five oxygen measurements with good (0) or probably good (4) quality control
    \item an oxygen measurement greater than $400\ \mu$mol/kg
    \item any pressure, temperature, and salinity measurements without good (0) quality control flag.
\end{itemize}
For Core Argo data with temperature and salinity only, we make similar, but not identical, exclusions for profiles (note that the Argo and SOCCOM data have different quality flag meanings):
\begin{itemize}
    \item any measurements of salinity less than $33$ PSU or greater than $37.2$ PSU
    \item a pressure gap of $200$ decibars or more between consecutive measurements
        \item a minimum pressure greater than $100$ decibars or a maximum pressure less than $1{,}450$ decibars

    \item temperature, salinity, or pressure quality control is not good (3, 4)
    \item missing pressure or location data (this keeps linearly interpolated locations while under ice part of the data)
    \item the pressure adjusted error was greater than $16$ decibars
    \item the closest SOCCOM profile was more than $500$ kilometers away
\end{itemize}

% \section{Product}\label{sec:product}
% \if\blind0{

% The predicted clusters, oxygen, temperature, and salinity values, as well as their uncertainties, are given and described in the Google Drive folder \url{https://drive.google.com/drive/folders/1_rG5woV45UHsLpX7sWl3C-lWRaNvZysX?usp=sharing}.

% }\fi

% \if\blind1{

% A link to the files of the product is included in the unblinded version here. 
% }\fi In Figure \ref{fig:cluster_month}, we plot the predicted clusters for each of the 12 months. 

\section{Additional Data Analysis Plots}\label{sec:add_pred_plots}

\begin{figure}
    \centering
    \includegraphics[width = .6\textwidth]{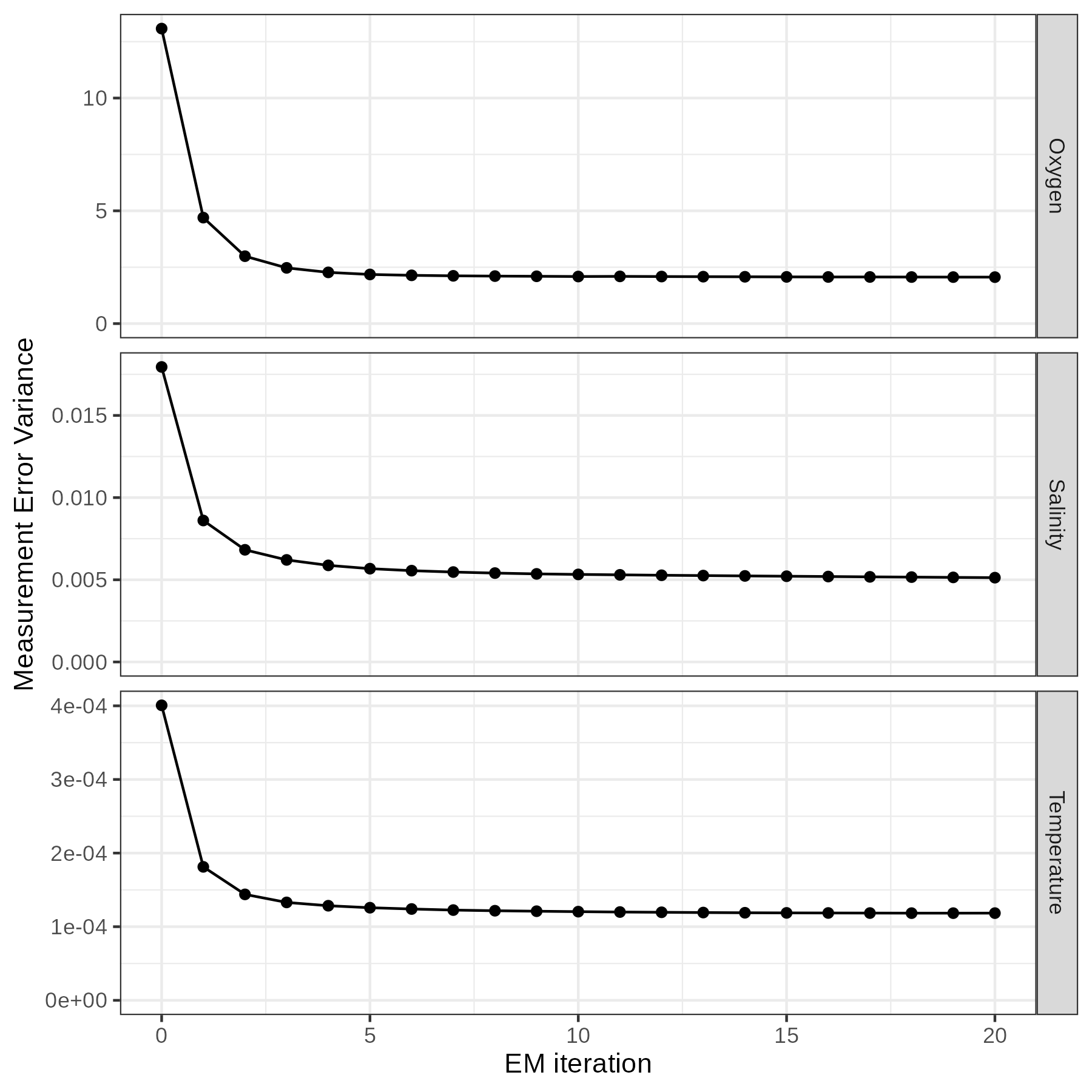}
    \caption{Estimated measurement error variables by Monte Carlo iteration for oxygen, salinity, and temperature. }
    \label{fig:meas_error}
\end{figure}

\begin{figure}
    \centering
    \includegraphics[width = .6\textwidth]{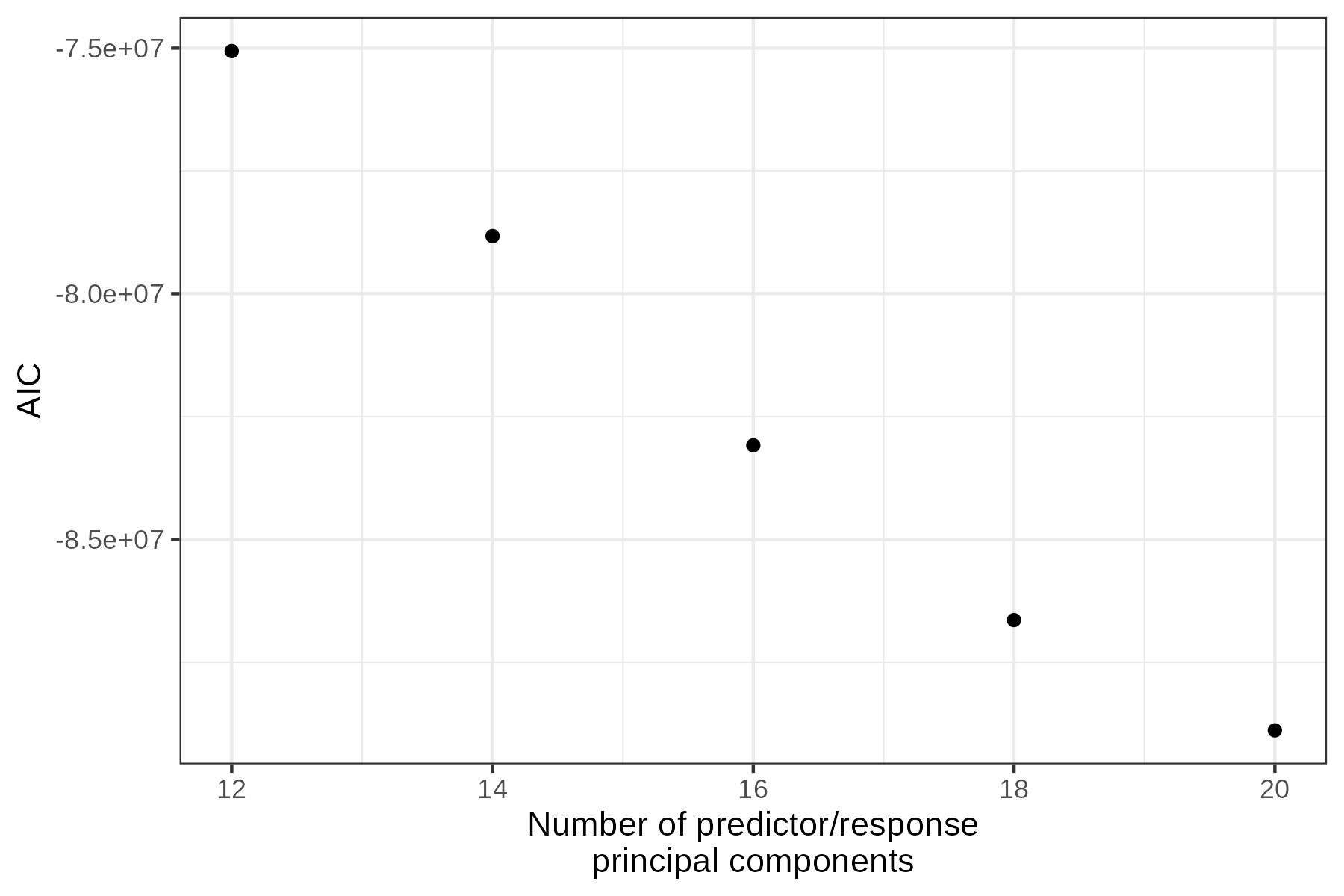}
    \caption{Akaike information criterion (AIC) for different number or principal component functions. }
    \label{fig:aic_bic}
\end{figure}

\begin{figure}
    \centering
    \includegraphics[width = .47\textwidth]{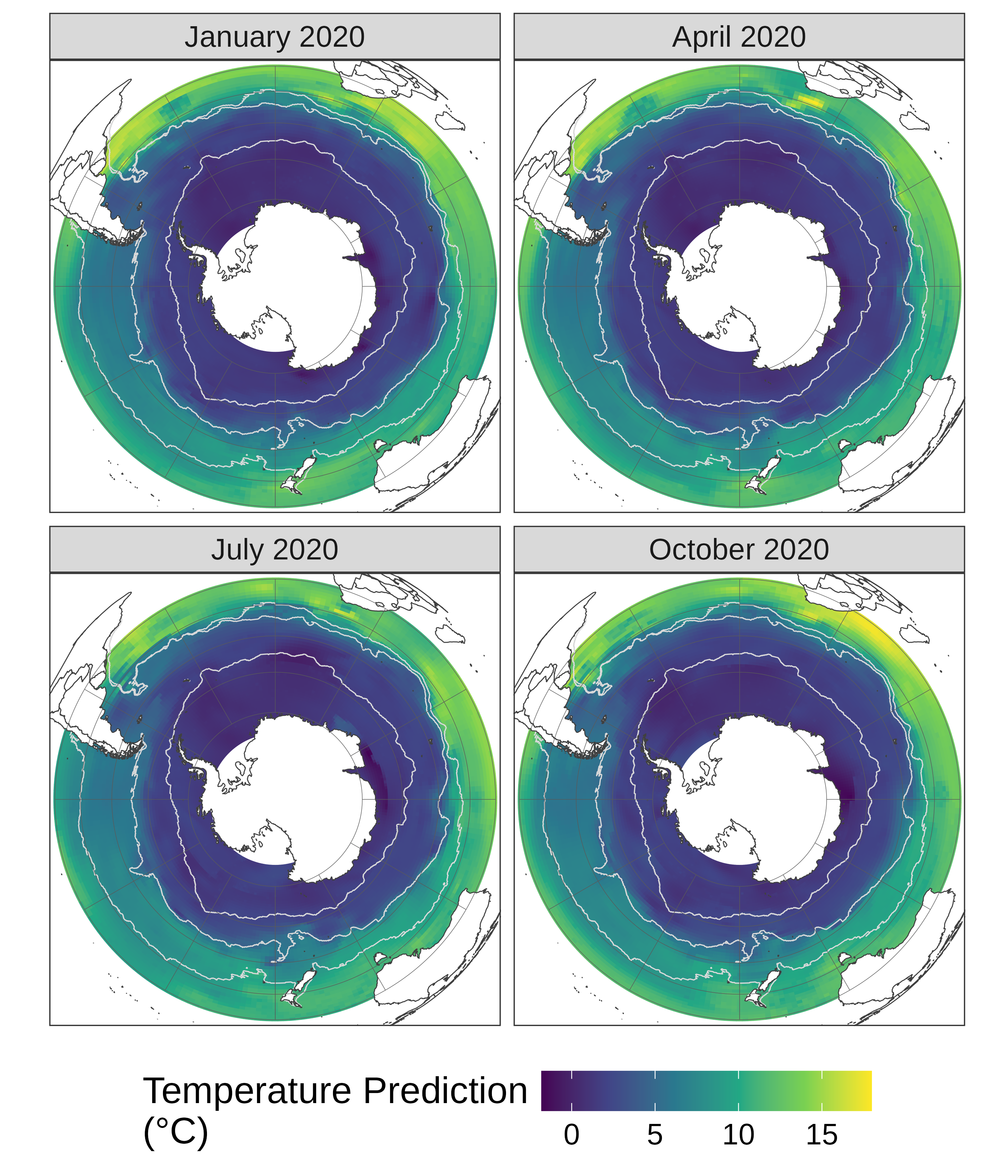}
    \includegraphics[width = .47\textwidth]{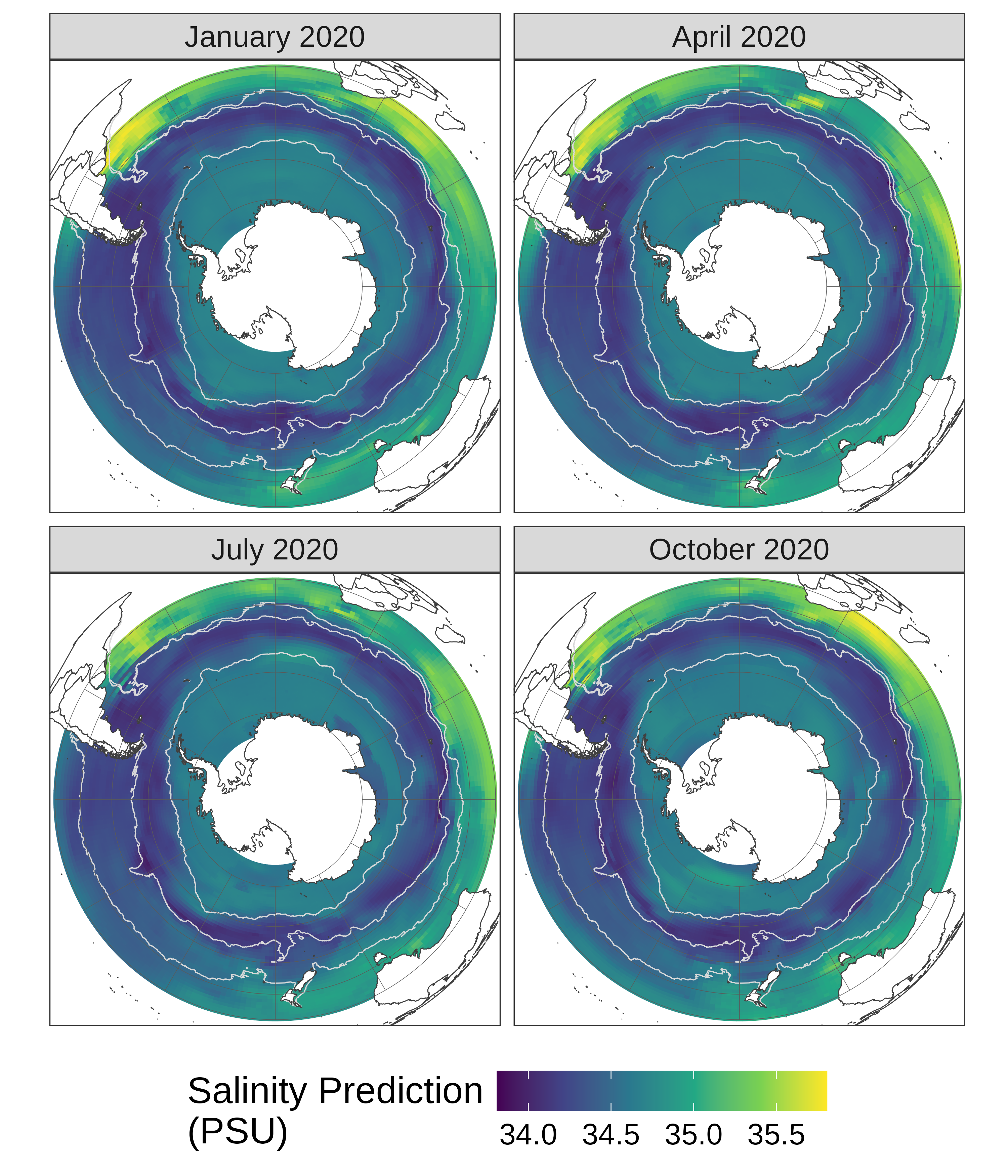}
    \caption{Temperature (left) and salinity (right) predictions for various $t$ at 300 decibar. We plot in white lines traditional estimates of fronts using the same criteria for fronts as \cite{bushinsky_oxygen_2017} based on the 2004-2018 \cite{roemmich_20042008_2009} estimate. }
    \label{fig:t_and_s}
\end{figure}

\begin{figure}
    \centering
    \includegraphics[width = .47\textwidth]{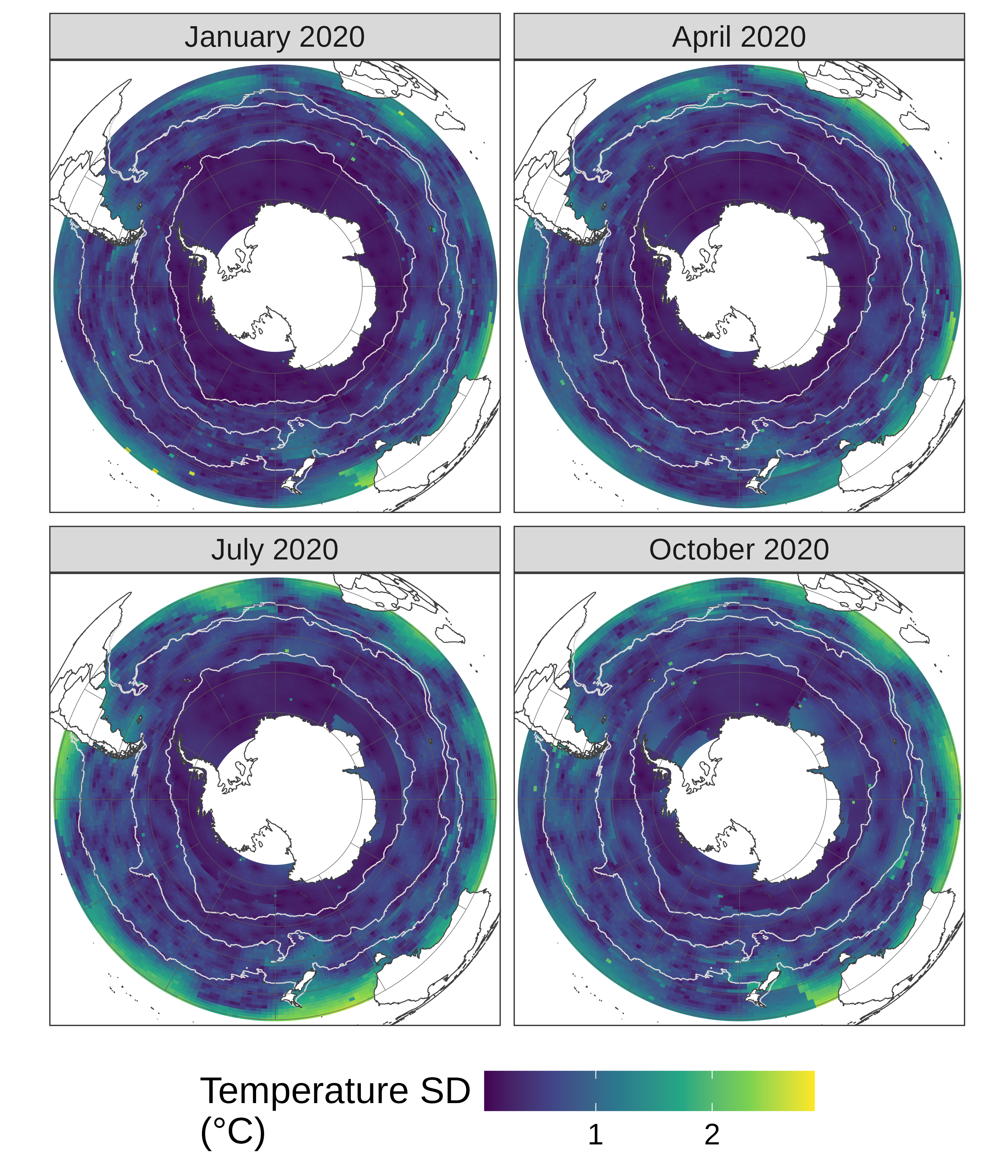}
    \includegraphics[width = .47\textwidth]{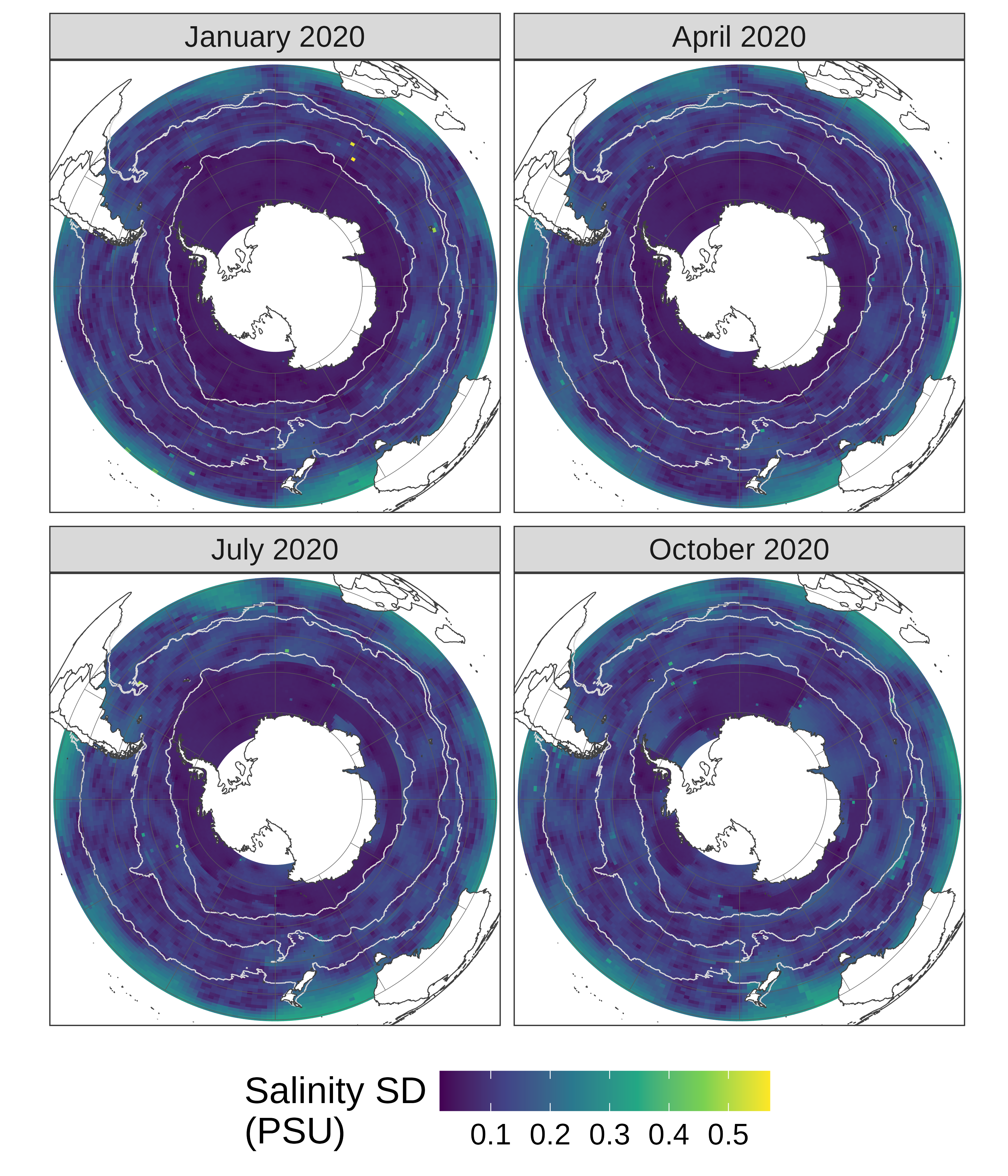}
    \caption{Temperature (left) and salinity (right) predicted standard deviations for various $t$ at 300 decibar. We plot in white lines traditional estimates of fronts using the same criteria for fronts as \cite{bushinsky_oxygen_2017} based on the 2004-2018 \cite{roemmich_20042008_2009} estimate. }
    \label{fig:t_and_s_sd}
\end{figure}

We show additional data analysis plots of interest in a variety of figures. 
In Figure~\ref{fig:meas_error}, we demonstrate the EM algorithm reduces the measurement error variances for the three variables, suggesting that the algorithm improves the fit to the data. 
We also run the model with the number of principal component functions for the response and predictors each set to the same levels: 12, 14, 16, and 20. 
The resulting AIC values are plotted in Figure \ref{fig:aic_bic}.
% In Figures~\ref{fig:mm_maps}~and~\ref{fig:mm_maps_time}, we demonstrate the seasonality dependence of the cluster memberships, showing fairly different clusters for January versus July. 
In Figures~\ref{fig:t_and_s} and \ref{fig:t_and_s_sd}, we plot predictions and predictive variances for temperature and salinity, respectively, which are obtained naturally from our analysis of temperature, salinity, and oxygen. 
%Similarly, we plot the first principal component of each estimated in Figure \ref{fig:pc_functions}: principal components of temperature and salinity, linear transformation functions, and principal components for $\mathcal{E}$.
%In Figure~\ref{fig:log_like}, we plot the estimated log-likelihood, as a function of the EM iteration, showing that our estimation strategy estimates parameters that increase the log-likelihood. 

\begin{figure}
    \centering
    \includegraphics[width = .98\textwidth]{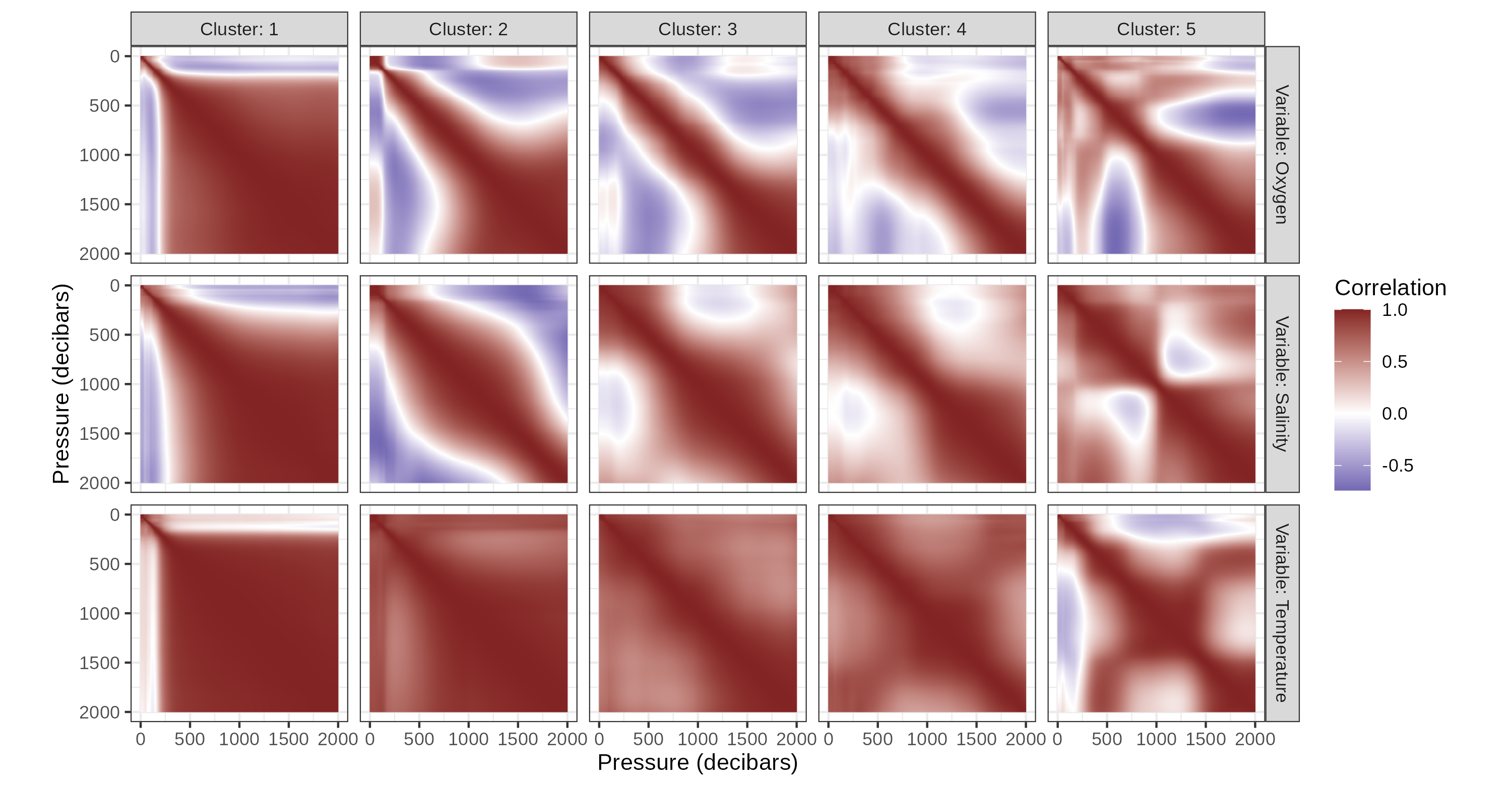}
    \caption{Estimated correlation within functional processes at different pressures by cluster.}
    \label{fig:correlation_by_var}
\end{figure}

\begin{figure}[ht]
    \centering
    \includegraphics[width = .9\textwidth]{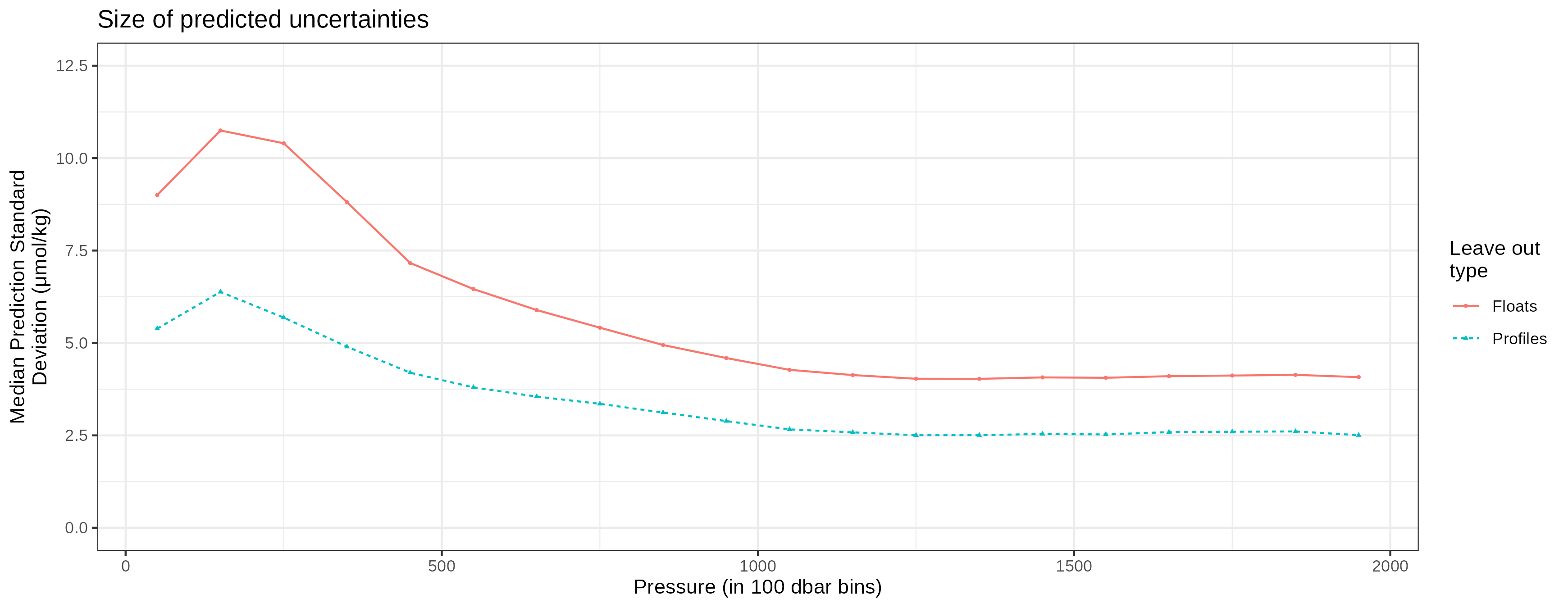}
    \caption{The median predicted standard deviations for leave-out experiments, by pressure.}
    \label{fig:leave_out_uncertainty}
\end{figure}

% \begin{figure}
%     \centering
%     \includegraphics[width = .9\textwidth]{images/log_likelihood.png}
%     %\includegraphics[scale=.13]{images/soft_var_nitrate_150_40_jul_core_predict.png}
%     \caption{Log-likelihood $\widehat{f(\mb{X}, \mb{Y})}$ for each MC iteration when considering $G = 5$, $Q_1 = Q_2 = 12$.}
%     \label{fig:log_like}
% \end{figure}    

\section{Comparison with \cite{jones_unsupervised_2019}}

We also compare with \cite{jones_unsupervised_2019}'s analysis of 6 clusters, who specifically focus on temperature data in the Southern Ocean.
%We compare to their analysis with 6 clusters.
Our January results for the first three clusters match well spatially with the first three clusters of \cite{jones_unsupervised_2019}.
For our fourth and fifth cluster, \cite{jones_unsupervised_2019} has three clusters that are more spatially divided into the Atlantic, Indian, and Pacific sectors. 
Our clusters are also determined by salinity and oxygen, which may result in less regional separation compared to when using temperature only. 
In looking at our seasonality plot, there are areas near Antarctica that are classified in cluster 2 instead of cluster 1 for some months; \cite{jones_unsupervised_2019} also identifies a number of profiles in the same areas with an outlying cluster. 
%In another work, \cite{rosso_water_2020} focus on the Kerguelen Sector of the Southern Ocean.
%Because of this limit, as well as a higher number of clusters used, our results are less directly comparable.  %they have so many more clusters, and more profiles in the north
Our results also demonstrate substantial space-time variability in the clusters, which \cite{jones_unsupervised_2019} does not analyze extensively.
%\moritz{Say someting here about ice moving?}

\section{Details on Importance Sampling}\label{S:sec:monte_carlo}

\textbf{Monte-Carlo E-step.} Our goal is to construct a proposal $h$ that is easy to sample from and that is close to the true conditional distribution $f(\mb{\alpha}, \mb{\eta}, \mb{Z} \vert \mb{X}, \mb{Y})$.
To do so, the key ingredient is that sampling from $f(\mb{\alpha}, \mb{\eta} \vert \mb{Z} = \mb{z}^{(\tau)}, \mb{X}, \mb{Y})$ is tractable; see Proposition \ref{prop:cond_dist1}.
It then only remains to specify a proposal distribution $\mathbb{P}_I$ for sampling the latent process of cluster memberships $\mb{Z} \vert \mb{X}, \mb{Y}$ and employ the following procedure:
\begin{enumerate}
    \item Sample $\mb{z}^{(\tau)}$ from $\mathbb{P}_I$ and compute $w^{(\tau)}$.
    \item Sample $\mb{\alpha}^{(\tau)}, \mb{\eta}^{(\tau)}$ from $f\big(\mb{\alpha}, \mb{\eta} \vert \mb{Z} = \mb{z}^{(\tau)}, \mb{X}, \mb{Y}\big)$.
\end{enumerate}
We now construct $\mathbb{P}_I$.
By Bayes' Theorem $\smash{\mathbb{P}(\mb{Z} = \mb{z} \vert \mb{X}, \mb{Y}) \propto f(\mb{X}, \mb{Y} \vert \mb{Z} = \mb{z})\mathbb{P}(\mb{Z} = \mb{z})}$.
Now replace $f(\mb{X}, \mb{Y} \vert \mb{Z} = \mb{z})$ with $\smash{f_I(\mb{X}, \mb{Y} \vert \mb{Z} = \mb{z}) := \prod_{i=1}^n f(\mb{X}_i, \mb{Y}_i | Z_i = z_i)}$, and set 
\begin{align}
    \mathbb{P}_I(\mb{Z} = \mb{z} | \mb{X}, \mb{Y}) &\propto f_I(\mb{X}, \mb{Y} \vert \mb{Z} = \mb{z})\mathbb{P}(\mb{Z} = \mb{z}) \nonumber\\
    &\propto \exp\left(\sum_{i = 1}^n\log(f(\mb{X}_i, \mb{Y}_i \vert Z_i = z_i)) + \xi\sum_{j_1\sim j_2} \frac{\mathbb{I}(z_{j_1} = z_{j_2})}{\mathrm{dist}(j_1, j_2)} \right) \label{eq:Potts},
\end{align}
where $\mb{z} = \{z_i\}_{i=1}^n$ and $i \sim j$ indicates that $i$ and $j$ are neighbors so that the second sum is over all possible neighboring pairs $j_1$ and $j_2$. 
Sampling from \eqref{eq:Potts}, known as the Potts model, has been studied extensively, see for example \cite{feng_2011} and references therein. 
We use a Gibbs sampler based on a single site updating scheme similar to \cite{feng_2011} to sample from $\mathbb{P}_I(\cdot \vert \mb{X}, \mb{Y})$. 
With this, the final proposal distribution is $\smash{h(\mb{\alpha}, \mb{\eta}, \mb{Z}) = \mathbb{P}_I(\mb{Z} = \mb{z} | \mb{X}, \mb{Y}) f(\mb{\alpha}, \mb{\eta} \vert \mb{Z} = \mb{z}, \mb{X},\mb{Y})}$.
Utilizing the fact that we can sample from $f(\mb{\alpha}, \mb{\eta} \vert \mb{Z} = \mb{z}, \mb{X}, \mb{Y})$, the importance sampling weights can be computed via Bayes' Theorem
\begin{align*}
    w^{(\tau)} 
    = \frac{\mathbb{P}\big(\mb{Z} = \mb{z}^{(\tau)}\vert\mb{X}, \mb{Y}\big)}{\mathbb{P}_I\big(\mb{Z} = \mb{z}^{(\tau)} \vert \mb{X}, \mb{Y}\big)} = C\frac{f\big(\mb{X}, \mb{Y} \vert \mb{Z} = \mb{z}^{(\tau)}\big) }{\prod_{i = 1}^n f\big(\mb{X}_i, \mb{Y}_i  \vert Z_i = z_i^{(\tau)}\big)}.
\end{align*}
The unknown constant $C$ cancels when self-normalizing the weights. 

\textbf{Model Selection.}
For model selection, we need a different proposal because the one used in the E-step can be sampled from but cannot easily be evaluated.
Again we only need a proposal for the latent field $\mb{Z}$ given the observed data.
Set
\begin{align}
    \mathbb{P}_{I_2}(\mb{Z} = \mb{z} \vert \mb{X}, \mb{Y}) = \prod_{i = 1}^n \frac{f(\mb{X}_i, \mb{Y}_i \vert Z_i = z_i)}{ f(\mb{X}_i, \mb{Y}_i)} \label{eq:model_selection_PI}
\end{align}
which corresponds to a Gaussian mixture model with independent observations. 
For any fixed $\mb{z}$, \eqref{eq:model_selection_PI} can easily be evaluated exactly.
Using Bayes' Theorem, we propose to approximate the observed data likelihood based on the samples $\mb{z}^{(1)}, \dots, \mb{z}^{(T)}$ from $\mathbb{P}_{I_2}$ via 
\begin{align*}
    f(\mb{X}, \mb{Y})  &= \sum_{\mb{z} \in \mathcal{Z}} f(\mb{X}, \mb{Y} \vert \mb{Z} = \mb{z}) \mathbb{P}(\mb{Z} = \mb{z}) \approx \frac{1}{T}\sum_{\tau = 1}^T \frac{f\big(\mb{X}, \mb{Y} \vert \mb{Z} = \mb{z}^{(\tau)}\big)\mathbb{P} \big(\mb{Z} = \mb{z}^{(\tau)}\big)}{ \mathbb{P}_{I_2} (\mb{Z} = \mb{z}^{(\tau)} \vert \mb{X}, \mb{Y})}
\end{align*}
Every term on the right hand side can be evaluated analytically with the exception of $\smash{\mathbb{P}\big(\mb{Z} = \mb{z}^{(\tau)}\big)}$ which can only be evaluated up to a constant.
However, this constant is constant across Monte-Carlo samples and hyper parameters to be selected and thus cancels when comparing the AIC criterion for different hyperparameters.

\section{Conditional Distributions}\label{app:cond_dist}

Here, we provide the conditional distributions required for our MCEM algorithm.
The first proposition gives the conditional distribution of the observed data given the latent field $\{Z(s_i)\}_{i = 1}^n$ in various scenarios.
In the following, for arbitrary matrices $A_1, \dots, A_M$, let bdiag($A_1, \dots, A_M$) denote the block diagonal matrix with the matrices $A_1, \dots, A_M$ on the diagonal.

\begin{proposition}\label{prop:cond_dist1}
    Let $\alpha$ be a normally distributed random vector of size $m_1$ with mean 0 and covariance $\Sigma_{\alpha}$. Let $\eta$ be another normally distributed random vector of size $m_2$ with mean 0 and covariance $\Sigma_{\eta}$.
    Furthermore, let $\epsilon_X$ and $\epsilon_Y$ by two random vectors with
    \begin{align*}
        \epsilon_Y \sim \mathcal{N}(0, \sigma_Y^2 I_{N_2}) ~~ \text{ and } ~~ \epsilon_X \sim \mathcal{N}\left(0, \mathrm{bdiag}(\sigma_{X_1}^2 I_{n^1}, \dots, \sigma_{X_K}^2 I_{n^K})\right) =: \mathcal{N}\left(0, D_X\right),
    \end{align*}where the parameters $\sigma_Y^2>0$ and $\sigma_{X_k}^2 > 0$ for $k=1, \dots, K$ are measurement error variances.
    Assume that $\alpha, \eta, \epsilon_X$ and $\epsilon_Y$ are mutually independent of each other.
    Let $N_1 = \sum_{k = 1}^K n^k$ and assume $\min(N_1, N_2) > \max(m_1, m_2)$.
    Furthermore, let $A_1, A_2, A_3$ be three matrices of dimensions $N_1 \times m_1$, $N_2 \times m_1$ and $N_2 \times m_2$ respectively. 
    Define
    \begin{align*}
        X &= A_1 \alpha + \epsilon_X, & Y &= A_2 \alpha + A_3 \eta + \epsilon_Y.
    \end{align*}
    Then
    \begin{align*}
        \begin{pmatrix}
            X \\ Y
        \end{pmatrix} \sim \mathcal{N}\left(0, \begin{pmatrix}
            D_X + A_1 \Sigma_{\alpha}A_1^{\top} & A_1 \Sigma_{\alpha}A_2^{\top} \\
            A_2\Sigma_{\alpha}A_1^{\top} & \sigma_Y^2 I_{n_2} + A_2\Sigma_{\alpha}A_2^{\top} + A_3\Sigma_{\eta}A_3^{\top}
        \end{pmatrix}\right).
    \end{align*}
\end{proposition}
\begin{proof}
    This follows directly from standard properties of the multivariate normal distribution.
\end{proof}
Next, we derive the conditional density $f(\mb{\alpha}, \mb{\eta} \vert \mb{X}, \mb{Y}, \mb{Z})$.
As observations from different clusters are assumed to be independent, it suffices to derive the distribution for specific cluster. 
Thus to ease notation, we drop cluster dependence.

\begin{proposition}\label{prop:cond_dist2}
    Let $B_i$ be the $n_i \times P$ matrix of basis evaluations for profile $i$ for the response. Similarly, let $\mb{B}_i$ be the respective $Kn_i \times KP$ be the respective matrix for the covariates, which is the block-diagonal matrix of evaluations for each predictor.
    Write $\tilde{\mb{B}} = \mathrm{bdiag}(B_1, \dots, B_n)$ and $\mb{B} = \mathrm{bdiag}(\mb{B}_1, \dots, \mb{B}_n)$. 
    For any matrix or vector $\Theta$ with appropriate dimension, write 
    \begin{align*}
        \mb{B}\Theta = \mathrm{bdiag}(\mb{B}_1\Theta,\dots \mb{B}_n\Theta) ~~~ \text{ and } \tilde{\mb{B}}\Theta = \mathrm{bdiag}(B_1\Theta,\dots B_n\Theta)
    \end{align*}
    as well as $D_X = \mathrm{bdiag}\left(D_{X^1}, \dots, D_{X^n}\right)$ where $
        D_{X^i}= \Var\left[\left(\epsilon_{i1}^1, \dots,\epsilon_{in_i}^1, \dots, \epsilon_{i1}^K, \dots,\epsilon_{in_i}^K \right)\right]$.
    For the response, let $\sigma^2_Y= \Var(\epsilon_{11}^Y)$.
    In addition, let $\Sigma_\alpha = \Var(\mb{\alpha})$ and $\Sigma_\eta = \Var(\mb{\eta})$ be defined by the spatial covariance. 
    Let
    \begin{align*}
        A = \begin{pmatrix}
            \Sigma_{\alpha} & 0 \\
            0 & \Sigma_{\eta}
        \end{pmatrix} \text{ and } U = \begin{pmatrix}
            D_X^{-1/2}\mb{B} \Theta_{\mathcal{X}} & 0 \\
            \sigma_Y^{-1}\tilde{\mb{B}}\Lambda & \sigma_{Y}^{-1}\tilde{\mb{B}}\Theta_{\mathcal{E}}
        \end{pmatrix}
    \end{align*}
    Within the same cluster, we have 
    \begin{align*}
        \mb{\alpha}, \mb{\eta} \vert \mb{X}, \mb{Y} &\sim \mathcal{N}\left((A^{-1} + U^{\top}U)^{-1} \begin{pmatrix}
            \mb{X} - \mb{B}_t \Upsilon_X \\ \mb{Y} - \tilde{\mb{B}}_t\Upsilon_Y
        \end{pmatrix}, (A^{-1} + U^{\top}U)^{-1}\right) \\
        &= \mathcal{N}\left((A^{-1} + U^{\top}U)^{-1} \begin{pmatrix}
            \mb{X}_0 \\ \mb{Y}_0
        \end{pmatrix}, (A^{-1} + U^{\top}U)^{-1}\right)
    \end{align*}

    Here, $\Theta_\mathcal{E}$, $\Lambda$, and $\Theta_{\mathcal{X}}$ are the respective matrices for the principal component/random effect function coefficients,  $\Upsilon_\mathcal{X} = [\Upsilon_{\mathcal{X}, r}]_{r=1}^{2R + 1}$ and $\Upsilon_\mathcal{Y}= [\Upsilon_{\mathcal{Y}, r}]_{r=1}^{2R + 1}$ are the respective vectors for the mean coefficients, ignoring notation on the cluster, and $\mb{B}_t = \mathrm{bdiag}(\mb{B}_1 \otimes \delta(t_1), \dots, \mb{B}_n \otimes \delta(t_n))$, $\tilde{\mb{B}}_t = \mathrm{bdiag}(B_1 \otimes \delta(t_1), \dots, B_n \otimes \delta(t_n))$, where \begin{align}\begin{split}\delta(t) &= (1, \sin(2\pi t/365.25), \dots, \sin(2\pi Rt/365.25),  \\
    &~~~~~~~~~~~\cos(2\pi t/365.25), \dots, \cos(2\pi Rt/365.25))\end{split}\label{eq:delta_mean_cov}\end{align}is the row vector of covariates for the mean functions.
\end{proposition}

\section{Description of Prediction Formulas}\label{sec:pred_add_info}

As in other parts of the supplement, we ignore the dependence on clusters in the following developments.
In Proposition \ref{prop:cond_dist1}, we found that, in each cluster, 
\begin{align}
    \mb{\alpha}, \mb{\eta} | \mb{X}, \mb{Y} &\sim \mathcal{N}\left((A^{-1} + U^\top U)^{-1} U^\top \begin{pmatrix}\mb{X}_0 \\ \mb{Y}_0\end{pmatrix} , (A^{-1} + U^\top U)^{-1} \right) \\
    &=  \mathcal{N}\left(\Sigma U^\top \begin{pmatrix}\mb{X}_0 \\ \mb{Y}_0\end{pmatrix} , \Sigma \right)\label{eq:conditional_distribution_scores_E_step}
\end{align}
where $\mb{X}_0$ and $\mb{Y}_0$ are centered versions of the data, $A$ and $U$ are defined as in Proposition \ref{prop:cond_dist2}.

We now consider new scores $\check{\mb{\alpha}}$ and $\check{\mb{\eta}}$ where we want to predict, potentially with additional predictor data $\check{\mb{X}}$. For simplicity, suppose that $\check{\mb{X}}$ is available; however, we discuss later how to form predictions when $\check{\mb{X}}$ is not available. Consider the matrix 
\begin{align*}
    \check{A}&= \begin{pmatrix} \Sigma_\alpha & \Sigma_{\alpha, \check{\alpha}} & 0 & 0 \\ \Sigma_{\alpha, \check{\alpha}}^\top & \Sigma_{\check{\alpha}} & 0 & 0 \\
    0 & 0 & \Sigma_{\eta} & \Sigma_{\eta,  \check{\eta}}\\ 0 &0  & \Sigma_{\eta, \check{\eta}}^\top & \Sigma_{\check{\eta}}
    \end{pmatrix}
\end{align*}
where $\Sigma_{\alpha, \check{\alpha}} = \mathbb{C}\textrm{ov}(\mb{\alpha}, \check{\mb{\alpha}})$, $\Sigma_{\check{\alpha}} = \mathbb{V}\textrm{ar}(\check{\mb{\alpha}})$, and likewise for the other matrices involving $\eta$. 
Also, note that $\check{A}^{-1}$ can be formed using a Vecchia approximation for the entire matrix 
$$
\begin{pmatrix}
    \Sigma_\alpha & \Sigma_{\alpha, \check{\alpha}} \\
    \Sigma_{\alpha,\check{\alpha}}^\top & \Sigma_{\check{\alpha}} 
\end{pmatrix}^{-1}
$$ 
and likewise for $\eta$ and $ \check{\eta}$. 
Recall the definition of $U$ in Proposition \ref{prop:cond_dist2}, and similarly define the matrices 
\begin{align*}
    U_{X, \alpha} &= D_X^{-1/2}\mb{B}\Theta_\mathcal{X}, &
    U_{Y, \alpha} &= \sigma_Y^{-1}\tilde{\mb{B}}\Lambda, &U_{Y, \eta} &=  \sigma_Y^{-1}\tilde{\mb{B}}\Theta_\mathcal{Y}
\end{align*}  Furthermore, let \begin{align*}
    U_{\textrm{all}} &= \begin{pmatrix} 
    U_{X, \alpha} & 0 & 0 & 0\\ U_{Y, \alpha} & 0 & U_{Y,\eta} & 0 \\
    0 & U_{\check{X}, \check{\alpha}} & 0 & 0
     \end{pmatrix}
\end{align*}where $U_{\check{X}, \check{\alpha}} = D_{\check{X}}^{-1/2}\mb{B}_{\check{X}}\Theta_\mathcal{X}$ is the respective matrix for the new data. The three groups of rows correspond to $\mb{X}$ and $\mb{Y}$, and $\check{\mb{X}}$, respectively, while the four groups of columns correspond to $\mb{\alpha}$, $\check{\mb{\alpha}}$, $\mb{\eta}$, and $\check{\mb{\eta}}$. 

\begin{proposition}\label{prop:cond_dist_pred}In this case, the conditional distribution of the scores can be written:
\begin{align*}
    \mb{\alpha}, \check{\mb{\alpha}}, \mb{\eta}, \check{\mb{\eta}} | \mb{X}, \mb{Y},\check{\mb{X}} \sim \mathcal{N}\left(\left(\check{A}^{-1} + U_{\rm{all}}^\top U_{\rm{all}}\right)^{-1} U_{\rm{all}}^\top \begin{pmatrix}{\mb{X}_0} \\ {\mb{Y}}_0 \\ \check{{\mb{X}}}_0\end{pmatrix} , \left(\check{A}^{-1} + U_{\rm{all}}^\top U_{\rm{all}}\right)^{-1} \right)
\end{align*} where ${\mb{X}}_0$, ${\mb{Y}}_0$, and $\check{{\mb{X}}}_0$ are centered versions of the data. 
\end{proposition}

The proof is similar to that of Proposition \ref{prop:cond_dist2}.
Likewise, when $\check{\mb{X}}$ is not available, we form $U_{\textrm{all}}$ with one group of rows deleted: \begin{align*}
    U_{\textrm{all}} &= \begin{pmatrix} 
    U_{X, \alpha} & 0 & 0 & 0\\ U_{Y, \alpha} & 0 & U_{Y,\eta} & 0 
     \end{pmatrix}
\end{align*}and compute the same formulas with only $\mb{X}$ and $\mb{Y}$ as the available data.

\section{Computation of Conditional Distributions}\label{sec:efficient_comp}

Consider the notation in Section \ref{app:cond_dist}, and note that our goal is to sample from $f(\mb{\alpha}, \mb{\eta} \vert \mb{X}, \mb{Y})$ efficiently.
All distributions should be considered conditional on $\mb{Z}$, but this notation is omitted as in Section \ref{app:cond_dist}. 
Write again
\begin{align*}
    \Sigma^{-1} = A^{-1} + U^{\top}U.
\end{align*}
In practice, $\Sigma^{-1}$ will be considerably sparse due to the independence assumption on the scores and because we apply Vecchia's approximation \citep{guinness_gaussian_2021} resulting in a sparse matrix $A^{-1}$.
Note that $U^{\top}U$ is (essentially) a block diagonal matrix.
However, the actual covariance matrix $\Sigma$ is not sparse. 
We next illustrate how to use sparseness of $\Sigma^{-1}$ instead.
\begin{enumerate}
    \item Instead of computing $\Sigma$ explicitly, we compute the conditional expectation in \eqref{eq:conditional_distribution_scores_E_step} using the conjugate Gradient solver as implemented in the \texttt{ConjugateGradient} class of the Eigen \texttt{C++} library \citep{eigenweb}. 
    This takes explicit advantage of the sparsity of $\Sigma^{-1}$.
    If $(\hat{\mu}^t_{\alpha}, \hat{\mu}^t_{\eta})^{\top}$ is output of running the conjugate Gradient solver after $t$ iterations, then we stop when 
    \begin{align*}
        \left\lVert \Sigma^{-1} \begin{pmatrix}
            \hat{\mu}_{\alpha}^t \\ \hat{\mu}_{\eta}^{t}
        \end{pmatrix} - U^{\top} \begin{pmatrix}
            \mb{X}_0 \\ \mb{Y}_0
        \end{pmatrix} \right\rVert \le 10^{-18}\left \lVert U^{\top} \begin{pmatrix}
            \mb{X}_0 \\ \mb{Y}_0
        \end{pmatrix} \right\rVert \approx 10^{-12},
    \end{align*}
    where the final approximation varies from EM step to EM step but only very slightly.
    
    \item To generate a mean-zero random normal vector with covariance $\Sigma$, we use Lanczos quadrature \citep[cf.\ Chapter 10;][]{golub_2013}.
    That is, we instantiate a standard normal random vector $\tilde{\mathcal{W}}$ and then normalize it to $\mathcal{W} = \tilde{\mathcal{W}} / \| \tilde{\mathcal{W}} \|$.
    Then, we compute a tridiagonal decomposition 
    \begin{align*}
        \Sigma^{-1} = V T V^{\top}
    \end{align*}
    where $V = \mathrm{nrow}(\Sigma^{-1}) \times K$ is orthogonal with the first column being equal to $\mathcal{W}$ and $T$ is tridiagonal $K \times K$ matrix.
    We then compute 
    \begin{align*}
        \Sigma^{1/2} \mathcal{W} \approx V T^{-1/2} V^{\top} \mathcal{W} = V T^{1/2} e_1,
    \end{align*}
    where $e_1$ is a vector with 1 for the first entry and zeros everywhere else.
    Here $K$ needs to be chosen sufficiently large, (note that if $K = \mathrm{nrow}(\Sigma^{-1})$, then the procedure is exact).
    We stop if 
    \begin{align*}
        \| \mathcal{W}^{T} - \mathcal{W}^{T-10} \|_{\infty} \le 10^{-6}.
    \end{align*}
    
    \item We need to compute the marginal likelihood of $\mb{X}$ and $\mb{Y}$ (once again conditional on cluster memberships) for the importance sampling weights and for model selection.
    Write 
    \begin{align*}
        D_{X, Y} &= \textrm{diag}\left(D_X^{1/2}, \sigma_YI_{\sum_{i=1}^n n_i}\right),\\
        \Sigma_{\textrm{data}} = \mathbb{V}\textrm{ar}\left(\begin{matrix}\mb{X} \\  \mb{Y}\end{matrix} \middle \vert  \Omega \right) &= D_{X, Y}UAU^\top D_{X, Y} + D_{X, Y}^2.
    \end{align*} Finally, write $n_{\textrm{data}} = (K+1)\sum_{i=1}^n n_i$ as the number of total observed measurements.
    Now, let $\mb{X}_0$ and $\mb{Y}_0$ be mean-centered versions of the data, so that \begin{align*}
        \ell(\Omega, \mb{X}, \mb{Y}) = -\frac{1}{2}\left(\log(2\pi)n_{\textrm{data}} +\log(|\Sigma_{\textrm{data}}|)+ \left(   \mb{X}_0^\top ,
            \mb{Y}_0^\top\right) \Sigma_{\textrm{data}}^{-1} \begin{pmatrix} \mb{X}_0 \\ \mb{Y}_0\end{pmatrix}\right)
    \end{align*}where $|\Sigma_{\textrm{data}}|$ denotes the determinant of $\Sigma_{\textrm{data}}$.
    Using the matrix determinant lemma, \begin{align*}
        |\Sigma_{\textrm{data}}| &= \left|A^{-1} + U^\top  U\right| \left|A^{-1}\right|^{-1} \left|D_{X,Y}^2\right|.
    \end{align*}
    The determinant of $A^{-1}$ can be computed efficiently due to Vecchia's approximation and final determinant (sum of a diagonal matrix) can be computed quickly as well. 
    For the quadratic from, using the Woodbury matrix identity, 
    \begin{align*}
        \begin{pmatrix}
                \mb{X}_0\\
                \mb{Y}_0
        \end{pmatrix}^\top \Sigma_{\textrm{data}}^{-1} \begin{pmatrix}    \mb{X}_0\\
                \mb{Y}_0 \end{pmatrix}&= \begin{pmatrix}
            \mb{X}_0\\
                \mb{Y}_0
        \end{pmatrix}^\top \left(D_{X,Y}^{-2} - D_{X,Y}^{-1}  U\left(A^{-1} + U^\top U\right)^{-1} U^\top  D_{X,Y}^{-1}\right) \begin{pmatrix}    \mb{X}_0\\
                \mb{Y}_0 \end{pmatrix} \\
             &=\left\lVert D_{X,Y}^{-1}\begin{pmatrix}    \mb{X}_0\\
                \mb{Y}_0 \end{pmatrix}\right\rVert^2 -  \mb{V}^\top\left(A^{-1} + U^\top U\right)^{-1} \mb{V},
    \end{align*} 
    where  
    $$
        \mb{V}=U^\top  D_{X,Y}^{-1} \begin{pmatrix}    \mb{X}_0\\
            \mb{Y}_0 \end{pmatrix}.
    $$
    The first term can be quickly computed.
    For the second term, we once again use Lanczos quadrature using the algorithm described in Chapter 9.3 in \cite{golub_2013} for quadratic forms. 
    We also use this approach to estimate the determinant of $A^{-1} + U^\top U$ combined with Hutchinson's trick \citep{hutchinson_trace}.
    
    After tailoring this likelihood computation to a particular clustering $\mb{Z}$, this solves the evaluation of $f(\mb{X}, \mb{Y} | \mb{Z})$ in the importance sampling weights.
    Evaluation of $f(\mb{X}_i, \mb{Y}_i | Z_i)$ can be done similarly.
\end{enumerate}

These three computational derivations allow working with matrices the size of the total number of principal component scores, rather than the total number of data points. 

\section{Maximization Step Updates}\label{sec:m_step_updates}

In this section, we outline the parameter updates during the M-step.
Throughout, we will define $\smash{\zeta_{i,g}^{(\tau)} = \bar{w}^{(\tau)}\mathbb{I}\big(Z_i^{(\tau)} = g\big)}$, and $\mathfrak{P}$ denote the relevant penalty matrices. Since we use self-normalizing weights, we have $\sum_{\tau = 1}^T \bar{w}^{(\tau)} = T$, which are occasionally used interchangeably in the ensuing formulas. 
We denote the sampled predictor scores for the $i$-th profile from the $\tau$-th Monte Carlo iteration as $\smash{\mb{\alpha}_{i}^{(\tau)} = \big(\alpha_{i, 1}^{(\tau)},\dots, \alpha_{i,Q_1}^{(\tau)}\big)^\top}$, which correspond to principal components from the cluster $Z_{i}^{(\tau)}$, and likewise for $\mb{\eta}_i^{(\tau)}$.
\begin{itemize}
    \item Use the notation for $\delta(t)$ defined in \eqref{eq:delta_mean_cov}, and let $\Upsilon^g_{\mathcal{Y}} = [\Upsilon_{\mathcal{Y}, r}^g]_{r=1}^{2R + 1}$ and $\Upsilon^g_{\mathcal{X}} = [\Upsilon_{\mathcal{X}, r}^g]_{r=1}^{2R + 1}$ be vectors. The means for the response are then updated via 
    \begin{align*}
        \Upsilon^g_{\mathcal{Y}} &= \left[ \sum_{\tau = 1}^T\sum_{i=1}^n  \zeta_{i,g}^{(\tau)} (B_i \otimes \delta(t_i))^\top (B_i \otimes \delta(t_i)) +\lambda_{\theta_{\mathcal{Y}}}(\mathfrak{P}\otimes I_{2R+1})\right]^{-1} \\
         &~~~~~~~~~~~~\left[\sum_{\tau = 1}^T\sum_{i=1}^n \zeta_{i,g}^{(\tau)} (B_i \otimes \delta(t_i))^\top\big( \mb{Y}_i - B_i \Lambda^g \mb{\alpha}_i^{(\tau)} - B_i \Theta^g_{\mathcal{E}}\mb{\eta}_i^{(\tau)}\big)\right].
    \end{align*}
    The updates for the means of the covariates are similar:
    \begin{align*}
       \Upsilon^g_{\mathcal{X}} &= \left[ \sum_{\tau = 1}^T\sum_{i=1}^n  \zeta_{i,g}^{(\tau)} (\mb{B}_i\otimes \delta(t_i))^\top (\mb{B}_i \otimes \delta(t_i)) +\lambda_{\theta_{\mathcal{X}}}(\mathfrak{P}\otimes I_{2R+1})\right]^{-1} \\
       &~~~~~~~~~~~~~\left[\sum_{\tau = 1}^T\sum_{i=1}^n \zeta_{i,g}^{(\tau)}(\mb{B}_i\otimes \delta(t_i))^\top \big( \mb{X}_i  - \mb{B}_i \Theta^g_{\mathcal{X}}\mb{\alpha}_i^{(\tau)}\big)\right].
    \end{align*}
    
    \item  We update the parameters in the matrices defining the principal components sequentially by column.
    The update for the $l$-th column of $\Lambda^g$ is 
    \begin{align*}
        \Lambda^g_{l}&= \left[\sum_{\tau = 1}^T \sum_{i = 1}^n \zeta_{i,g}^{(\tau)}\big(\alpha_{i,l}^{(\tau)}\big)^2B_i^{\top}B_i + \lambda_{\Lambda}\mathfrak{P}\right]^{-1} \\
        &~~~~\left[\sum_{\tau = 1}^T \sum_{i = 1}^n\zeta_{i,g}^{(\tau)} \alpha_{i,l}^{(\tau)} B_i^\top \bigg(\mb{Y}_i - (B_i \otimes \delta(t_i)) \Upsilon^g_{\mathcal{Y}} - B_i \Theta^g_{\mathcal{E}}\mb{\eta}_i^{(\tau)} - \sum_{l^\prime \neq l}\alpha_{i, l^\prime}^{(\tau)}B_i \Lambda^g_l \bigg) \right].
    \end{align*}
    The update for the $l$-th column of $\Theta_{\mathcal{X}, g}$ is 
    \begin{align*}
        \Theta^g_{\mathcal{X}, l} &= \left[\sum_{\tau = 1}^T \sum_{i = 1}^n \zeta_{i,g}^{(\tau)}\left(\alpha_{i, l}^{(\tau)}\right)^2\mb{B}_i^{\top}D_{X_i}^{-1}\mb{B}_i+ \lambda_{\Theta_{\mathcal{X}}}\mathfrak{P}\right]^{-1} \\
        &~~~~\left[\sum_{\tau = 1}^T \sum_{i = 1}^n \zeta_{i,g}^{(\tau)} \alpha_{i, l}^{(\tau)} \mb{B}_i^{\top}D_{X_i}^{-1} \bigg(\mb{X}_i - (\mb{B}_i \otimes \delta(t_i))\Upsilon^g_{\mathcal{X}}- \sum_{l^\prime \neq l}\alpha_{i,l^\prime}^{(\tau)}\mb{B}_i (\Theta^g_{\mathcal{X}, l^\prime}) \bigg) \right].
    \end{align*}
    The update for the $j$-th column of $\Theta_{\mathcal{E},g}$ is 
    \begin{align*}
        \Theta^g_{\mathcal{E},l} &= \left[\sum_{\tau = 1}^T \sum_{i = 1}^n \zeta_{i,g}^{(\tau)}\left(\eta_{i, l}^{(\tau)}\right)^2B_i^{\top}B_i+ \lambda_{\Theta_{\mathcal{E}}}\mathfrak{P}\right]^{-1} \\
        & ~~~~\left[\sum_{\tau = 1}^T \sum_{i = 1}^n \zeta_{i,g}^{(\tau)}\eta_{i,l}^{(\tau)} B_i^{\top}\bigg(\mb{Y}_i - (B_i\otimes \delta(t_i)) \Upsilon^g_{\mathcal{Y}} - B_i \Lambda^g \mb{\alpha}_i^{(\tau)} - \sum_{l^\prime \neq l}\eta_{i, l^\prime}^{(\tau)}B_i (\Theta^g_{\mathcal{E}, l^\prime}) \bigg) \right].
    \end{align*}
    
    \item The update for the measurement error of the response is 
    \begin{align*}
        \sigma_{Y}^2 = \frac{1}{T\sum_{i=1}^n n_i} \sum_{\tau = 1}^T \sum_{i = 1}^n\sum_{g = 1}^G \zeta_{i,g}^{(\tau)} \left\lVert \mb{Y}_i -(B_i\otimes \delta(t_i)) \Upsilon^g_{\mathcal{Y}} - B_i \Lambda^g \mb{\alpha}_i^{(\tau)} - B_i \Theta^g_{\mathcal{E}} \mb{\eta}_i^{(\tau)}  \right\rVert^2.
    \end{align*}
    Furthermore, for $\Theta_{\mathcal{X}}^g$ and $\theta^g_{\mathcal{X}}$, let $\Theta_{\mathcal{X}, -T}^{g}$ and $\Upsilon_{\mathcal{X}, -T}^{g}$  denote the rows corresponding to temperature. 
    Then, defining $\sigma_{T}^2 = \Var(\epsilon_{11}^T)$, we obtain an update via
    \begin{align*}
        \sigma_{T}^2 = \frac{1}{T\sum_{i=1}^n n_i} \sum_{\tau = 1}^T \sum_{i = 1}^n\sum_{g = 1}^G \zeta_{i,g}^{(\tau)} \left\lVert \mb{T}_{i} - (B_i\otimes \delta(t_i))\Upsilon^g_{\mathcal{X}, -T} - B_i \Theta^g_{\mathcal{X},-T} \mb{\alpha}_i^{(\tau)}  \right\rVert^2
    \end{align*}
    where $\mb{T}_{i}$ are the measurements for the $i$th temperature profile.
    The corresponding update for salinity is similar.
    
    \item Let $\mb{\alpha}_{g,l}^{(\tau)}$ be the random vector of observations of the random field of the $l$-th principal component score for the covariates in cluster $g$ from the $\tau$-th sample from the importance sampling procedure.
    Let $R_{g, l, \alpha}^{(\tau)}$ denote the corresponding correlation matrix. 
    Furthermore, let $n_g^{(\tau)} = \sum_{i=1}^n \mathbb{I}\left(Z_i^{(\tau)} = g\right)$ denote the number of sites belonging to cluster $g$ for sample $Z^{(\tau)}$.
    Correspondingly, let $\sigma^2_{g,l,\alpha}$ denote the marginal variance of the $l$-th predictor score for the $g$-th group. 
    Then the update for $\sigma^2_{g,l,\alpha}$ is 
    \begin{align*}
        \sigma_{g,l,\alpha}^2 = \frac{1}{\sum_{\tau = 1}^T w^{(\tau)}}\sum_{\tau = 1}^T w^{(\tau)} \frac{1}{n_g^{(\tau)}}\left(\mb{\alpha}_{g,l}^{(\tau)}\right)^{\top} \left(R_{g, l, \alpha}^{(\tau)}\right)^{-1}\mb{\alpha}_{g,l}^{(\tau)}.
    \end{align*}
    
    The marginal variances of $\mb{\eta}$ for each group and principal component are updated in the same way. 
    
    \item The number and kind of spatial (spatio-temporal) correlation parameters depends on the choice of correlation function. 
    For the updates, we use numerical optimization via Fisher scoring as in \cite{guinness_gaussian_2021}.
    
    \item We update $\xi$ by numerically finding the root of the gradient of the pseudo-likelihood with respect to $\xi$, that is, if $d_{ij} = \mathrm{dist}(s_i, s_j)$
    \begin{align*}
        \frac{1}{T} \sum_{\tau = 1}^T \sum_{i=1}^n \sum_{g=1}^G \zeta_{i,g}^{(\tau)}\left( \sum_{j \in \partial i} Z_{jg}^{(\tau)}/d_{ij} - \frac{\sum_{g^* = 1}^G \left\{ \sum_{j \in \partial i} Z_{jg^*}^{(\tau)}/d_{ij}\right\}\textrm{exp}\left( \xi  \sum_{j \in \partial i} Z_{jg^*}^{(\tau)}/d_{ij}\right)}{\textrm{exp}\left(\xi  \sum_{j \in \partial i} Z_{jg^*}^{(\tau)}/d_{ij}\right)}\right)
    \end{align*}
    where $Z_{jg}^{(\tau)} = \mathbb{I}\left(Z_j^{(\tau)} = g\right)$.
\end{itemize}
Finally, an orthogonalization step is necessary to comply with the orthogonality constraints on the principal component functions. 
The details are outlined in Section \ref{sec:ortho_bsplines} next.

\section{Avoiding the Use of Orthogonal B-splines}\label{sec:ortho_bsplines}

In previous functional data work \citep[for example,][]{zhou_pca,liang_modeling_2020}, orthogonal B-splines are proposed for estimation of the principal components. 
First, for a B-spline basis $B(p)$ (in a row vector), a corresponding orthogonal B-spline basis can be written $\tilde{B}(p) = B(p) R^{-T}$, where $R^{-T} = (R^{-1})^T$ is a full rank matrix, with $RR^T$ is defined by the matrix of inner products $\int_0^{2000} B(p)^\top B(p) \textrm{d} p$.
Therefore, creating orthogonalized B-splines corresponds to an invertible change of basis. 
As a result, the estimated fitted functions from any penalized least squares estimate should be identical.
However, they may be represented by different bases. 

For both our work and work that uses orthogonalized B-splines, an orthonormalization step is used to ensure that the resulting functions are orthonormal. 
Consider the univariate setting, and following \cite{zhou_pca}, let $\hat{\Sigma}$ be the empirical covariance matrix of the scores. Also, let $\Theta$ be an estimated matrix of principal component functions for one group. Then the orthogonalization step is based on the eigendecomposition $
    R \Theta \hat{\Sigma} \Theta^\top R^\top = O D O^\top $ and $
    \Theta_{new} =R^{-1} O$ where $D$ is diagonal and $O$ is orthogonal. 
We can see that \begin{align*}
    \int \Theta_{new}^\top B(t)^\top B(t) \Theta_{new} \, \mathrm{d}t %&= O^\top R^{-\top} \int B(t) ,  B(t) \, \mathrm{d}t R^{-1} O \\
    &= O^\top R^{-\top}R^\top RR^{-1} O = I_{Q_2}
\end{align*}ensuring that the resulting functions are orthonormal.
When using orthogonal B-splines, the same step is taken, but in this case $\int_0^{2000} \tilde{B}(p)^\top \tilde{B}(p) \textrm{d} p$ and therefore $R$ are both the identity matrix. 

Regular B-splines reduces the number of non-zero entries required for one variable from approximately $P\sum_{i=1}^n n_i /2$ to $4 \sum_{i=1}^n n_i$, so we use them to obtain the same estimates.  
The complete orthonormalization step then consists of orthonormalizing separately $\Theta_{g,\mathcal{X}}$ and $\Theta_{g,\mathcal{Y}}$ for each $g$ as described above and as in \cite{zhou_pca}, while also adjusting $\Lambda_g$ appropriately. 

\section{Details for the Initialization} \label{seq:details_init}
Here we provide more details on the initialization procedure. 
First, we run $k$-means for interpolated values of the data to obtain an initial cluster assignment.
After this initialization, we use the model introduced in \cite{zhou_pca} extended to a Gaussian mixture model.
We now assume that the pairs $(\mathcal{X}_{s_i}, \mathcal{Y}_{s_i})$ are $n$ i.i.d.\ functional random variables.
Furthermore, we model the unobserved latent variables $\{Z_i\}$ as i.i.d. uniformly distributed on $\{1, \dots, G\}$.
Write
\begin{align*}
    \mathcal{Y}_i = \mu_{t_i,\mathcal{Y}}^{Z_i} + \sum_{l = 1}^{Q_1} \beta_{il} \phi_{Z_i, l}, \quad \mathcal{X}_i = \mu_{t_i,\mathcal{X}}^{Z_i} + \sum_{l= 1}^{Q_1} \alpha_{il} \psi_{Z_i , l}.
\end{align*}
Let $\mb{\alpha}_i = (\alpha_{i1}, \dots, \alpha_{iQ_1})$ and $\mb{\beta}_i = (\beta_{i1}, \dots, \beta_{iQ_1})$.
Furthermore, let $D_{Z_i, \alpha} = \Var(\mb{\alpha}_i \vert Z_i)$ and $D_{Z_i, \beta} = \Var(\mb{\beta}_i \vert Z_i)$ be the diagonal matrices of marginal variances of the principal component scores.
To model the correlation between $\mathcal{X}_i$ and $\mathcal{Y}_i$, we use a linear regression model for the principal component scores, that is 
\begin{align*}
    \left(\begin{matrix}
        \mb{\beta}_i \\ \mb{\alpha}_i 
    \end{matrix} ~ \middle\vert ~ Z_i\right)  \sim \mathcal{N}\left(0, \begin{pmatrix}
        D_{Z_i , \beta} & C_{Z_i}D_{Z_i ,\alpha} \\
        D_{Z_i ,\alpha}C_{Z_i}^{\top} & D_{Z_i , \alpha}
    \end{pmatrix}\right).
\end{align*}
Using the similar notation as in the main document, we approximate 
\begin{align*}
    (B(p_1), B(p_2)) \Theta^g_{\mathcal{X}, l} &\approx \psi_{g, l}(p_1, p_2)\quad 
    \gamma_{\mathcal{X},r}^g(p_1, p_2) = (B(p), B(p))\Upsilon_{\mathcal{X}, r}^g \\
    B(p) \Theta_{\mathcal{Y}, l}^g &\approx \phi_{g,l}(p)\quad 
    \gamma_{\mathcal{Y}, r}^g(p) = B(p)\Upsilon_{\mathcal{Y},r}^g.
\end{align*}
The parameters are then estimated similarly to the procedure in the main document, with the only change being that the MCEM procedure is simpler as all spatial correlation is ignored so that obtaining the samples $\big\{\mb{\alpha}^{(\tau)}_i, \mb{\beta}_i^{(\tau)}, Z_i^{(\tau)} \big\}_{i = 1}^n$ for $\tau = 1, \dots, n$ is straightforward.
Most M-step updates are analogous to the ones outlined in Section \ref{sec:m_step_updates}, and the M-step updates for $\{D_{g, \alpha}\}$, $\{D_{g, \beta}\}$ and $\{C_g\}_{g =1}^G$ can be found in \cite{zhou_pca}.

After estimating the above model by running the MCEM algorithm for a few iterations, we use the final estimates for the covariates as well as the mean estimate for the response of the independent model as initial values  for the spatio-temporal model.
Additionally, we set $\Lambda_{g} = \tilde{\Theta}_{g, \mathcal{Y}}C_g$ and 
\begin{align*}
    \Theta_{g, \mathcal{E}} = \tilde{\Theta}_{g, \mathcal{Y}} P.
\end{align*}
where $P$ is the matrix consisting the eigenvectors of the matrix $D_{g, \beta} - C_g D_{g,\alpha}C_{g}^{\top}$.

\section{Clustering Accuracy Simulation Study}

We test the performance of the Monte Carlo estimation strategy in a simulation study. 
The domains are $s_i \in [0,1]\times [0,1]$ and $p_{ij} \in [0,1]$, respectively. 
Locations $s_i$ are drawn uniformly on $\mathcal{S}$, and the $p_{ij}$'s are observed on a uniform grid. 
For the scores, we use an isotropic exponential covariance function $\mathbb{C}\textrm{ov}\left(\alpha_{g,l}(s), \alpha_{g,l}(s^\prime)\right) = \sigma^2_{g,\alpha, l} \textrm{exp}\left(- \left\lVert s - s^\prime \right\rVert/\kappa_{g,\alpha, l}\right)$ with $\sigma^2_{g,\alpha, l}$ and $\kappa_{g, \alpha,l} > 0$.
The study is based on 100 independently-simulated datasets.

To assess our clustering strategy, we consider a setting comparable to \cite{liang_modeling_2020}, who provide an extensive comparison of clustering methodologies for univariate functional data. 
We choose $G = 2$ groups, $3$ principal components per group, and $n = 200$ locations.
The mean was taken as the same for both groups as $\mu(p) = e^{2-2p}\cos(5(p - 0.1))$ (ignoring seasonality), while the $l$-th PC of the $g$-th group is taken as the $(g + 2l)$-th function of an orthogonal spline basis of dimension $11$ with uniformly-spaced knots. 
We set the measurement error variance to be $1.0$.
We use a five-nearest-neighbors graph to describe the dependence structure of the Markov random field and set $\xi = 0.5$.
The score variances were taken to be the same for both groups, with values of $1/3$, $1/6$, and $1/12$, respectively.
The spatial range parameters $\kappa_{g, l}$ are defined by the matrix: $
    \mb{\kappa} =  \begin{pmatrix}0.10 & 0.05 & 0.07  \\ 
    0.07 & 0.10 & 0.05\end{pmatrix} \in \mathbb{R}^{G \times Q_1}$.
We consider two settings for the number of measurements per location ($n_i = 20$ and $n_i = 100$).

We consider the Gibbs approach of \cite{liang_modeling_2020} with our importance sampling (IS) approach.
We take $G$ and the number of principal components (PCFs) as known. 
As \cite{liang_modeling_2020} compare their approach with others, we focus on theirs as a contrast.

\begin{figure}
    \centering
    \includegraphics[width = .9\textwidth]{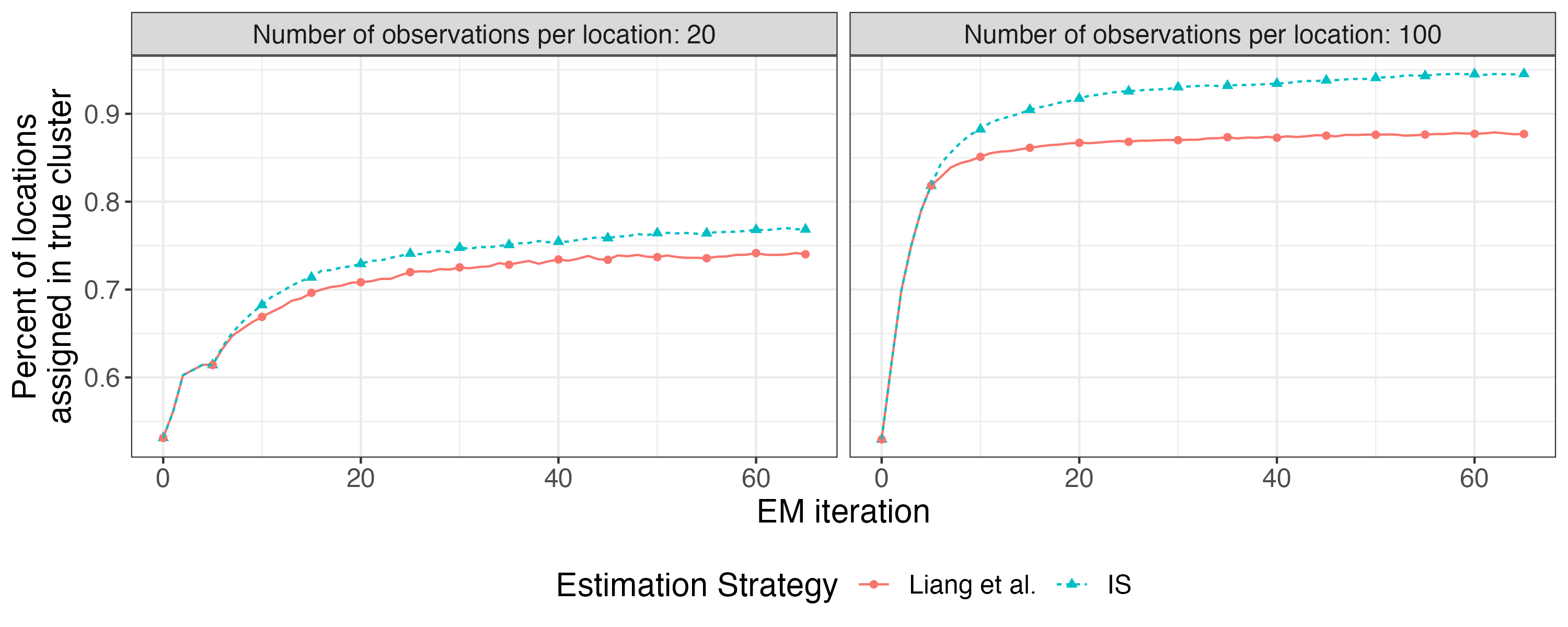}
    \caption{Percent of profiles correctly identified, averaged over simulations. The first 5 iterations are under the independent model for all models. IS refers to importance sampling.
    }
    \label{fig:simu_one_var}
\end{figure}

In Figure~\ref{fig:simu_one_var}, we plot the percent of locations whose cluster memberships are correctly identified averaged over the simulations. 
In the dense setting ($n_i = 100$), the Gibbs sampler from \cite{liang_modeling_2020} only slightly improves upon its initialization. 
This is because, across Gibbs samples and EM iterations, the sampled clusters do not change enough, indicating inadequate exploration of the set of possible cluster assignments. 
In contrast, importance sampling appears to perform comparatively well in both scenarios with significant improvements over the initialization. 
Thus, ignoring spatial correlation of the functional data in our proposal distribution appears to be appropriate. 
We have considered a variety of simulation settings (varying $G$, the type and number of PCFs, and $\mb{\kappa}$), and often Gibbs sampling performs well when $m_i < 50$, while it essentially does not work when $m_i \geq 75$. 

\bibliographystyle{apalike2}
\bibliography{cokriging.bib}